\documentclass[
preprint,    
amsmath,amssymb,
aps,
]{revtex4-2}
\usepackage{graphicx}
\usepackage{epstopdf,epsfig}
\usepackage{comment}
\usepackage{hyperref}
\usepackage{multirow}
\hypersetup{
    colorlinks = true,
    urlcolor   = blue,
    citecolor  = black,
}
\usepackage{latexsym}
\usepackage{float}
\usepackage{caption}
\usepackage{subcaption}

 \usepackage{adjustbox}
 \usepackage{vcell}
 \usepackage{bm}
 \usepackage{amsmath}
 \usepackage{amsfonts}
 \usepackage{amssymb}
 
\newcommand{\RomanNumeralCaps}[1]
\linenumbers

\def\bx{\bm{x}}
\def\bv{\bm{v}}
\def\barbv{\bar{\bm{v}}}
\def\bvprime{\bm{v}^\prime}
\def\barbomega{\bar{\bm{\omega}}}
\def\bomegaprime{\bm{\omega}^\prime}
\def\bomega{\bm{\omega}}
\def\bw{\bm{w}}
\def\bu{\bm{u}}
\def\barbu{\bar{\bm{u}}}
\def\buprime{\bm{u}^\prime}

\def\total{\mbox{d}}
\def\bcdot{\bm{\cdot}}
\def\btimes{\bm{\times}}
\def\bddot{\bm{:}}
\def\barv{\bar{v}}
\def\baromega{\bar{\omega}}
\def\bme{\bm{e}}
\def\bOmega{\bm{\Omega}}
\def\bnabla{\bm{\nabla}}
\def\barbOmega{\bar{\bm{\Omega}}}
\def\bOmegaprime{\bm{\Omega}^\prime}

\def\baprime{\bm{a}^\prime}
\def\balphaprime{\bm{\alpha}^\prime}
\def\bD{\bm{D}}
\def\bDo{\bm{D}^{\omega}}

\def\alphaprime{\alpha^\prime}
\def\bvast{\bm{v}^{\ast}}

\def\bk{\bm{k}}
\def\Re{\mbox{Re}}
\def\St{\mbox{St}}

\begin{document}

	\title{Dynamics of Particle-laden Turbulent Suspensions: Effect of particle roughness}
	
	\author{S. Ghosh} 
		\affiliation{Department of Chemical Engineering, 	Indian Institute of Technology Bombay, Mumbai 400 076, India }
	\author{V. Kumaran}
		\email{kumaran@iisc.ac.in}
		\affiliation{Department of Chemical Engineering, Indian Institute 	of Science, Bangalore 560 012, India}
	\author{P. S. Goswami}
		\email{psg@iitb.ac.in}
		\affiliation{Department of Chemical Engineering, 	Indian Institute of Technology Bombay, Mumbai 400 076, India }

\begin{abstract}

    The Fluctuating Force Fluctuating Torque (F3T) model is developed and evaluated for the dynamics of a 
    turbulent particle-gas suspension of rough spherical particles in a turbulent Couette flow in the limit where
    the viscous relaxation time of the particles and the time between collisions are much larger than
    the integral time for the fluid turbulence.
    The fluid force/torque exerted on the particles comprise a steady part due to the 
    difference in the particle velocity/angular velocity and the fluid mean velocity/rotation rate, and a fluctuating 
    part due to the turbulent velocity/vorticity fluctuations. The fluctuations are modeled as Gaussian white noise whose
    variance is determined from the fluid velocity and vorticity fluctuations.
    The smooth and rough inelastic collision models are considered for particle-particle and particle-wall 
    collisions. 
    The results show that inclusion of roughness is important for accurately predicting
    the particle dynamics; the second moments of the velocity fluctuations for rough particles are
    higher than those for smooth particles by a factor of 2-10, while the second moments of the angular
    velocity fluctuations are higher by 1-2 orders of magnitude. The F3T model quantitatively predicts
    the number density, mean and root mean square velocity and angular velocity profiles, and the distribution functions
    for the particle velocity and angular velocity, even though a Gaussian model is used for the highly non-Gaussian
    distributions for the force and  torque fluctuations.
\end{abstract}
	\maketitle
\section{Introduction}\label{sec:Introduction}
Particle-laden turbulent flows are an essential part of industrial applications in petrochemical industries, pneumatic conveying of solids in chemical and pharmaceutical industries, drying operations, and the motion of fuel droplets inside internal combustion engines. Such flows are also frequently encountered in natural processes like the formation of aerosols and the motion of pollutants in the air and ocean, and geophysical phenomena such as avalanches,
sand, and dust storms. Though the advent of high-speed computation has enabled the investigation of such flows through direct numerical simulation for small systems, the 
simulation of industrial-scale geometries is still not feasible. Therefore, it is essential to devise effective modeling
strategies for these flows. Particle-laden flows are composed of two phases, a turbulent gas phase and a particle phase which is dispersed in the gas. A strong two-way coupling exists between particles and the gas phase. In order to investigate the physics of the flow, stresses, heat transfer, and mass transfer, it is essential to examine the dynamics of both the phases and their coupling. In addition to the mean flow, the fluctuations in the fluid and particle phases play an important role in the dynamics. The particle velocity fluctuations are generated due to the 
turbulent fluctuations in the fluid and particle collisions. The fluid turbulent velocity fluctuations could
be augmented or damped due to the presence of the particles. The two-way interaction between the continuous gas phase and discrete particle phase enhances the complexity of the analysis.

At low mass loading, the presence of particle phase does not alter the fluid phase turbulence, but the turbulence strongly influences the particle dynamics \citep{gore1989effect, elghobashi1993two,louge1991role, FLM471218}. \citet{wang2019modulation} have studied the effect of particle inertia on turbulence modulation and on the turbulence regeneration cycle for particle laden turbulent Couette flow. \cite{ghosh2022dynamics} have demonstrated that there is insignificant modulation of turbulent fluctuations due to the presence of particle in a Couette flow when the mass loading is $ 0.2$ or less. Therefore, the particle dynamics can be modeled using one-way coupling at lower mass loading, where the effect of the turbulence on the particle dynamics is incorporated.
\citet{tanaka2008classification} introduced the dimensionless momentum number for representing turbulence modification. Through the compilation of experimental data, they found that turbulence attenuation takes place when particle momentum number is within a certain range; there is turbulence augmentation outside this range. 
\cite{moin_2020}  has found that the  dynamics of particles with low and moderate Stokes number  depends strongly on the complex spatio-temporal details of streaks, ejections, and sweeps in the near-wall region;
however, at higher Stokes number, the particle concentration depends on the variance of the particle wall normal velocity fluctuation. A similar observation has also been reported by \cite{goswami2010particle1} in their DNS study and in the experimental investigation of \cite{khalitov2003effect}. This indicates that the particle concentration distribution is uncorrelated with turbulent structures near the wall, and the effect of  turbulent fluctuation on the particle phase can be modeled as a random noise in the limit of high Stokes number. 

There are a large number of studies on the Direct Numerical Simulations (DNS) of particle laden 
flows.  The majority of these are focused on the dynamics of particle and fluid phases for Poiseuille flows \citep{vance2006properties, marchioli2008statistics, jie2022existence}.
\citet{subramaniam2018towards} provided a review of the modeling methodology in Eulerian-Lagrangian framework. If the particles are smaller than the Kolmogorov scale, the point-particle approximation is used in DNS (PP-DNS). The point-particle Eulerian-Lagrangian modeling methodology employs a suitable drag law, improved interphase mass, momentum and energy transfer models and incorporates inter-particle interaction.
\citet{garg2007accurate,garg2009numerically} have devised a technique for accurate prediction of interphase momentum transfer. \citet{horwitz2016accurate} proposed a correction for perturbed velocity field to calculate Stokes drag, in the absence of explicit inter-particle interaction. The effect of inter-particle interaction is incorporated in pairwise interaction extended point-particle or PIEP model by \citep{akiki2017pairwise}. The inclusion of inter-particle collision requires a suitable deterministic collision model for the particles.
\citet{elghobashi1994predicting} used a Stokes number based on large-eddy turnover time and particle volume fraction to parameterize turbulence. DNS is very accurate in predicting the dynamics of both the fluid and particle phases, because the governing equations are fully resolved down to the smallest turbulence scales. However, due to the high computational cost of DNS, suitable models that are less computationally intensive are required for particle laden flows.

For suspensions with particles of diameter 50 $\mu$m or higher, the modeling
is complicated by two factors. The first is the complexity of turbulent gas flow and the effect of turbulent fluctuating velocity on the particles. The second is the inertia of the particles and the fluctuations arising out of particle-particle collisions. For particles having smaller size (1-3 $\mu$m), the Stokes number (ratio of particle inertia to the fluid inertia) is small. This implies that the particle relaxation time is small compared to the characteristic flow time-scale, and the particles follow the local fluid streamlines. For larger particles of size ($\sim$50$\mu$m), the Stokes number becomes large, though the Reynolds number may not be relatively small. The drag force on the particles can be approximated by Stokes law
or a modified Stokes law that incorporates fluid inertia. The particles do not follow local streamlines at high Stokes number, but they move across streamlines because of their inertia. 
Particle velocity fluctuations are generated due to fluid turbulence and inter-particle collisions. If the correlation time for the fluid fluctuations is much smaller than the particle relaxation time, the net effect of these velocity fluctuations of the particle phase can be considered as a fluctuating force that is not correlated in time. The modeling of the fluctuating
force and torque on the particles (due to the fluid vorticity fluctuations), and the effect of particle roughness in 
collisions, is the focus of the present study. Since we are interested in the effect of the turbulent fluctuations
on the particle phase, one-way coupling is used here, where the effect of fluid fluctuations on the particles is 
included, but the effect of particle forces on the fluid turbulence is neglected.

The effect of energy dissipation due to inter-particle interactions has been extensively studied in the context 
of both continuum modelling and event-driven simulations for granular flows. The fluid forces are negligible compared to the instantaneous particle collisions for large particles of size greater than about 100$\mu$m or more in air. The analogy between motion of particles in granular material and the motion of ideal gas molecules has been made through the kinetic theory approach. Attempts were made to develop constitutive equations like the Chapman-Enskog procedure for hard sphere gases \citep{chapman1970mathematical}. The formulations of the balance laws and constitutive relations for smooth inelastic particles are of two types. The first is the modification of the Navier-Stokes equations for a fluid where the energy equation incorporates a term for the energy loss due to the inelastic collisions \citep{jenkins1983theory,lun1984kinetic,sela1996kinetic, sela1998hydrodynamic}, while the second type of formulation is based on the moment expansion model \citep{jenkins1985kinetic,kumaran1998vibro} where the higher moments of velocity was incorporated. There have been derivations of kinetic equations up to Burnett order starting from the Boltzmann equation using an expansion with the Knudsen number, and inelasticity of the particle collisions was incorporated as a small parameter \citep{sela1996kinetic,sela1998hydrodynamic}. \citet{FLM397872} derived a set of conservation equations and constitutive relations for the dynamic properties for rapid granular flow of slightly inelastic, slightly rough particles appropriate for both dense and dilute regime by taking moments of linear and angular velocities of the particles in transport equation. In this study, the collision rule for rough inelastic colliding spheres was formulated. \citet{luding1995granular} showed how behaviour of a system of rough inelastic spherical granular particles inside a vibrated two-dimensional box changes with the change of particle-particle and wall-particle frictional parameters. They introduced a collision model where the tangential co-efficient of restitution and Coulombic friction coefficient between two rough colliding spheres were related. \citet{kumaran2005kinetic} used normal and tangential co-efficient of restitution for analysing granular shear flow of rough inelastic spherical particles at high-Knudsen number limit where frequency of wall-particle collision is much higher than that of particle-particle collision. \citet{reddy2009structure} analysed the structure and dynamics of linear shear flow of rough inelastic disks at high area fractions using event-driven simulation of inelastic, rough, hard-particle collision. A stochastic modeling approach to describe the collision between the particle and isotropic and anisotropic rough wall has been reported \citep{konan2009stochastic,konan2011detached,radenkovic2018stochastic}.
\citet{nott2011boundary} derived boundary conditions at a rigid wall for a granular material consisting of rough inelastic spheres. Inelasticity and roughness were considered for particle-particle and particle-wall collisions.

The effect of the torque exerted by the particles on the fluid has been analysed. The shape of the particle and the local difference between the fluid vorticity and the particle angular velocity contribute to the net torque  exerted on
the particle by the fluid. \citet*{yin2003modelling} developed a model considering the non-spherical particles to simulate flow and combustion in biomass fired furnaces. The difference of fluid rotation rate at the particle center and particle angular velocity, and the non-spherical shape of the particles leading to the non-coincidence of the center of pressure and the center of mass were considered as the sources of torque acting on the particles. Through DNS simulations of Lagrangian point-particles in turbulent channel-flow, \citet*{andersson2012torque} incorporated the concept of torque-coupling along with the conventional two-way force-coupling. They observed significant reduction in the turbulence modulation in comparison with conventional two-way force-coupling results, for 5:1 prolate spheroids. \citet{zhao2011particle}, through their two-way coupled DNS simulation with torque coupling for spherical particles, investigated how the particles of different rotational inertia interact with local fluid field.

There have been relatively few studies on the effect of fluid phase velocity fluctuation on granular flow. \citet*{louge1991role} 
considered the particulate phase as heavy colliding grains with fluid drag acting on each particle, and investigated the effect of particle collision on turbulent suspension in a vertical pipe. They considered inter-particle collisions, and neglected the effect of fluid turbulent velocity fluctuations on the particles. \citet{kumaran1993properties,kumaran1993properties1} 
studied the effect of fluid drag on gas-particle suspension where the particles were settling in a fluid, and the turbulent velocity fluctuations were not present. \citet{tsao1995simple} 
studied the same for shear flow. \citet{kumaran2003stability} incorporated the turbulent velocity fluctuation in the kinetic theory framework by assuming the fluctuating velocity can be modelled as Gaussian white noise, and studied the effect of this fluctuations on the stability of the linear shear flow of a granular material. \citet*{variano2011rotational} modelled the Lagrangian autocorrelation function of angular velocity of particles in homogeneous, isotropic turbulence as exponentially decaying functions. They validated their model through experiments using Stereoscopic Particle Imaging Velocimetry (SPIV) and concluded that the angular velocity statistics are the same
as those obtained from the Ornstein-Uhlenbeck process.
\citet{goswami2010particle1}, in their work of Direct Numerical Simulation of particle laden turbulent channel flow at low Reynolds Number and high Stokes number, found that in presence of one-way coupling, the acceleration distribution of the particles can be represented as a Gaussian distribution in Stokesian regime. They concluded that particle phase velocity fluctuations are not correlated with fluid velocity fluctuation, and the effect of turbulent velocity fluctuation on the particle can be modelled as Gaussian white noise when the fluid correlation time is much smaller than the particle
relaxation time. Based on this, the Fluctuating Force Simulation was formulated \citep{goswami2010particle}. 
\citet*{choi2004lagrangian} studied the behaviour of the Lagrangian time scale in near wall region and analyzed the dependence of dispersion of fluid particle on its initial position for a turbulent channel flow. They also compared various interpolation schemes to determine the Lagrangian velocity and acceleration statistics along a fluid particle and studied Lagrangian velocity structure functions. \citet{squires1991lagrangian} studied the properties of dispersion tensor, eddy diffusivity tensor, Lagrangian and Eulerian autocorrelation function of velocities in decaying isotropic turbulence and in turbulent homogeneous shear flow. An extensive comparison of integral time scales  in Eulerian and Lagrangian reference frames was reported. 

The dynamics of the particle laden turbulent flows can be described either by an Eulerian-Eulerian framework or by a Eulerian-Lagrangian framework. In Eulerian-Eulerian or two-fluid approach, both the fluid phase and the particle phase are considered to be inter-penetrating continua. For high loading of the particles, when the integral fluid scale is higher than the inter-particle distance (small Knudsen number $\mbox{Kn}<0.1$), this approach is relevant. The Phase-space Probability Density Function (PDF) based approach as been reviewed by \citet{reeks2014transport,reeks2021development}. 
The main objective was to find out a suitable master-equation to account for the statistical nature of the processes e.g. transport, mixing, and deposition of particles in turbulent flow. There are two formalisms in the PDF approach. The first formalism, known as Kinetic Models (KM), consider the probability density of a particle having a certain velocity $\bm{v}$ and position $\bm{x}$ at a given time ($t$)  \citep{reeks1991kinetic,reeks1992continuum,buyevich1971statistical,buyevich1972statistical1,buyevich1972statistical2}. 
In the second formalism, known as the Generalised Langevin Models or GLM, the probability distribution function is expressed as a function of particle velocity, position, time and additionally velocity of the carrier phase flow local to a particle \citep{haworth1987pdf,simonin1993eulerian,minier2014guidelines}.
The Eulerian-Lagrangian framework involves tracking each particle of the dispersed phase individually and coupling those with the fluid phase descriptions in Eulerian grids.  Therefore, a \textcolor{black}{detailed} understanding of the fluctuating force and \textcolor{black}{the} fluctuating torque exerted on the particles is necessary to model the dynamics of the particle phase. Several investigations have been reported to characterize the fluid phase fluctuation through the analysis of probability density function (pdf) of velocity fluctuations. Many of those studies are for isotropic homogeneous turbulence \citep{andres2019statistics,bandi2009probability, rani2014stochastic, perrin2015relative, meyer2013rotational,  mathai2016translational}. 
 Particle-particle and wall-particle interactions were incorporated in the Fluctuating Force Simulation (FFS) developed by \citep{goswami2010particle} where particle-particle and particle-wall collisions are modeled using event-driven simulations, whereas the effect of fluctuating fluid velocity was modelled based on the Gaussian random white noise.

Here, the effect of particle roughness is incorporated into the Fluctuating Force Fluctuating Torque (F3T) model. 
Previous studies have considered the interaction between smooth particles, where there is no transmission
of particle momentum in the direction tangential to the surfaces at contact in particle-particle and particle-wall
collisions. Here, the angular velocity of the particles is incorporated as an additional dynamical variable, and the 
transmission of momentum tangential to the surface of contact is included using the rough particle collision model.
Whereas the fluctuating force due to the turbulent fluid velocity fluctuations was incorporated in previous studies,
the fluctuating torque on the particles due to the fluid vorticity fluctuations is also included in the present study.
We examine
whether particle roughness and the fluctuating torque causes a significant difference in the mean and mean square of
the particle velocities. Since the focus is on the particle statistics, a one-way coupled model is used here, where
the effect of fluid fluctuations on the particle phase is included, but the reverse effect of particle force and torque
on the fluid turbulence is not incorporated.

The next section focuses on the modeling of the fluctuating fluid velocity field and vorticity field on the particle
phase in a turbulent Couette flow. The particle-particle and particle-wall collisions are described here. The simulation procedure, and the process for deriving the amplitudes of the 
fluctuating force and torque from the fluid turbulence is discussed in section 3. The results for the variation in the 
particle statistics due to the particle roughness is the subject of section 4, and the main conclusions are presented
in section 5.


\section{Modelling Methodology}\label{sec:Modelling Methodolgy}
The effect of fluid turbulence on the particle fluctuations in a particle laden turbulent Couette flow 
at Reynolds number 750 was described by \citet{goswami2010particle} using the `fluctuating force' model. 
The particle volume fraction was in the range $10^{-4}-7 \times 10^{-4}$. 
The turbulence modification is small for this range of volume fractions (\citet{muramulla2020disruption}).
The particle relaxation time and collision time are comparable, and both are 
much larger than the fluid time scale. The force was modeled as Gaussian white noise
in the limit where the correlation time for the turbulent velocity fluctuations 
is much smaller than
the relaxation or collision times of the particles, and the variance of the noise distribution was
determined from the second moment of the fluid velocity fluctuations. One-way coupled simulations were
used, where the particle dynamics is affected by the turbulent fluctuations, but the fluid turbulence is not affected by the presence of the particles. The particle phase mean and fluctuating velocities, and the velocity distributions for smooth, elastic spherical particles were compared with results from Direct Numerical Simulations (DNS). Even though particle volume fractions were considered to be sufficiently low that the presence of particle phase does not change the turbulence intensity of the fluid phase, it was reported that the particle-particle interactions play a significant role in determining the distribution of particle velocities. 
In addition to the fluctuating force, we have introduced the fluctuating torque model to capture the effect of 
fluid vorticity fluctuations and collisions between rough particles on the particle velocity and angular velocity distributions. The fluctuating torque model is developed in the following sections. Two particle collision models are used for particle-particle and particle-wall collisions, the `smooth-particle' model where there is no transmission of momentum parallel to the surfaces at contact, and the two-parameter `rough particle' model where the relative velocity tangential to the surfaces of contact is $- \beta$
times that before collision, where $\beta$ is the tangential restitution coefficient.

\subsection{DNS simulations:}
\label{sec:Fluid}
Direct numerical simulation (DNS) is performed for determining the fluid turbulence statistics for
calculating the diffusion tensors in equations \ref{eq:diffvel} and \ref{eq:diffangvel}, and for
calculating the particle phase statistics for comparison with the F3T simulations. In the latter
calculation, the coupling is one-way, and the particle force and torque on the fluid is not 
included in the calculations.

\begin{figure}
\centering
	\includegraphics[width=1.0\textwidth]{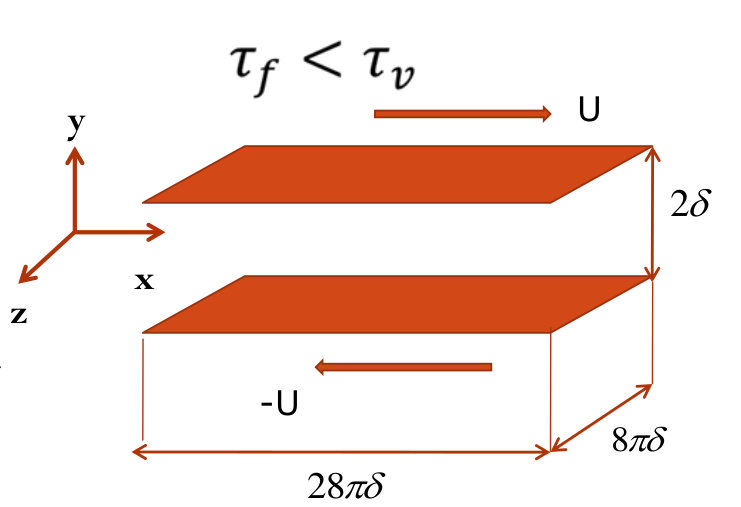}
\caption{Configuration and co-ordinate system for the turbulent Couette flow in a channel.}
	\label{fig:schematic}
\end{figure}

The simulation box, shown in figure \ref{fig:schematic}, is of dimension $10\pi\delta$  in the flow $(x)$ direction, $2\delta$ in the wall-normal ($y$) direction and $4\pi\delta$ along span-wise ($z$) 
direction, where $\delta$ is the half-width of the channel. The upper and the lower walls move in opposite directions
with velocities $+U$ and $-U$ respectively. The Reynolds number based on channel half-width $\delta$ and wall velocity $U$ ($Re_{\delta}$) is 750. The ratio of the wall velocity $U$ and the friction velocity $u_\ast$ is $(U/u_\ast) = 14.2$, and the ratio of the 
channel half-width and the wall unit is $(\rho \delta u_\ast/\mu) = 52.7$.
A pseudo-spectral method is used to solve the Navier-Stokes equation, where Fourier collocation is used in the $x$ and $z$ directions and Chebyshev collocation in the wall-normal $y$ direction. The primitive form of Navier-Stokes equation is solved using Kleiser–Schumann algorithm; the details of the DNS method and validations are reported in \cite{goswami2010particle1}.
The number of grids used in $x$, $y$, and $z$ directions are 120, 55, and 90 respectively, and the grid resolution is $13.64 (\mu/\rho u_\ast)$ and $7.79 (\mu/\rho u_\ast)$ in the stream-wise and span-wise
directions respectively. 
The grid spacing in the wall-normal direction varies from $1.69 \times 10^{-4} \delta$ at
the wall to $0.058\delta$ at the center, which ensures the
resolution is sufficient enough to capture the dissipation at the smallest scales.
The flow geometry and the coordinate system are shown in the figure \ref{fig:schematic}.

The mean and the mean square fluctuating velocity and vorticity of the fluid are compared with the DNS results of \citep{komminaho1996very} in appendix \ref{sec:validation}.

\subsection{Fluctuating Force Fluctuating Torque Simulation}
\label{sec:FFS}
In the earlier fluctuating force model, the force on the particle is separated into
two components, a deterministic component due to the difference between the particle
velocity and the fluid mean velocity, and a stochastic component due to the 
turbulent fluid velocity fluctuations. The latter is modeled as Gaussian white
noise in the limit where the correlation time for the fluid velocity fluctuations
is small compared to the viscous relaxation time or the collision time for the 
particles. The effect of the fluctuating force is modeled as a diffusion in velocity
space in the Boltzmann-Fokker-Planck description of the distribution function.
The diffusion coefficient is calculated from the velocity
autocorrelation function for the fluid velocity fluctuations. 

In the present model, in addition to the linear velocities, the angular velocities are included in the description of the particle dynamics. The torque on a particle consists of a 
deterministic component due to the mean vorticity of the fluid, and a Gaussian white
noise due to the effect of the fluid vorticity fluctuations. 
The particle velocity and angular velocity are expressed as the sum of the mean,
$\barbv$ and $\barbomega$, and the fluctuations about the mean, $\bvprime$
and $\bomegaprime$. The mean quantities, which are defined as time averages,
are functions of only the wall-normal co-ordinate $y$. The mean velocity is
in the stream-wise $x$ direction, $\barbv = \barv_x \bme_x$, and the mean 
angular velocity of the particles generated due to the mean shear in the $x-y$
plane is in the span-wise direction, $\barbomega = \baromega_z \bme_{z}$. Here,
$\bme_x$ and $\bme_z$ are the unit vectors in the $x$ and $z$ directions.

Linear approximations are used for the acceleration and angular acceleration of
the particle,
\begin{eqnarray}
 \frac{\total \bv}{\total t} & = & \mbox{} 
 \mbox{} - \frac{\bv - \bu}{\tau_v}, \label{eq:vel1} \\
 \frac{\total \bomega}{\total t} & = & \mbox{} 
 - \frac{\bomega - \tfrac{1}{2} \bOmega(\bx)}{\tau_r}, \label{eq:angvel1}
\end{eqnarray}
The relaxation times $\tau_v$ and $\tau_r$ are derived from the Stokes'
law for the drag force on a particle proportional to the difference in 
the particle and fluid velocity, and the torque on a rotating particle
for low particle Reynolds number,
\begin{eqnarray}
 m \frac{\total \bv}{\total t} & = & \mbox{} 
 \mbox{} - 3 \pi \mu d (\bv - \bu), \label{eq:vel2} \\
 I \frac{\total \bomega}{\total t} & = & \mbox{} 
 - \pi \mu d^3 (\bomega - \tfrac{1}{2} \bOmega(\bx)). \label{eq:angvel2} 
\end{eqnarray}
Here, $m$ and $I$ are the mass and moment of inertia of the particle
respectively. For a spherical particle with uniform mass density $\rho_p$,
the mass and moment of inertia are $m=(\pi \rho_p d^3/6)$ and $I = (m d^2/10) = 
(\pi \rho_p d^5/60)$. Therefore, the relaxation times expressed in terms
of the particle mass and diameter are,
\begin{eqnarray}
 \tau_v & = & \frac{m}{3 \pi \mu d} = \frac{\rho_p d^2}{18 \mu}, \label{eq:tauv} \\
  \tau_r & = & \frac{I}{\pi \mu d^3} = \frac{\rho_p d^2}{60 \mu}. \label{eq:taur}
\end{eqnarray}
Therefore, $\tau_r = 0.3 \tau_v$ for spherical particles for low particle
Reynolds number.

The Stokes number is the ratio of relaxation time and the flow time scale $(\delta/U)$ is,
\begin{eqnarray}
 \St & = & \frac{\tau_v}{(\delta/U)} = \frac{\rho_p d_p^2 U}{18 \mu \delta} = 
 \frac{\mbox{Re}}{18} \frac{\rho_p}{\rho} \left( \frac{d_p}{\delta} \right)^2, 
 \label{eq:st}
\end{eqnarray}
where $\rho$ is the fluid density and $\Re = (\rho U \delta/\mu)$ is the 
flow Reynolds number. The Stokes number is $50$ in the present simulations.

The fluid velocity $\bu$ and vorticity $\bOmega$ at the particle location are 
also expressed as the sum of the mean and fluctuating part, $\bu = \barbu +
\buprime$ and $\bOmega = \barbOmega + \bOmegaprime$. Equations 
\ref{eq:vel1} and \ref{eq:angvel1} are expressed as,
\begin{eqnarray}
 \frac{\total \bv}{\total t} & = & \mbox{} 
 \mbox{} - \frac{\bv - \barbu}{\tau_v} + \frac{\buprime}{\tau_v} =
 \mbox{} - \frac{\bv - \barbu}{\tau_v} + \baprime, \label{eq:vel} \\
 \frac{\total \bomega}{\total t} & = & \mbox{} 
 - \frac{\bomega - \tfrac{1}{2} \barbOmega(\bx)}{\tau_r} 
 + \frac{\bOmegaprime}{2 \tau_r} = \mbox{} - \frac{\bomega - \tfrac{1}{2} 
 \barbOmega(\bx)}{\tau_r} + \balphaprime. \label{eq:angvel}
 \end{eqnarray}
The acceleration is separated into two components, a deterministic component 
proportional to the difference between the particle velocity and the mean
fluid velocity $(\bv - \barbu)$, and a stochastic component $\baprime =
(\buprime/\tau_v)$, where $\buprime$ is the fluid turbulent velocity 
fluctuation. Here, it is assumed that the correlation time for the turbulent
velocity fluctuations is much smaller than the viscous relaxation time or
the collision time of the particles. In this case, a Langevin formulation
can be used \cite{vklangevin,pope_2000,goswami2010particle}, where the effect of the turbulent
fluctuations is modeled as Gaussian white noise with zero mean and second
moment related to the correlation in the fluctuating velocity,
\begin{eqnarray}
 \overline{a_i^\prime(t)} & = & 0,
 \nonumber \\
 \overline{a_i^\prime(t) a_j^\prime(t^\ast)} & = & 2 D_{ij} \delta(t-t^\ast),
 \label{eq:noisevel}
\end{eqnarray}
where indicial notation is used for the components of the vectors, and
the diffusion tensor $\bD$ is symmetric but not isotropic. The average $\overline{\cdot}$
is an average over the probability (Gaussian)
distribution of the acceleration $\baprime$. The diffusion $\bD$
tensor is related to the autocorrelation function for the fluid turbulent
velocity fluctuations,
\begin{eqnarray}
 D_{ij} & = & \frac{1}{\tau_v^2} \int_0^\infty \total t^\ast \overline{u_i^\prime(t^\ast)
 u_j^\prime(0)}. \label{eq:diffvel}
\end{eqnarray}
It should be noted that the diffusion tensor is a function of the cross-stream
$y$ position, because the intensity of the velocity fluctuations depends on
the cross-stream location. The diffusion tensor is anisotropic, that is, the 
diffusivities in the three co-ordinate directions are different. In addition, the component
$D_{xy}$ of the diffusion coefficient is non-zero, because the velocity
fluctuations in the stream-wise and cross-stream directions are correlated.

In the fluctuating force fluctuating torque model, the angular acceleration
$\balphaprime$ due to the fluid vorticity fluctuations is included in the 
equation \ref{eq:angvel} for the particle angular velocity, in a manner
similar to the acceleration in the equation \ref{eq:vel} for the particle velocity.
This is modeled as Gaussian white noise,
\begin{eqnarray}
 \overline{\alphaprime_i(t)} & = & 0,
 \nonumber \\
\overline{\alphaprime_i(t) \alphaprime_j(t^\ast)} & = & 2 D^{\omega}_{ij}
 \delta(t-t^\ast). \label{eq:noiseangvel}
\end{eqnarray}
The diffusion coefficient $\bD^{\omega}$ is related to the autocorrelation in the 
fluid turbulent vorticity fluctuations $\bOmegaprime$,
\begin{eqnarray}
 D^{\omega}_{ij} & = & \frac{1}{4 \tau_r^2} \int_0^\infty \total t^\ast 
 \overline{\Omega_i^\prime(t^\ast) \Omega_j^\prime(0)}.
\label{eq:diffangvel}
\end{eqnarray}
The diffusion tensor $\bD^{\omega}$, which is a function of the cross-stream $y$
co-ordinate, is symmetric and anisotropic. 

The `Langevin' equations, \ref{eq:vel} and \ref{eq:angvel}, are formally equivalent 
to the Boltzmann-Fokker-Planck formulation for the distribution function \citep{goswami2010particle1} for
the velocity and angular velocity fluctuations. The distribution function $f(\bx, \bvprime, 
\bomegaprime, t)$ is defined such that $n(\bx, t)
f(\bx, \bvprime, \bomegaprime, t) \total \bx \, \total \bvprime \, \total 
\bomegaprime$ is the 
number of particles in the differential volume $\total \bx \, \total \bvprime \,
\total \bomegaprime$, where $n(\bx, t)$ is the number density
of the particles at location $\bx$. The distribution function is normalised,
that is, the integral over the velocities and angular velocities of the particles
is $1$. The Boltzmann-Fokker-Planck equation for the distribution function is,
	\begin{eqnarray}
		\label{eq2.1}
		\frac{\partial (nf)}{\partial t} + (\barbv + \bvprime) \bcdot \bnabla (n f)
		- \frac{\total \barv_x}{\total y} \frac{\partial (v_y^\prime n f)}{\partial v_{x}^\prime} \hspace{1in} \mbox{} & & \nonumber \\ \mbox{} - \frac{1}{\tau_v} \bnabla_{\bvprime} \bcdot ((\barbv + \bvprime - \barbu) n f)  - \bD \bddot \bnabla_{\bvprime} \bnabla_{\bvprime} (n f) \hspace{.6in}  \mbox{} & & \mbox{} \nonumber \\ \mbox{} - \boxed{\frac{1}{\tau_r} \bnabla_{\bomegaprime} \bcdot ((\barbomega + \bomegaprime - \tfrac{1}{2}\barbOmega) n f)} - 
		\boxed{\bD^{\omega} \bddot \bnabla_{\bomegaprime} \bnabla_{\bomegaprime} (nf)} & = & \frac{\partial_c (n f)}{\partial{t}}. \label{eq:boltzmann}
\end{eqnarray}
The above equation without the terms in boxes is the Boltzmann-Fokker-Planck
equation for smooth particles, and the terms in boxes account for the effect of
particle rotation. The first term on the left is the time derivative of the 
distribution function. The second term on the left is the convection in the 
spatial co-ordinates due to the particle velocity. The third term on the left 
represents the effect of mean shear---when a particle moves in the cross-stream
direction at constant velocity, the fluctuating velocity changes because the 
mean velocity at the final and initial locations are different, and this results
in a change in the distribution function. The fourth term on the left is the 
change in the distribution function due to the deterministic acceleration of
the particle resulting from a difference in the particle velocity and the fluid
mean velocity, which is equivalent to the first term on the right in equation
\ref{eq:vel} in the Langevin description. The fifth term on the left is the diffusion
in the velocity co-ordinates due to the fluid turbulent velocity fluctuations
which are modeled as Gaussian white noise (last term on the right in \ref{eq:vel}).
The sixth and seventh (boxed) terms on the left are due to particle rotation,
and these were not included in the earlier fluctuating force model. The sixth term
on the left is due to the deterministic part of the angular acceleration (first term 
on the right in \ref{eq:angvel}) due to the difference between one half of the 
fluid mean vorticity and the particle angular velocity, 
while the seventh term on the left is due to the turbulent fluid vorticity fluctuations
(second term on the right in \ref{eq:angvel}). 

The term on the right in equation \ref{eq:boltzmann} is the collision integral due to rough 
inter-particle collisions, which is formally written as,
\begin{eqnarray}
\label{eq_collision_int}
\frac{\partial_c (nf)}{\partial{t}} & = & \tfrac{1}{4} n n^\ast (\sigma + \sigma^\ast)^2 
\int_{\bw \bcdot \bk < 0} \total \bk \int \total \bvast \int \total \bomega^\ast
\left[f(\bv_b^\prime, \bomega_b^\prime)f(\bv_b^\ast, \bomega_b^\ast)\right.
\nonumber \\ & & \mbox{} \left. - f(\bv^\prime, \bomega^\prime) f(\bv^\ast, 
\bomega^\ast) \right] \bw \bcdot \bk. \label{eq:collint}
\end{eqnarray}
Here, $n$ and $n^\ast$ are the number densities, $\sigma$ and $\sigma^\ast$ are the diameters
of the colliding particles; 
$\bv^\ast, \bomega^\ast$ are the velocity and angular velocity of
the second particle colliding with the particle with velocity and angular velocity
$(\bv^\prime, \bomega^\prime)$; $(\bv_b^\prime, \bomega_b^\prime)$ and 
$(\bv_b^\ast, \bomega_b^\ast)$ are the pre-collisional velocities of a pair of
particles such that the post-collisional velocities are $(\bv^\prime, 
\bomega^\prime)$ and $(\bv^\ast, \bomega^\ast)$ respectively; $\bw = \bv^\prime
- \bv^\ast$ is the velocity difference; $\bk$ is the unit vector along the 
line joining the centers of the particles at collision directed from particle
with velocity $(\bv^\ast, \bomega^\ast)$ to that with velocity $(\bv^\prime, 
\bomega^\prime)$, as shown in figure \ref{fig:collision}. 
Note that the integral over the solid angle $\total \bk$
in equation \ref{eq:collint} is carried out over $\bw \bcdot \bk < 0$, that is, when the particles approach prior to a collision. Here, we have considered the dilute limit for small volume fraction where the pair distribution function is $1$, and neglected the variation in the mean velocity across a distance comparable to the particle diameter. The first term on the right in the square brackets in \ref{eq:collint}
is the `gain' term for the differential volume $\total \bv^\prime \, \total \bomega^\prime$ in 
velocity space, where a collision between two particles with pre-collisional velocities 
$(\bv_b^\prime, \bomega_b^\prime)$ and $(\bv_b^\ast, \bomega_b^\ast)$ results in 
one particle having post-collisional velocity $(\bv^\prime, \bomega^\prime)$. The
second term in the square brackets in \ref{eq:collint} is the `loss' term for the differential
volume $\total \bv^\prime \, \total \bomega^\prime$ due to the collisional change
in velocity of a particle with velocity $(\bv^\prime, \bomega^\prime)$.

\subsection{Collision Rule for Rough Inelastic Hard Particles}
\label{sec:coll}

The smooth and rough particle collision models for the collision between hard spherical particles,
previously used in the kinetic theory of molecular gases \cite{pidduck,chapman1970mathematical}
and in kinetic theories for granular flows \cite{lun1984kinetic,kumaran2005kinetic,kumaran2006rough}.

 \begin{figure}
 \centering
 \includegraphics[width=0.4\textwidth]{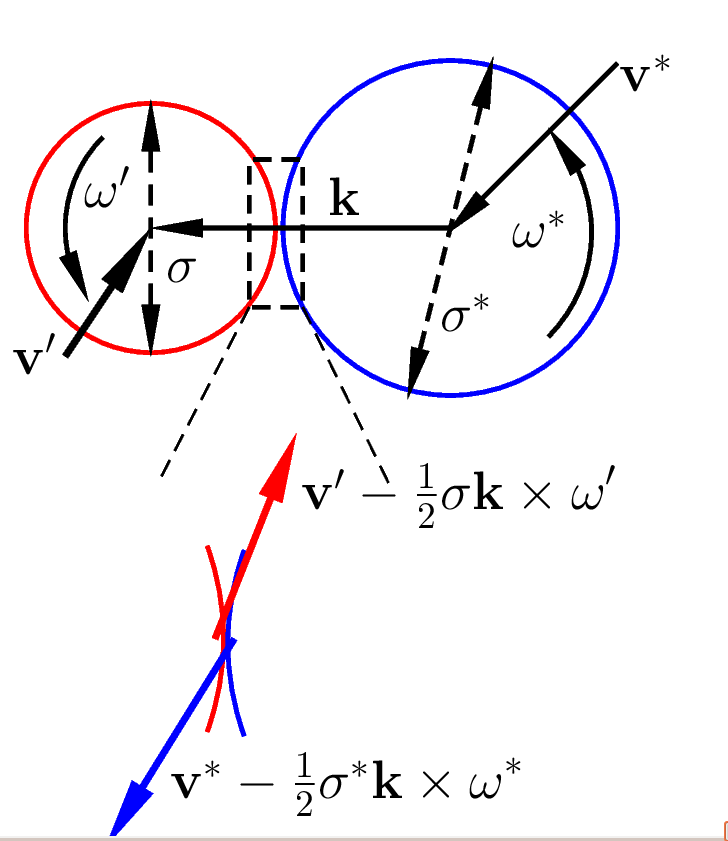}
 \caption{\label{fig:collision} Pre-collisional velocity and angular velocity for a binary 
  collision between two particles with diameters $\sigma$ and $\sigma^\ast$, velocities
  $\bm{v}^\prime$ and $\bm{v}^\ast$ and angular velocities $\bm{\omega}^\prime$ and 
   $\bm{\omega}^\ast$ respectively.}
  \end{figure}
 
Consider a collision between two spherical particles with diameter, mass, moment of inertia, velocity and angular velocity $\sigma, m, I, \bv^\prime,\bomega^\prime$ and $\sigma^\ast, m^\ast, I^\ast, \bv^\ast,\bomega^\ast$ respectively, as shown in figure \ref{fig:collision}. The pre-collisional relative velocity 
between the surfaces of the two particles at the point of contact, $\bm{g}$ is,
\begin{equation}
	\label{eq:collrule1}
	\bm{g} = \bv^\prime - \bv^\ast - \tfrac{1}{2} \bk \btimes (\sigma \bomega^\prime + \sigma^\ast \bomega^\ast).
\end{equation}
In a rough inelastic collision, the post- and pre-collisional
relative velocities, $\bm{g}_a$ and $\bm{g}$, are related as follows,
\begin{equation}
 \bm{g}_a \bcdot \bk = - e \bm{g} \bcdot \bk, \: \: \bm{g}_a \btimes \bk = - \beta \bm{g} \btimes \bk,
 \label{eq:collrule2}
\end{equation}
where $e$ and $\beta$ are the tangential and normal coefficients of restitution.
The post-collisional velocities $\bv_a^\prime$ and $\bv_a^\ast$ are 
\begin{equation}
 \bv_a^\prime = \bv^\prime - \frac{\bm{J}}{m}, \: \:
  \bv_a^\ast = \bv^\ast + \frac{\bm{J}}{m^\ast}, \label{eq:collrule3}
\end{equation}
and the post-collisional angular velocities $\bomega_a^\prime$ and $\bomega_a^\ast$ are,
\begin{equation}
 \bomega_a^\prime = \bomega^\prime - \frac{\sigma (\bk \btimes \bm{J})}{I}, \: \: 
  \bomega_a^\ast = \bomega^\ast - \frac{\sigma^\ast (\bk \btimes \bm{J})}{I^\ast},
  \label{eq:collrule4}
\end{equation}
where the impulse $\bm{J}$ is,
\begin{eqnarray}
 \bm{J} & = & \frac{m m^\ast}{m + m^\ast} \{ - (1+e) \bk \bk \bcdot (\bv^\prime - \bv^\ast)
 - \beta^\ast [(\bm{I} - \bk \bk) \bcdot (\bv^\prime-\bv^\ast)
 \nonumber \\ & & \mbox{} + \tfrac{1}{2} \bk \btimes
 (\sigma \bomega^\prime + \sigma^\ast \bomega^\ast)] \}, \label{eq:collrule5}
\end{eqnarray}
and $\beta^\ast$ is,
\begin{equation}
 \beta^\ast = \frac{(1+\beta) K}{1+K}, \: \: K = \frac{2(m + m^\ast)}{m m^\ast} 
 \left( \frac{\sigma^2}{2 I} + \frac{\sigma^{\ast 2}}{2 I^\ast} \right)^{-1}. \label{eq:collrule6}
\end{equation}

Since the particles are considered monodisperse in our simulations, $m=m^\ast$, $\sigma=
\sigma^\ast$ and $I = I^\ast$. For particle-wall collisions, we take the limit $m^\ast
\rightarrow \infty$, $I^\ast \rightarrow \infty$ and $\sigma^\ast \rightarrow \infty$
for the wall, and the unit vector $\bk$ is along the perpendicular to the wall.

\subsection{Particle simulation}
\label{simu_method}

A variable time step molecular dynamics procedure is used for solving the
equations for the particle linear and angular velocity, \ref{eq:vel} and \ref{eq:angvel}.
The change in velocity and angular velocity during particle collisions, which are considered 
instantaneous, are governed by equations \ref{eq:collrule1}-\ref{eq:collrule6}.
The time step ($\Delta t$) is set equal to the lesser of $0.025 (\delta/U)$ or $0.0005\tau_v$.
The collision times between pairs of approaching particles are calculated
assuming ballistic trajectories. 
If the collision time $\Delta t_c$ is shorter than the simulation time step $\Delta t$, the positions and velocities of all the particles are updated by the time $\Delta t_c$ up to the collision time. Then, the velocities and angular velocities of the colliding particles are changed in accordance with the collision rules \ref{eq:collrule1}-\ref{eq:collrule6}. The positions and velocities are then advanced by the reminder of the time step $\Delta t - \Delta t_c$. In order to speed up computation, neighbour lists are used to detect
particles which could potentially collide within a time step.

To model the effect of the turbulent flow field on the particles, a random force and torque has been applied on the particle for every fixed time step.  
The second moments of the random force and the torque are calculated  from the autocorrelation function of fluid velocity and vorticity fluctuations respectively in an Eulerian reference frame as discussed in the next section. The procedure for calculating the stochastic noise amplitude from the diffusion tensors $\bD$ and $\bDo$ is explained in the appendix.
It is to be noted that intensity of both the fluid velocity and vorticity fluctuations are functions of  wall-normal distance.


The velocity of each particle is intialized with the fluid velocity interpolated at the initial position of each particle. This ensures that the initial non-dimensional velocities are within the range of $-U$ to $+U$, which are the 
velocities of the bottom and top plates. The particle angular velocities are initialised randomly within the range of $\pm 1 (U/\delta)$. At each step, we  calculate the fluid drag on the 
particle due to the difference in particle instantaneous velocity and 
the local fluid velocity, and the torque due to the difference between the particle angular
velocity and one half of the local fluid vorticity. 
\subsection{Parameter regime}
The numerical values of the salient parameters are provided in table \ref{tab:table1}. The ratio of
densities is $2 \times 10^3$, which is a typical ratio for glass particles ($\rho_p = 2.5 \times 10^3 
\mbox{kg/m}^3$) settling in air ($\rho = 1.25 \mbox{kg/m}^3$). The volume fraction, $\phi = 10^{-4}$,
and the mass loading $0.2$ are sufficiently small that the turbulence modification is not
significant (\cite{muramulla2020disruption}).
The flow Reynolds number is 750, the friction velocity is $0.07$ times the plate velocity $U$, and
the friction Reynolds number is 52.7. The Kolmogorov scale is $0.014 \delta$, which is $\Re^{-3/4}$ 
times the height of the channel, $2 \delta$. The particle diameter, $0.0135 \delta$, is smaller 
than the Kolmogorov scale. The viscous relaxation time and the rotational relaxation time,
equations \ref{eq:tauv} and \ref{eq:taur}, are expressed in terms of the Reynolds number, the density
ratio and the ratio of the particle diameter and channel half-width. The flow time scale
is $(\delta/U)$. In section figures \ref{fig:e_mean_part_stats} and \ref{wall_beta_mean_part}, 
it is shown that the maximum 
difference between the particle velocity and the fluid velocity is about $0.5 U$. Based on 
this the particle Reynolds number $(\rho d_p (0.5 U)/\mu)$, which is the ratio of fluid inertia 
and viscosity on the particle scale, is about 5. This is much smaller than the Reynolds number of
about 24 for the formation of a separation bubble at the rear of the particle; the inertial 
correction to Stokes' law is about 30\% at $\Re_p = 5$. The particle Stokes number, which
is the ratio of the viscous relaxation time $\tau_v$ and the flow time $(\delta/U)$ is 50.
The rotational Stokes' number based on the rotational relaxation time $\tau_r$ is 15.
\begin{table}
 \begin{tabular}{lcr}
 Volume fraction & $\phi$ & $10^{-4}$ \\
   Density ratio & $(\rho_p/\rho)$ & $2 \times 10^3$ \\
  Flow Reynolds number $\Re$ & $(\rho U \delta/\mu)$ & 750 \\
  Friction velocity $u_\ast$ & $\sqrt{\tau_w/\rho}$ & $0.07 U$ \\
  Friction Reynolds number $\Re_\ast$ & $(\rho u_\ast \delta/\mu)$ & $52.7$\\
  Kolmogorov scale & $\Re^{-3/4} \times 2 \delta$ & $0.014 \delta$ \\
  Particle diameter & $d_p$ & $0.0135 \delta$ \\
  Particle diameter/wall unit & $(\rho d_p v_\ast/\mu)$ & $0.71$ \\
  Viscous relaxation time $\tau_v$ & $(\rho_p d_p^2/18 \mu)$ & \\
  Rotational relaxation time $\tau_r$ & $(\rho_p d_p^2/60 \mu)$ & \\
  Collision time $\tau_{pp}$ & $(n (\pi d_p^2) v_p^\prime)^{-1}$ & \\
  Particle Reynolds number $\Re_p$ & $(\rho_p d_p (0.5 U)/\mu)$ & 5 \\
  Flow time scale & $(\delta/U)$ & \\
  Stokes number $\St$ & $(\tau_v/(\delta/U))$ & $50$ \\
  Rotational Stokes number $\St_r$ & $(\tau_r/(\delta/U))$ & $15$ \\
 \end{tabular}
\caption{\label{tab:table1} Dimensional and dimensional parameters used in the 
present simulations. Here, $\rho$ and $\mu$ are the fluid density and viscosity,
$U$ is the plate velocity, $\delta$ is the channel half-width, $\tau_w$ is the wall
shear stress, $n$ is the particle number density, $\rho_p$ is the mass density of 
the material comprising the particles and $d_p$ is the particle 
diameter.}
\end{table}

The time between collisions of a inter-particle, commonly used in kinetic theories for
gases and granular flows, is $\tau_{p} \sim (n \pi d_p^2 v_p^\prime)^{-1}$, where $n$
is the number density of the particles. In a laminar flow,
particles will follow the fluid streamlines in the absence of gravitational and other forces, and they will 
collide only if the distance between particle centers is less than one particle diameter. The relative
velocity in this case is $v_p^\prime = (U d_p/\delta)$, the product of the strain rate $(U/\delta)$ and
the particle diameter $d_p$. In a turbulent flow, the magnitude of the particle velocity fluctuations
due to fluid turbulence can be estimated from equations \ref{eq:vel}, \ref{eq:noisevel} and \ref{eq:diffvel}.
The diffusion coefficient can be estimated as $D_{ij} \sim (u^\prime)^2  (\delta/U) /\tau_v^2$. When this is substituted
into \ref{eq:vel} and \ref{eq:noisevel}, the magnitude of the particle fluctuating velocity is 
$v_p^\prime \sim ((u^\prime)^2 (\delta/U)/\tau_v)^{1/2} \sim (u^\prime/\sqrt{\St})$. If the 
turbulent velocity fluctuations are comparable to the mean velocity, the relative
particle velocity due to turbulent fluctuations, $(U/\St^{1/2}) \sim 0.14 U$ is larger than
that due to the mean velocity gradient $(U d_p/\delta) \sim 0.0135 U$.

The time between collisions, scaled by the flow time scale $(\delta/U)$ are shown in figure 
\ref{fig:coll_freq_bar}. The particle-particle and particle-wall collision times are in the 
range $20-100 (\delta/U)$, which is comparable to the viscous relaxation time $50 (\delta/U)$.
Thus, the simulations are carried out in the regime where the collision time and the viscous
relaxation time are comparable, and both are much larger than the flow time scale $(\delta/U)$.
\section{Diffusion tensors}
To obtain the elements of translational and rotational diffusion tensors which are used in Fluctuating-Force-Fluctuating-Torque-Simulations (F3TS), we perform direct numerical simulation (DNS) for unladen fluid phase
without the reverse torque due to the particles. The diffusivities are used to compute the fluctuating random force and torque on the particles as described in the appendix. 



\begin{figure}
 \begin{subfigure}{0.48\textwidth}
  \includegraphics[width=1.0\textwidth]{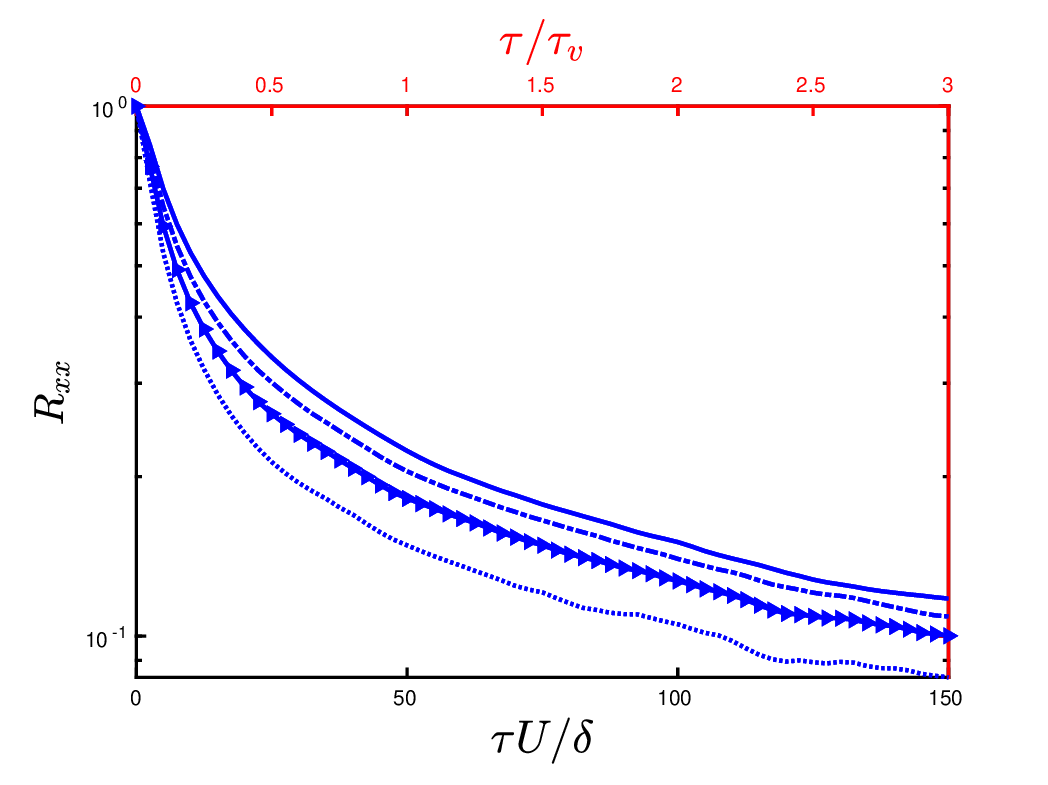}
    \caption*{(a)}
\end{subfigure}
\begin{subfigure}{0.48\textwidth}
 	\includegraphics[width=1.0\textwidth]{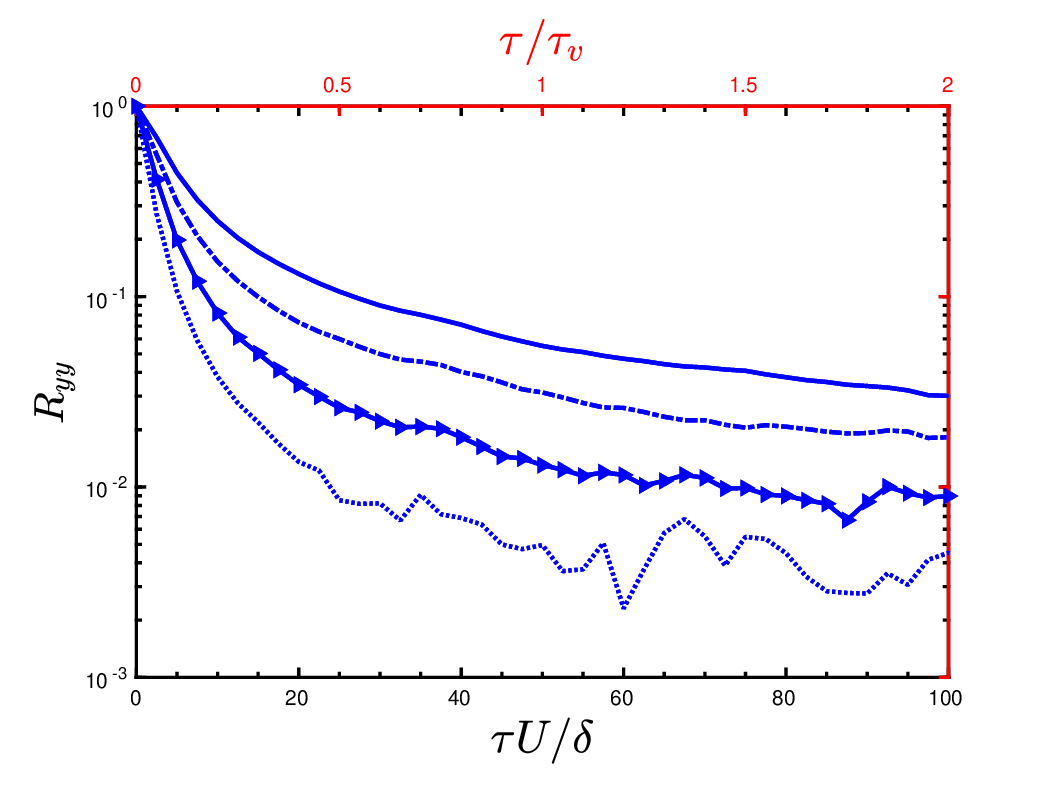}
 	\caption*{(b)}
 	\end{subfigure}
 	\begin{subfigure}{0.48\textwidth}
 	\includegraphics[width=1.0\textwidth]{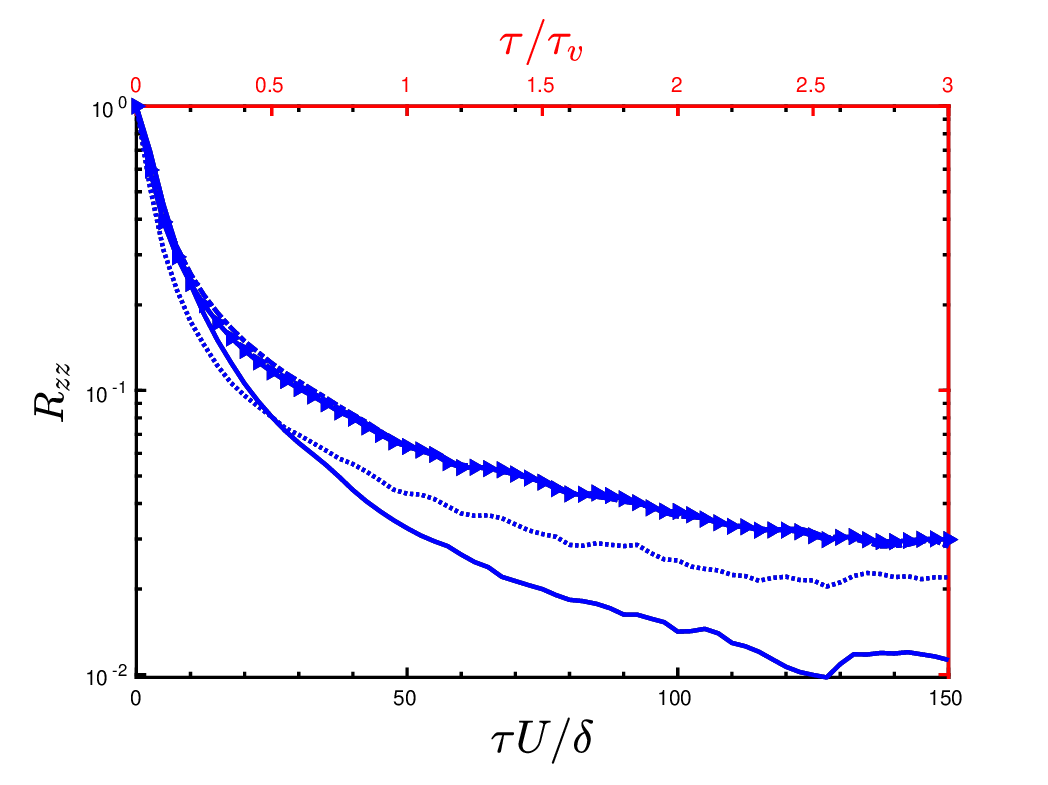}
  	\caption*{(c)}
  	\end{subfigure}
  \begin{subfigure}{0.48\textwidth}
  \includegraphics[width=1.0\textwidth]{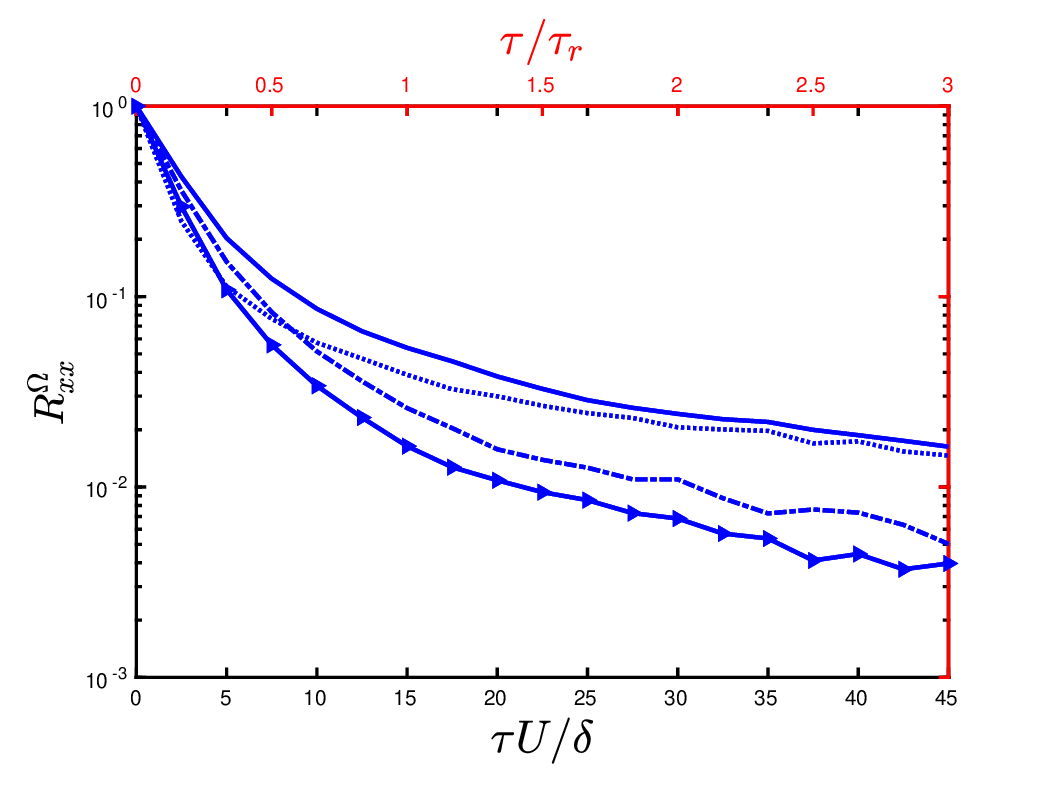}
    \caption*{(d)}
\end{subfigure}
\begin{subfigure}{0.48\textwidth}
 	\includegraphics[width=1.0\textwidth]{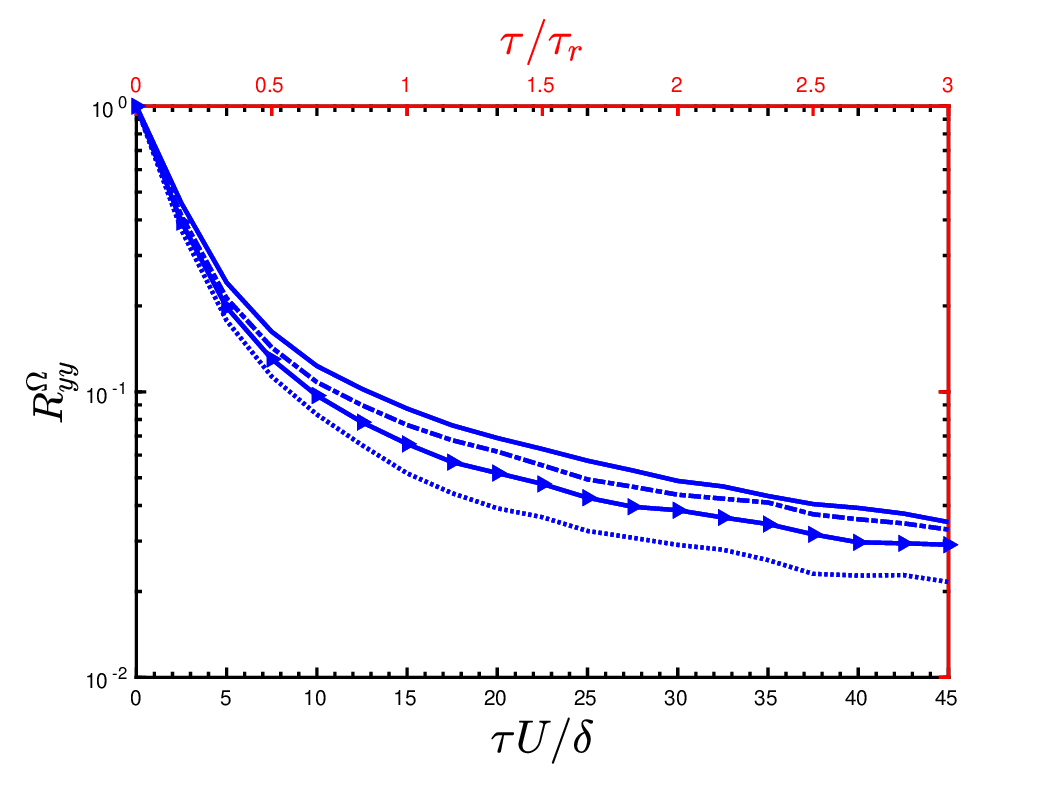}
 	\caption*{(e)}
 	\end{subfigure}
 	\begin{subfigure}{0.48\textwidth}
 	\includegraphics[width=1.0\textwidth]{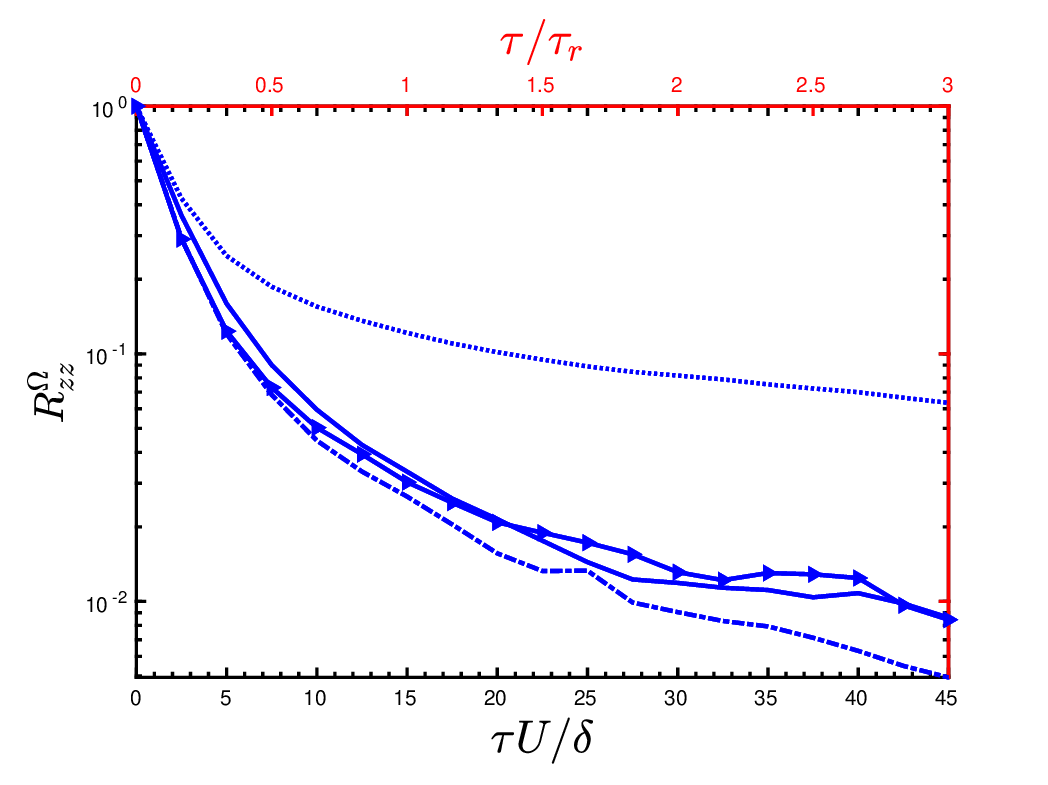}
  	\caption*{(f)}
  	\end{subfigure}
 	\caption{Zone-wise average decay of fluid velocity auto-correlation (\textit{a}) $R_{xx}$, (\textit{b}) $R_{yy}$, (\textit{c}) $R_{zz}$ \protect and zone-wise average decay of vorticity auto-correlation (\textit{d}) $R^\Omega_{xx}$, (\textit{e}) $R^\Omega_{yy}$, (\textit{f}) $R^\Omega_{zz}$ determined from the DNS simulations that resolve all
 	the turbulence scales. Time is scaled by the flow time scale $(\delta/U)$ in the bottom abscissa, and the equivalent time scaled by 
 	$\tau_v$ or $\tau_r$ is shown in the top abscissa for $\St = 50$ and $\St_r = 15$.
The zones are, (i) $0.1 < y/\delta < 0.2$ ($\cdots$), (ii) $0.2 < y/\delta < 0.4$ ($\triangleright$), (iii) $0.4 < y/\delta < 0.6$ ($\cdot$) and (iv) $0.6 < y/\delta < 1.0$ shown by ($-$). 
 	}
 	\label{fig:autocorr}    
 \end{figure}

The translational and rotational diffusion coefficients defined in \ref{eq:diffvel}-\ref{eq:diffangvel}
can be defined in terms of the fluid velocity autocorrelation functions $R_{ij}$, $R^\Omega_{ij}$,
figure \ref{fig:autocorr}, 
\begin{eqnarray}
	D_{ij}&=&\frac{\overline{ u'_i(0)u'_j(0)}}{\tau^2_v}\int_{0}^{\infty} \total t' \, 
	\frac{\overline{u'_i(t')u'_j(0)}}{\overline{ u'_i(0)u'_j(0)}}
    = \frac{\overline {u'_i(0)u'_j(0)}}{\tau^2_v}\int_{0}^{\infty}\total t' \, R_{ij}
	\label{eq3.1},
\end{eqnarray}
\begin{eqnarray}
D^\omega_{ij}& =& \frac{\overline{\omega_i^\prime (0)\omega_j^\prime (0)}}{4 \tau^2_r}\int_{0}^{\infty} 
\total t' \, \frac{\overline{\Omega_i^\prime (t') \Omega_j^\prime (0)}}{\overline{\Omega_i^\prime(0) \Omega_j^\prime(0)}}
= \frac{\overline{\Omega'_i(0)\Omega'_j(0)}}{4\tau^2_r}\int_{0}^{\infty} \total t' \, R^{\Omega}_{ij}.
	\label{eq3.2}
\end{eqnarray}
From symmetry, $D_{xz}$, $D_{yz}$, $D^{\omega}_{xz}$ and $D^{\omega}_{yz}$ are zero.

The autocorrelation functions can be defined in different reference frames,
\begin{enumerate}
 \item The Eulerian autocorrelation function, which is at a fixed point in space.
 \item The Lagrangian autocorrelation function in a reference frame situated on a particle.
\end{enumerate}
Though the force and torque exerted on the particles is captured by the Lagrangian autocorrelation
function, this is difficult to model because it does depend on the specific particle trajectory. 
In particular, the Lagrangian autocorrelation function depends on the viscous relaxation time of
the particle which determines the persistence of the particle velocity, and the time between
collisions which depends on the particle loading. The autocorrelations in the Eulerian reference
frame do not depend on particle properties or loading, making these more convenient for modeling.
When the fluid correlation time is much smaller than the viscous relaxation time of the particles,
the particle does not move very far before the fluid velocity fluctuations are decorrelated, and we 
would expect the Eulerian and Lagrangian autocorrelation functions to be similar. 
Here, we first examine the similarity between the Eulerian and Lagrangian
correlations before proceeding to the fluctuating force fluctuating torque simulations.

The temporal decay of the velocity and vorticity auto-correlation functions 
in an Eulerian reference frame at  different wall-normal positions ($y^+$) is 
shown in figure~\ref{fig:autocorr}. Here, the channel is divided into four zones,
from the wall to the center, $0.1 \delta < y < 0.2 \delta$, $0.2 \delta < y < 0.4 \delta$,
$0.4 \delta < y < 0.6 \delta$ and $0.6 \delta < y < \delta$. The auto-correlation functions are 
evaluated by averaging over 800 time frames 
with different time origins from unladen DNS run with 
a sampling time interval of $9.25 (\mu/\rho u_{\ast}^2)$. 
It is observed that the velocity auto-correlation function for stream wise velocity 
($R_{xx}$) at the center and the span-wise vorticity ($R^{\Omega}_{{zz}}$)
close to the wall are the slowest decaying autocorrelation functions. These autocorrelation 
functions decrease by an order of magnitude within a time comparable to 1.5-2 times the viscous
relaxation time of the particles. All the other components of the Eulerian autocorrelation
functions decrease by an order of magnitude within a time period which is a fraction of the 
particle viscous relaxation time.
In viscous sublayer, the slow decay of correlation function is an indicative of elongated span-wise
rolls in the high shear zone. An important finding in figure \ref{fig:autocorr} is that the decay
in $R_{xx}$ and $R^{\Omega}_{zz}$ is not well fitted by an exponential function, and
appears to have a stretched exponential form. The initial decay of the other autocorrelation functions are
all well fitted by an exponential form, and these decay within a time period shorter than the viscous
relaxation time. 

\begin{figure}		
	\includegraphics[width=0.49\textwidth]{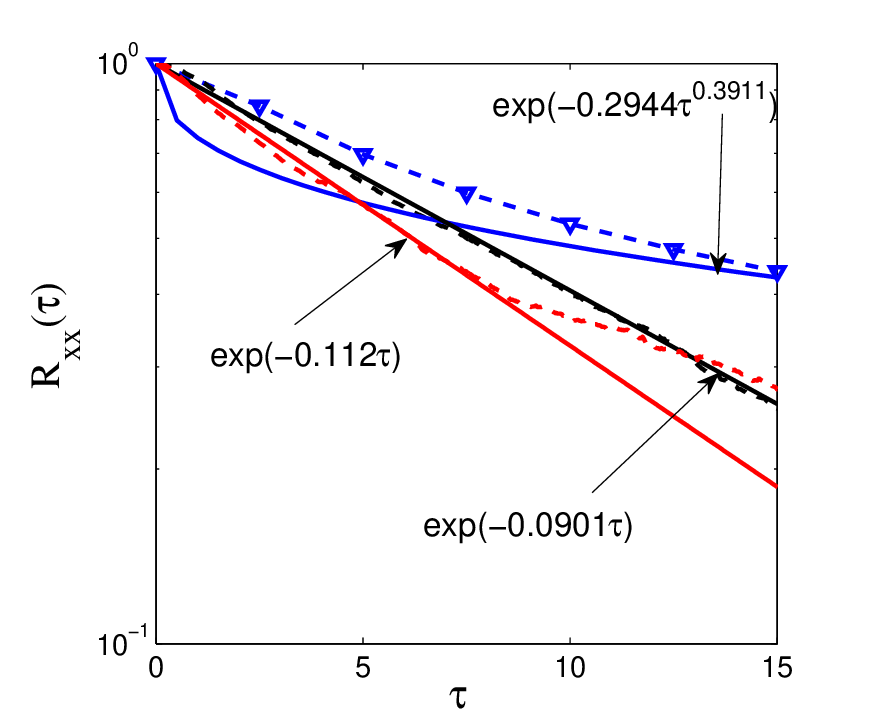} 
	\includegraphics[width=0.49\textwidth]{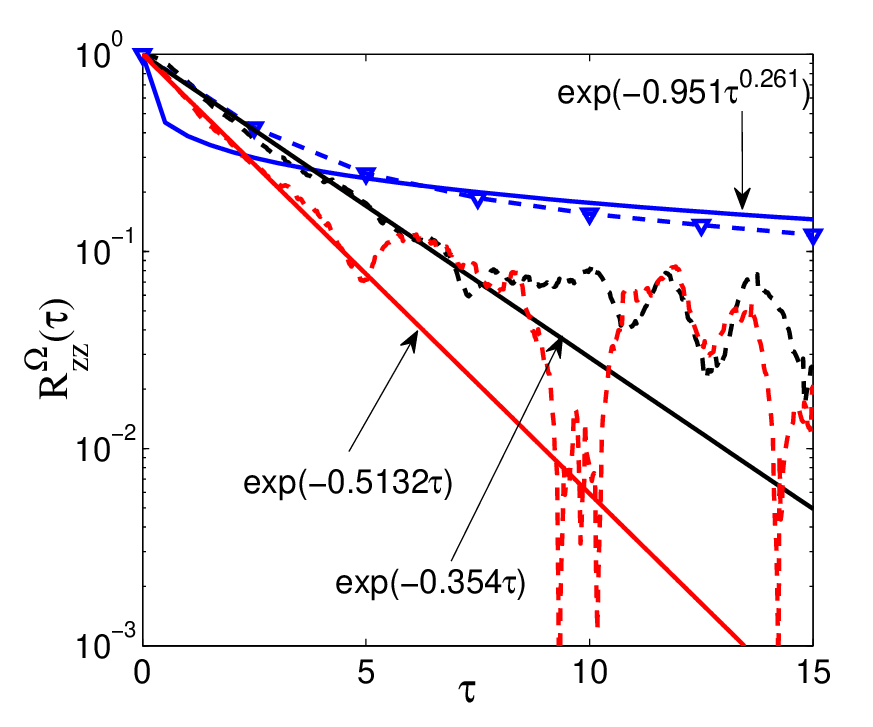}
	\caption{Comparison of zone-wise averaged fluid phase correlations (a) $R_{xx}$ in $0.6 < y/\delta < 1.0$ and (b) $R^{\Omega}_{zz}$ $0.1 < y/\delta < 0.2$ computed from DNS simulations that resolve all the turbulence scales in the Eulerian frame $\nabla$('--' blue), particle-Lagrangian frame for $\beta=-1.0$ ('--' black) and particle-Lagrangian frame for $\beta=+1.0$ ('--'red) with their corresponding fitting functions(thick lines).}	
	\label{fig:fluid_corr_fitting}
\end{figure}

Due to the slow decay of the Eulerian autocorrelation functions, $R_{xx}$ and $R^{\Omega}_{zz}$, it appears
that these can not be modeled as random Gaussian noise. However, the Lagrangian correlations in 
a reference frame moving with the particle do exhibit a rapid exponential decay, due to the particle motion. 
 This is in contrast to the Eulerian correlations, where $R_{xx}$ and $R^\Omega_{ zz}$ 
 in figure \ref{fig:autocorr} which
 could not be fitted with exponential functions, but required stretched exponential fits.
 This is expected; though the Eulerian autocorrelation function may decay very slowly, 
 the Lagrangian autocorrelation function in a reference frame moving with the particle,
 shown in figure \ref{fig:fluid_corr_fitting}, decay faster
 because of the particle motion. This is illustrated in figure \ref{fig:fluid_corr_fitting}, where the 
 Eulerian correlation at a fixed location is compared with the Lagrangian correlation 
 in a reference frame moving with the particles for smooth and rough particles across 
 two different zones in the channel.
The solid lines in these figures are exponential fits to the initial decay of the autocorrelation
functions. 
For the autocorrelation function $R_{xx}$ \ref{fig:fluid_corr_fitting}(a), the decay of 
the Lagrangian autocorrelation function is much faster than that of the Eulerian autocorrelation function at the center of the channel, and the initial decay is well fit by an exponential form. 
The decay of the Lagrangian span-wise vorticity autocorrelation function in \ref{fig:fluid_corr_fitting}(b) is noisy, in comparison to the Eulerian correlation function 
which is relatively smooth. However, the initial decay of the Lagrangian span-wise vorticity correlation is faster than its Eulerian counter-part, and it is well fit by an exponential form.

\begin{figure}
	\centerline{\includegraphics[width=1.2\textwidth]{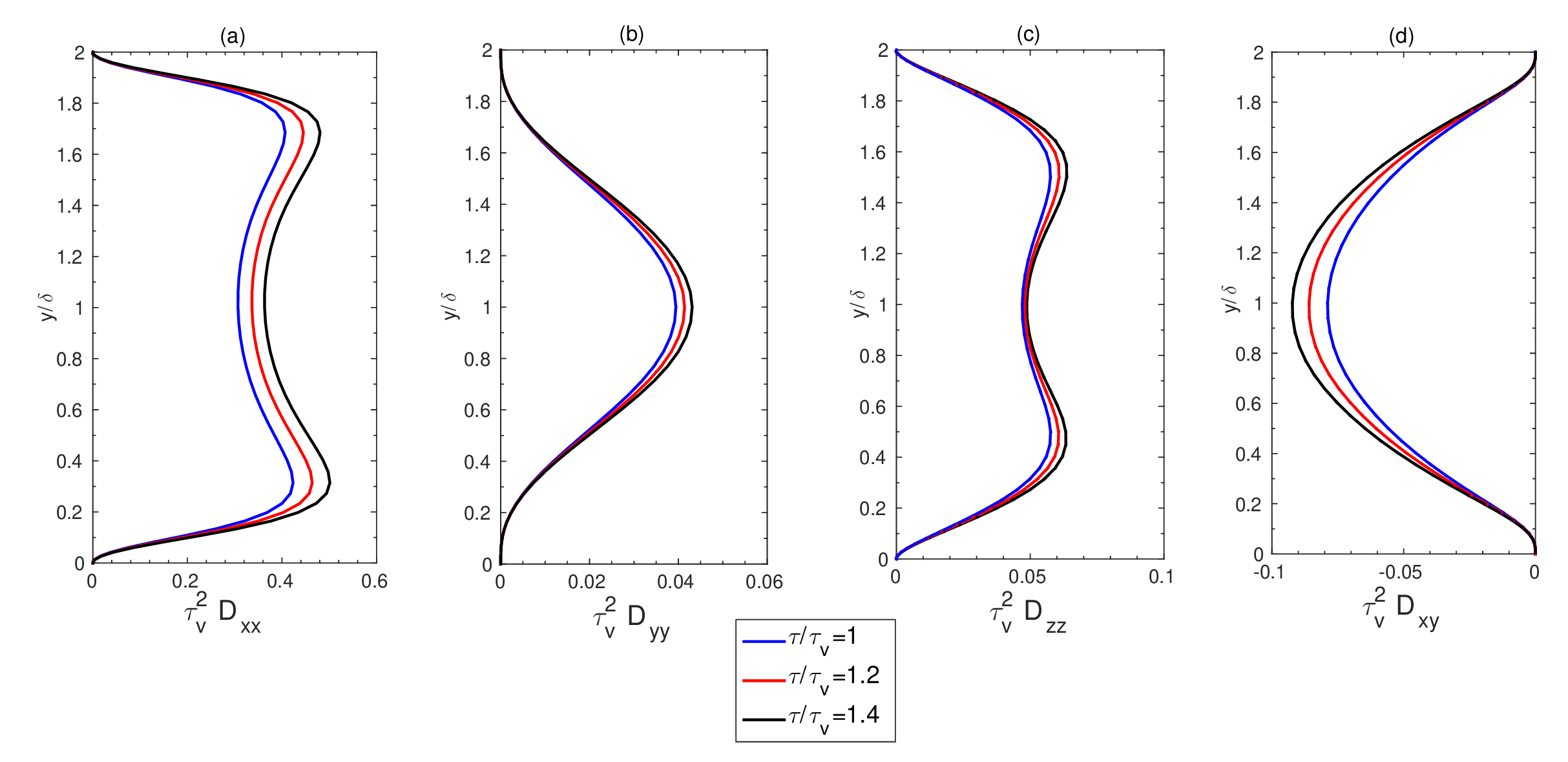}}
	\caption{Variation of translational diffusivity (multiplied by square of viscous relaxation time) with channel-width ${y/\delta}$ (a) ${\tau_v^2D_{xx}}$ (b) ${\tau_v^2D_{yy}}$ (c) ${\tau_v^2D_{zz}}$ (d) ${\tau_v^2D_{xy}}$ evaluated over
	different time intervals $(\tau / \tau_v) = 1$ (blue), $(\tau/\tau_v) = 1.2$ (red) and $(\tau / \tau_v) = 1.4$ (black).
}
	\label{fig:vel_diff_at_tau}
\end{figure}
\begin{figure}
	\centerline{\includegraphics[width=1.1\textwidth]{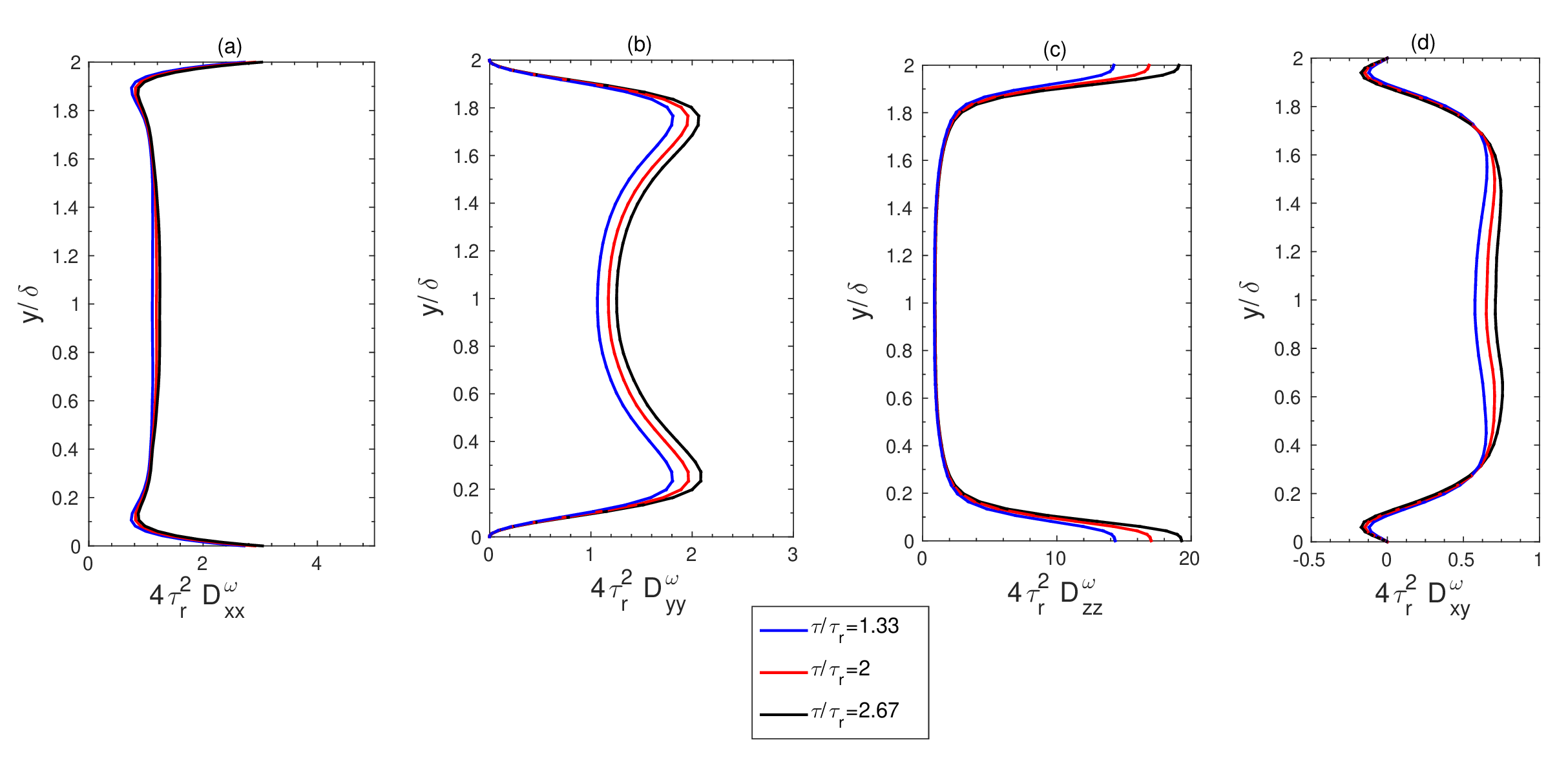}}
	\caption{Variation of Rotational Diffusivities (multiplied by square of rotational relaxation time) with channel-width ${y/\delta}$ (a) ${4\tau_r^2 D^{\omega}_{xx}}$ (b) ${4\tau_r^2 D^{\omega}_{yy}}$ (c) ${4\tau_r^2 D^{\omega}_{zz}}$ (d) ${4\tau_r^2D^{\omega}_{xy}}$ evaluated at different correlation time intervals $(\tau / \tau_v) = 1.33$ (blue), $(\tau/\tau_v) = 2.00$ (red) and $(\tau / \tau_v) = 2.67$ (black).
}
	\label{fig:rot_diff_at_tau}
\end{figure}

An important issue is the upper cut-off used for the time integrals in equations ~\ref{eq3.1} \& \ref {eq3.2}. In case of slowly decaying Lagrangian correlation functions, \citet{squires1991lagrangian} considered the integral time as the time for which correlation function decays to the  value of $\mbox{e}^{-1}$. In order to estimate the Lagrangian acceleration decorrelation time, \citet{goswami2010particle1} used a estimation of integral time scale by taking a value 
more than 7 times the value at which the velocity auto-correlation function decays to $\mbox{e}^{-1}$.
In a similar approach, in the present study, the upper limit of integration for velocity auto-correlation is set to $\tau=185$ wall units
($50 (\delta/U))$ or $1.0 \tau_v$, which is about 7 times the value at which correlation function decays to $\mbox{e}^{-1}$.  
For the vorticity correlation function, the upper limit is 
set to to $\tau=74$ wall units, which is beyond 7 times the value at which correlation function
decays to $\mbox{e}^{-1}$ and equivalent to $1.33\tau_r$. The effect of variation in the upper limit of integration on magnitude of the diffusivity is shown in figures~\ref{fig:vel_diff_at_tau} and \ref{fig:rot_diff_at_tau}. 
The translational and rotational diffusivities computed with three integration, times which
are 6 and 7 times the time taken for the correlation functions $R_{ij}$
and $R^{\Omega}_{ij}$ to decrease to $\mbox{e}^{-1}$,
are  shown in figures~\ref{fig:vel_diff_at_tau} and \ref {fig:rot_diff_at_tau}. 
It is evident from figure~\ref{fig:vel_diff_at_tau} that there is a larger variation in the stream wise diffusion, by about 10\%, when the upper limit of integration is increased from $\tau=222$ to  $\tau=259$. Similarly, it is also observed from figures~\ref{fig:rot_diff_at_tau} (c) 
that the near wall values of ${D^{\Omega}_{zz}}$ 
increases by about 11\% when upper limit of the integration is changed from  $\tau=111$ to  $\tau=148$. 


In figures \ref{fig:vel_diff_at_tau} and \ref{fig:rot_diff_at_tau}, the diffusivities were
calculated by integrating the autocorrelation functions different cross-stream locations.
The diffusivities can also be modeled as the product of the equal-time correlation and an
integral time scale $\tau_{ij}^{I}$ and $\tau_{ij}^{\Omega}$,
From equation \ref{eq3.1}, the translational diffusivity $D_{ij}$ is related to the corresponding average integral time-scale, $\tau^I_{ij}$,
\begin{equation}
       D_{ij}\tau_v^2=\overline{u_i'(0)u_j'(0)}
       \tau^I_{ij}.
    \label{sra1}
\end{equation}
Similarly, from equation \ref{eq3.2}, the rotational diffusivity $D^\omega _{ij}$ is related to the corresponding average integral time-scale for rotation ( $\tau^\Omega_{ij}$),
\begin{equation}
    D^\omega _{ij}\tau_r^2=\frac{1}{4}\overline{\Omega_i'(0)\Omega_j'(0)} \tau^\Omega_ {ij}.
    \label{sra2}
\end{equation}
There is a factor $\tfrac{1}{4}$ in the above equation because the fluid rotation rate is one half of the vorticity.
It should be noted that the integral times are anisotropic, and the suffix 'ij' carries the information on directionality of the velocity and vorticity fluctuations. The integral times in different zones in different reference frames 
in the channel are tabulated in
tables \ref{table:1} and \ref{table:2} of in appendix \ref{sec:integraltimes}. These tables show that there is relatively little variation in the integral time with cross-stream distance. Table \ref{table:1} shows the the Eulerian and Lagrangian integral times are comparable even when a stretched exponential fitting has to be used for the autocorrelation function. In comparison, the mean square of the fluctuating velocity and vorticity in figures \ref{fig:vel_diff_at_tau} and 
\ref{fig:rot_diff_at_tau} exhibit a much larger variation with cross-stream distance. Therefore, the average of the Eulerian integral time that is independent of the cross-stream location, but is dependent on the direction of the velocity/vorticity fluctuations, is used in 
equations \ref{sra1} and \ref{sra2} to calculate the diffusivity. 
%
%

\begin{figure}	
\begin{subfigure}{0.49\textwidth}
\centering
\includegraphics[width=1.0\textwidth]{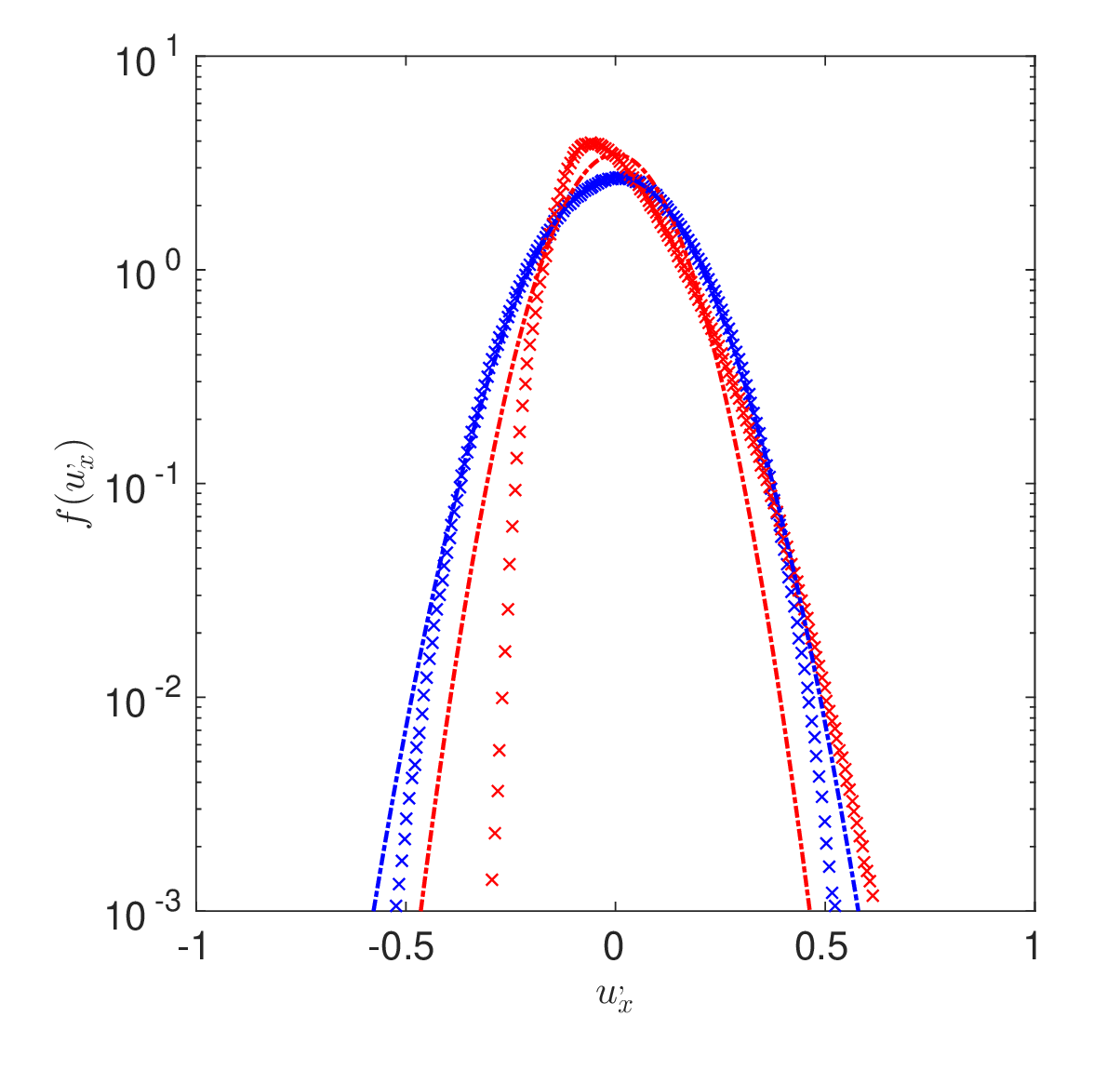}
\caption*{(a)}
\end{subfigure}
\begin{subfigure}{0.49\textwidth}
\includegraphics[width=1.0\textwidth]{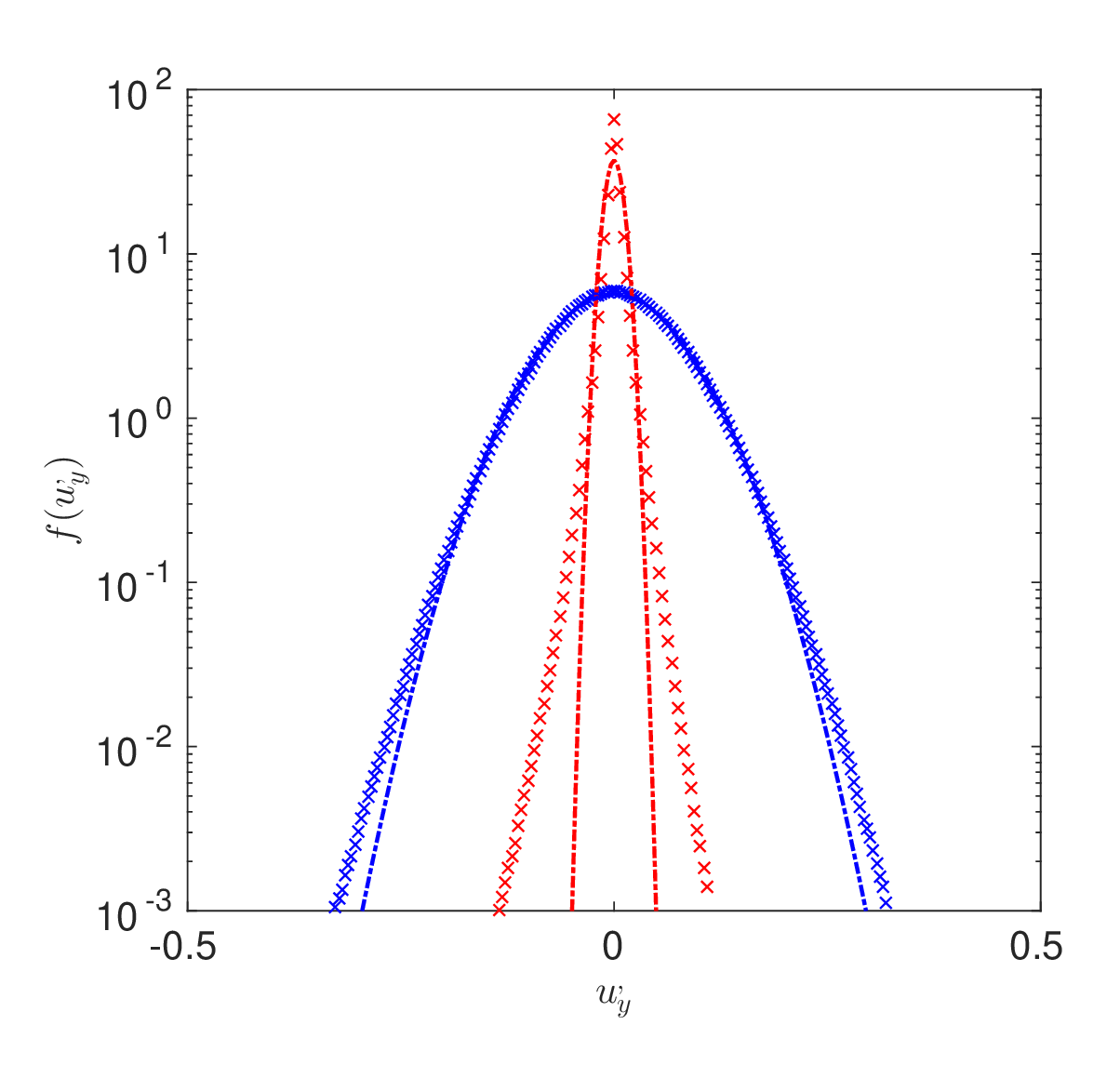}
\caption*{(b)}
\end{subfigure}
\begin{subfigure}{0.5\textwidth}
\centering
\includegraphics[width=1.0\textwidth]{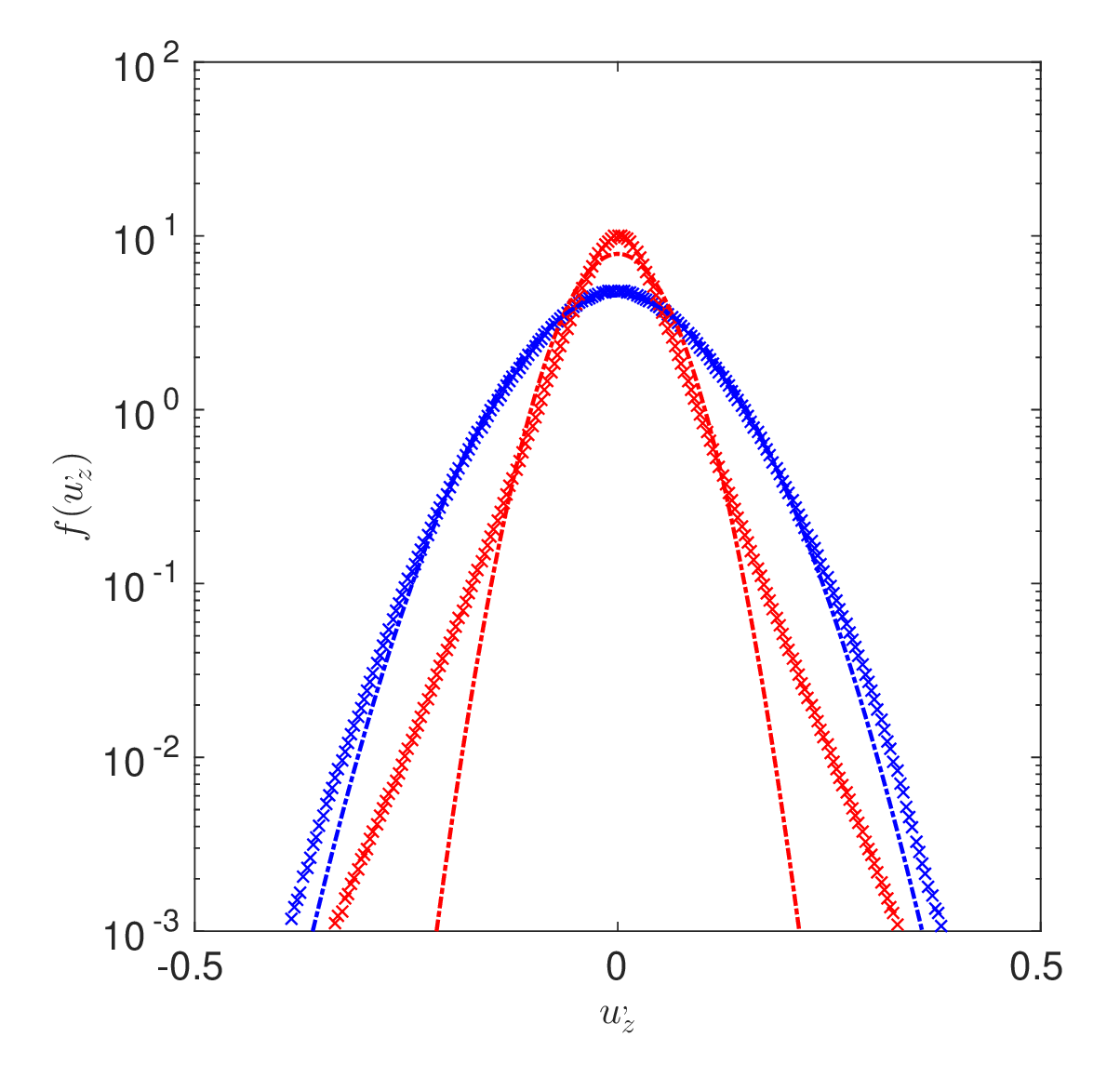}
\caption*{(c)}
\end{subfigure}
\caption{Fluid velocity distribution function along three different directions (a) $f(u'_x)$, (b) $f(u_y')$ and (c) $f(u_z')$;  each pdf is computed at three different 'y' positions in Couette: at the center $y^+=52.7$ or $y/\delta=1.0$ (blue), and near the wall  $y^+=4.74$ or $y/\delta=0.09$ (Red). The symbols are from DNS simulations that resolve
all the turbulence length scales, and the lines are the corresponding Gaussian fits.}	
\label{fig:fluid_vel_dist}
\end{figure}

The distributions of the fluid velocity and vorticity fluctuations at different
cross-stream locations are shown in figures \ref{fig:fluid_vel_dist} and \ref{fig:fluid_vort_dist}, along with the Gaussian fits. 
It is evident that these distributions are not
Gaussian. However, in the F3T simulations, it is assumed that the 
distributions are Gaussian. It is shown in section \ref{fig:fluid_vort_dist} that
the results of the F3T simulations are in quantitative agreement with the DNS 
simulations despite the approximation in the form of the distribution.


\begin{figure}	
\begin{subfigure}{0.49\textwidth}
\centering
\includegraphics[width=1.0\textwidth]{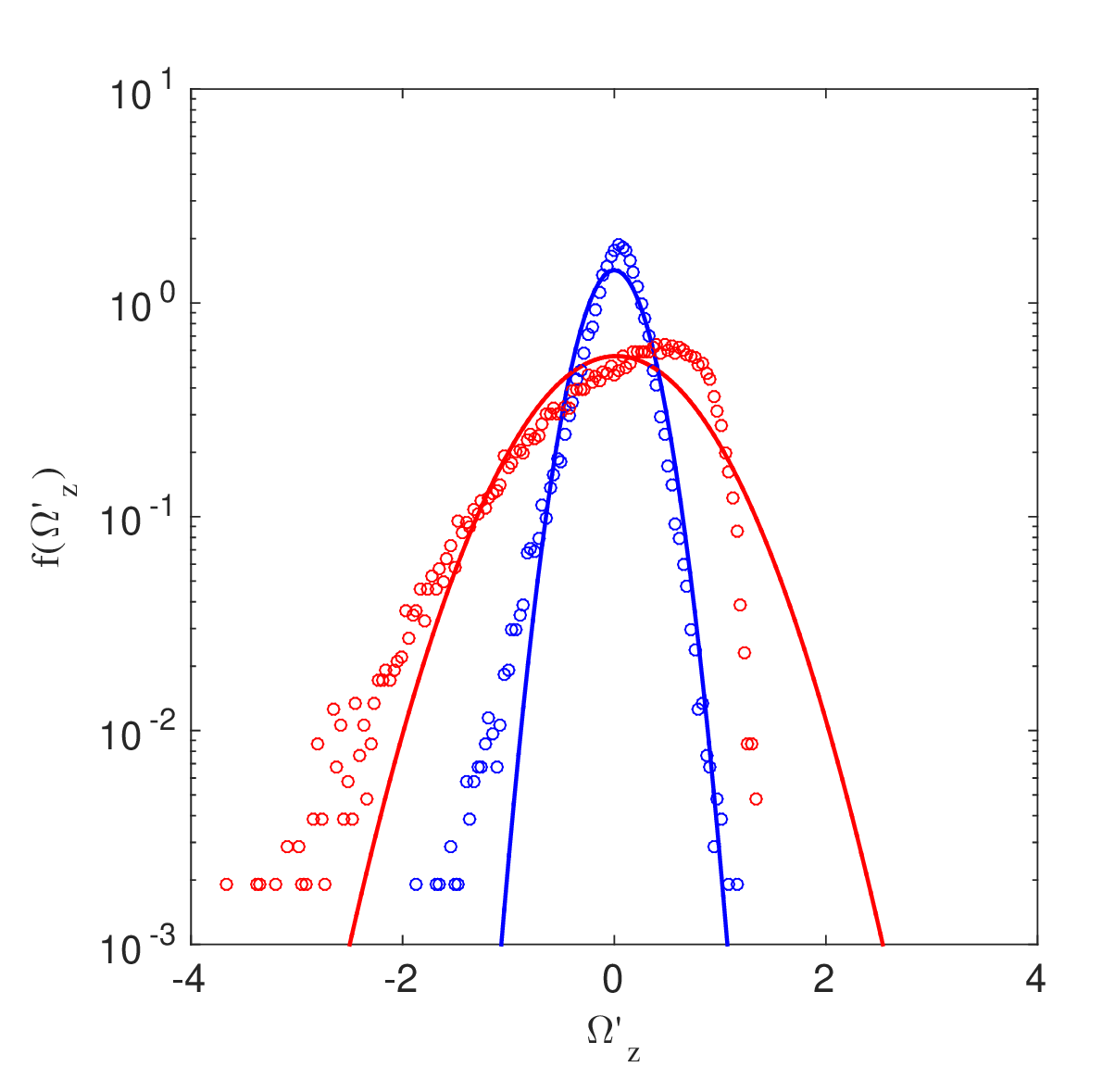}
\caption*{(a)}
\end{subfigure}
\begin{subfigure}{0.49\textwidth}
\includegraphics[width=1.0\textwidth]{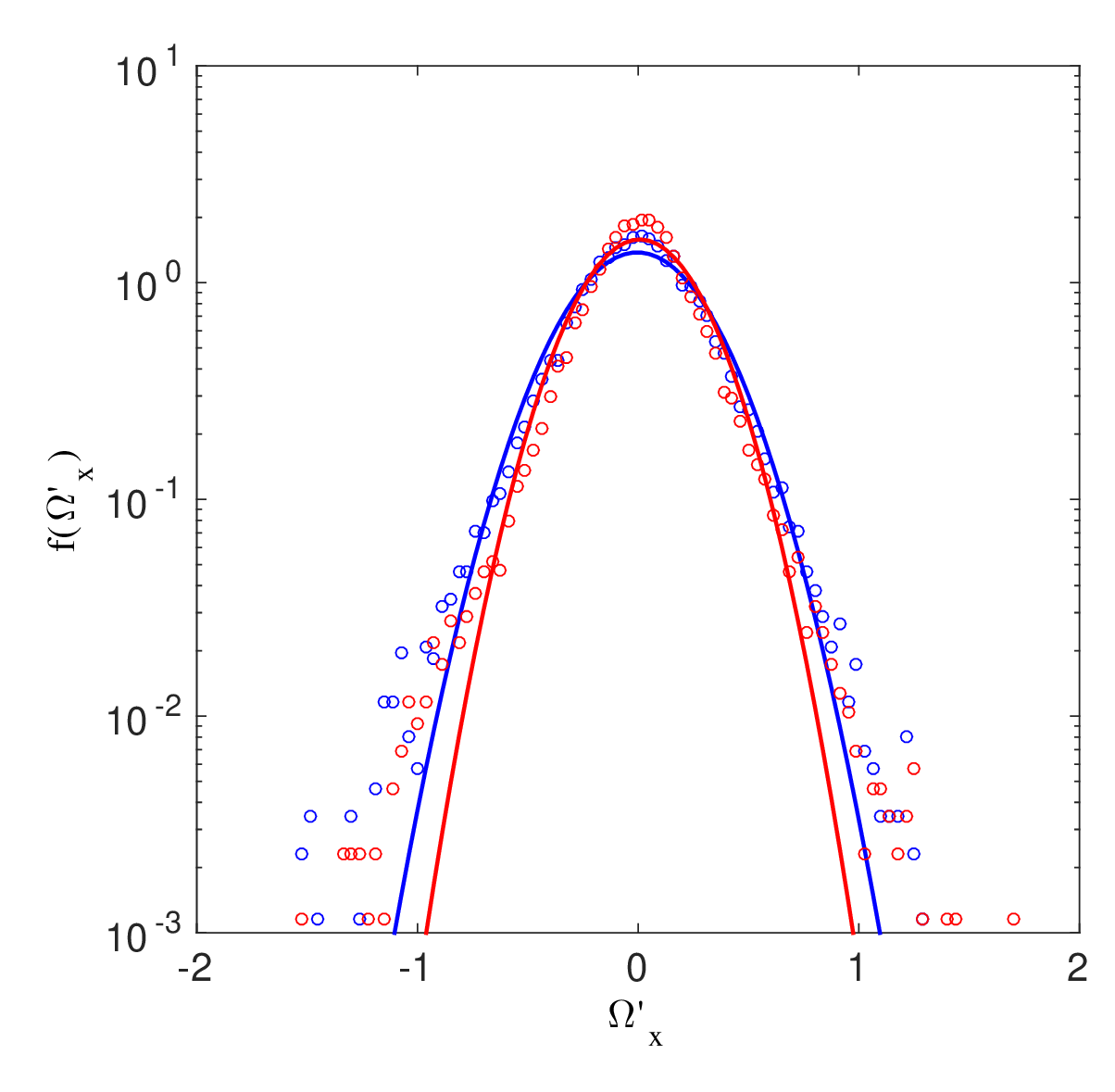}
\caption*{(b)}
\end{subfigure}
\begin{subfigure}{0.5\textwidth}
\includegraphics[width=1.0\textwidth]{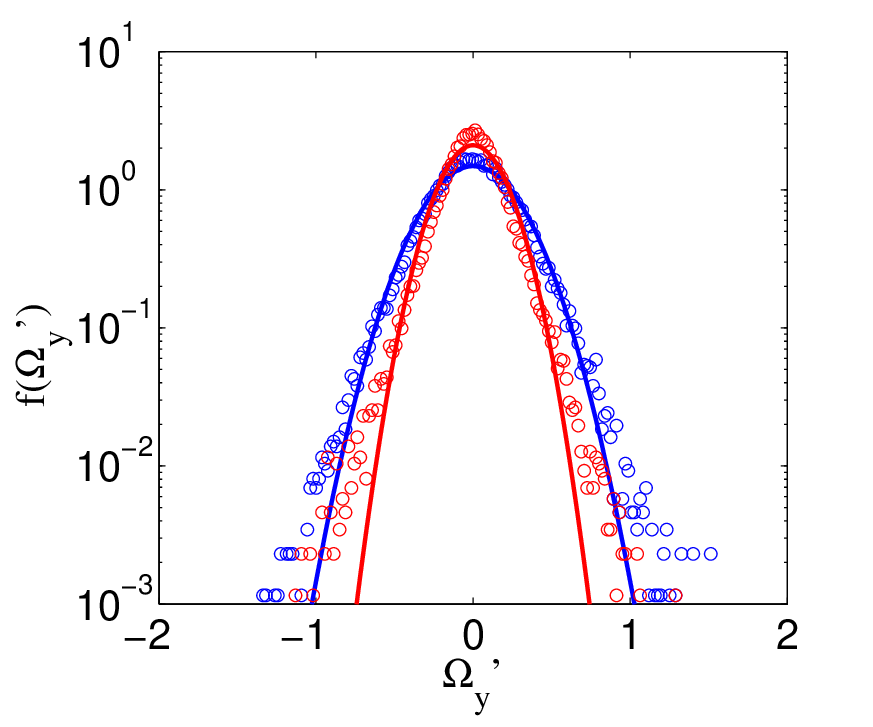}
\caption*{(c)}
\end{subfigure}
\caption{Fluid angular velocity distribution function along three different directions (a) $f(\Omega_z')$, (b) $f(\Omega_x')$ and (c) $f(\Omega_y')$;  each pdf is computed at two different 'y' positions in Couette: at the center $y^+=52.7$ or $y/\delta=1.0$ (blue), and near the wall  $y^+=4.74$ or $y/\delta=0.09$ (red). The symbols are from 
DNS simulations that resolve all the turbulence scales and the lines are the corresponding Gaussian fits}	
\label{fig:fluid_vort_dist}
\end{figure}

%

\section{Particle statistics:}
\label{fig:fluid_vort_dist}
The velocity statistics for the particles from  Fluctuating Force-Fluctuating Torque simulations (F3TS), where the linear
and angular acceleration on the particles are modeled as Gaussian white noise, are compared with the results of simulations
where the fluid velocity and vorticity from DNS simulations are used to determine the force and torque on the particles. 
In the DNS simulations, one-way coupling is used and the effect of particles on the fluid turbulence is neglected.
It is to be noted that the lift force and the error in the drag due to the disturbance velocity field 
(\citet{mehrabadi2018direct}) does not significantly influence the 
overall particle dynamics for the Reynolds number and flow parameters used here 
(\citet{muramulla2020disruption}), and so these were neglected in the present calculation. 
 All the simulations were performed in dilute limit with 7750 spherical particles of size 39 $\mu$m
 in a simulation box of size $10 \pi \delta \times 2 \delta \times 4 \pi \delta$ shown in 
 figure \ref{fig:schematic}. The particle volume 
 fraction is $10^{-4}$ and material  density 
of the particle is chosen to be 2000 kg/m$^3$. Simulations have also been carried out for a larger
simulation box of size $28 \pi \delta \times 2 \delta \times 8 \pi \delta$ for the same volume fraction
with 43400 particles. The variation in the particle statistics is only about 2\% when the box size
is increased by a factor of 5.6, indicating that there is statistical convergence.
For all the simulations presented here, the particle Stokes number 
based on fluid integral time scale $\delta/U$ unit is 50. The ratio of the channel half-width $\delta$ and the particle diameter $d_p$ is $74.36$.
\subsection{Smooth inelastic particles}
\label{sec:inelastic}
\begin{figure}
	\begin{subfigure}{1.0\textwidth}
\centering
    \includegraphics[width=0.65\textwidth]{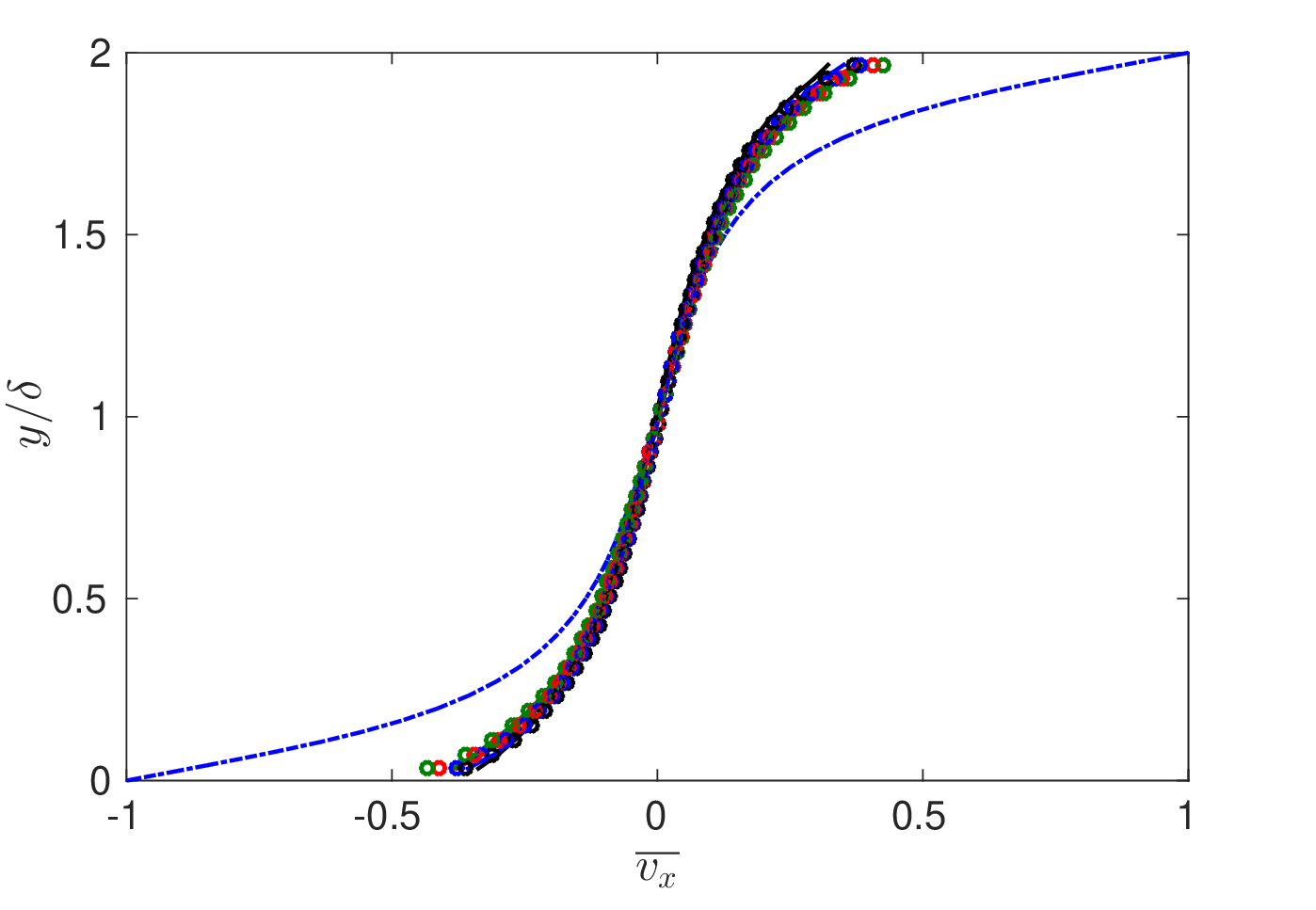}
    \caption*{(a)}
\end{subfigure}
\begin{subfigure}{0.49\textwidth}
 	\includegraphics[width=1\textwidth]{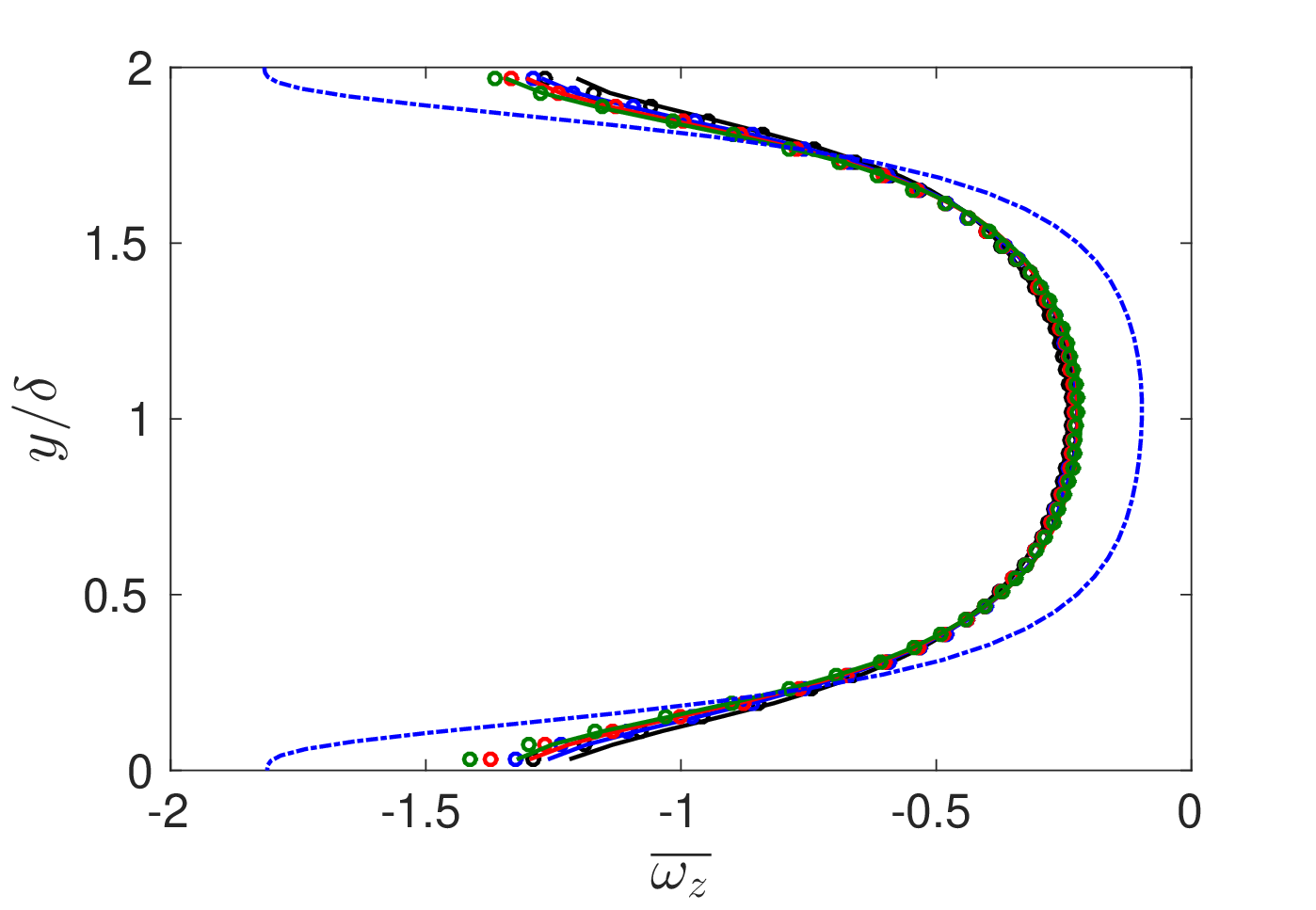}
 	\caption*{(b)}
 	\end{subfigure}
 	\begin{subfigure}{0.49\textwidth}
 	\centering
 	\includegraphics[width=1\textwidth]{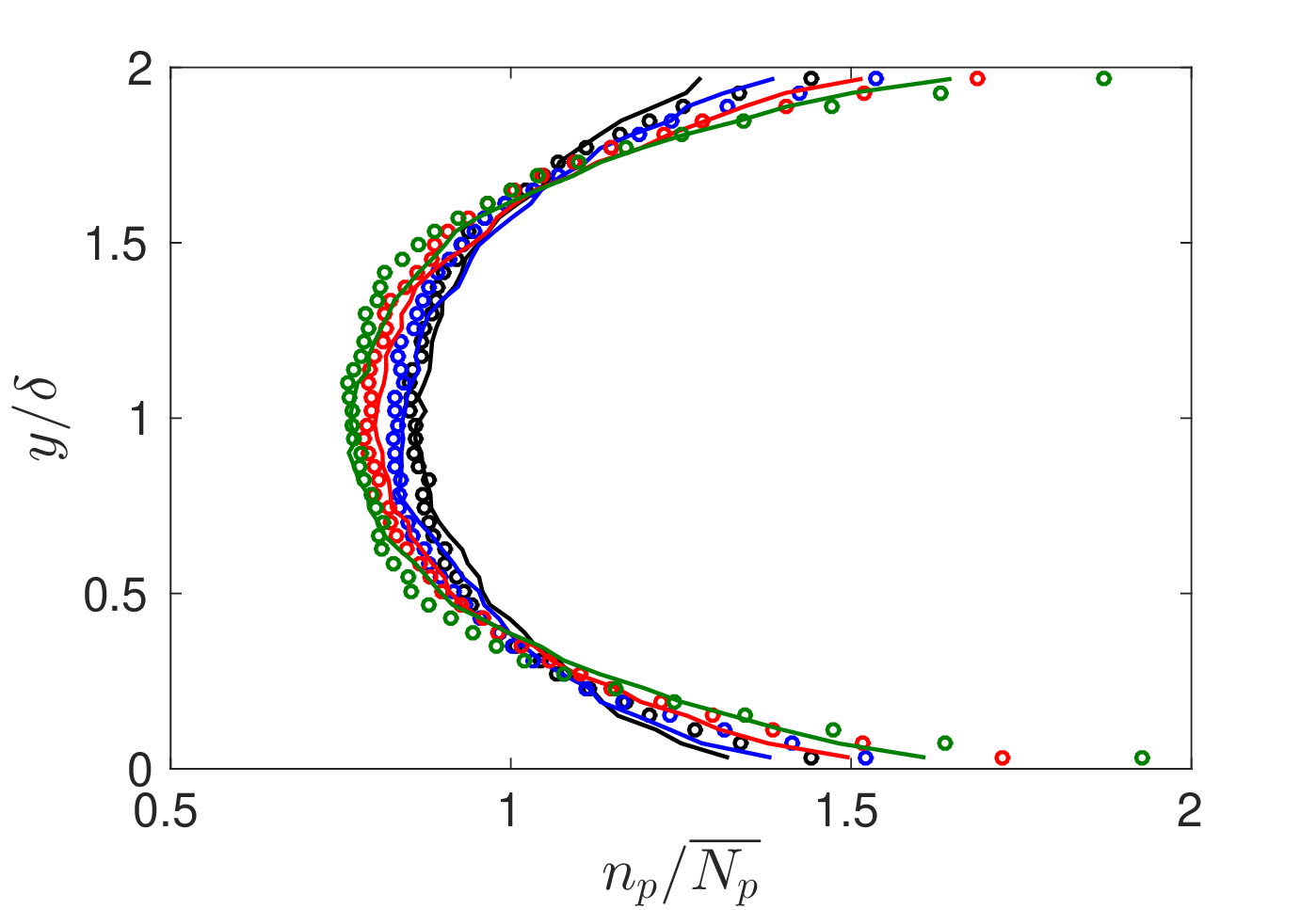}
  	\caption*{(c)}
  	\end{subfigure}
	\caption{Variation of particle phase statistics along channel-width $y/\delta$ (a) mean velocity $\overline{v_x} $, (b) mean angular velocity $\overline{\omega_z}$, (c) average particle number concentration for smooth particles with $\beta_{pp} = \beta_{pw} = -1$
	and different values of the normal coefficient of restitution $e$.
	The lines are the results
  from DNS simulations that resolve all the turbulence length scales without any sub-grid modeling, and the symbols
  are the results of the F3T model. The different colours are {\bf ---} $e=1.0$, \textcolor{blue}{\bf ---} $e=0.9$,
  \textcolor{red}{\bf ---} $e=0.8$ and \textcolor{green}{\bf ---} $e=0.7$, \textcolor{blue}{- $\cdot$ - $\cdot$ -} fluid profile.
  } 
	\label{fig:e_mean_part_stats}
\end{figure}

The particle volume fraction, mean velocity and mean angular velocity for smooth inelastic particles are shown
in figure \ref{fig:e_mean_part_stats}. The particle mean velocity profile and vorticity profiles are qualitatively 
similar to that of the fluid mean velocity with a higher gradient near the wall, as shown in figure 
\ref{fig:e_mean_part_stats}(a) and (b).
However, there are quantitative differences between the particle and fluid velocity; in particular, the particle 
mean velocity is non-zero at the walls, and the velocity profile exhibits significant slip. There is also a difference
between the particle angular velocity and one half of the fluid vorticity; the fluid rotation rate is higher
than the particle rotation rate at the wall, and lower at the center. It should be noted that there is no
change in the angular velocity of the particles in collisions, and the particle angular velocity is solely determined
by the torque exerted on the particles due to the fluid vorticity. The variation in the particle angular velocity
is smaller than that for the fluid rotation rate across the channel due to the smoothing effect of the cross-stream
motion of the particles. Figure~\ref{fig:e_mean_part_stats} (c) shows that the particle concentration is lower
at the center and higher at the walls. Here, the number density of particles has been non-dimensionalised by 
the width-averaged number density.

Figure~\ref {fig:e_mean_part_stats} shows that both the results from  F3TS simulations for the particle mean velocity,
angular velocity and number density are all in quantitative agreement with the results of the DNS simulations. 
A variation in the coefficient of restitution
causes only a small variation in the velocity and number density profiles. There is a very small increase in the particle mean 
velocity and the magnitude of the mean angular velocity at the wall when the coefficient of restitution is decreased. The 
particle number density at the wall exhibits a slightly larger variation across the channel when the coefficient of restitution
is decreased. Even these small variations seem to be correctly captured by the F3TS simulations.

\begin{figure}
	\begin{subfigure}{0.49\textwidth}
    \includegraphics[width=1.0\textwidth]{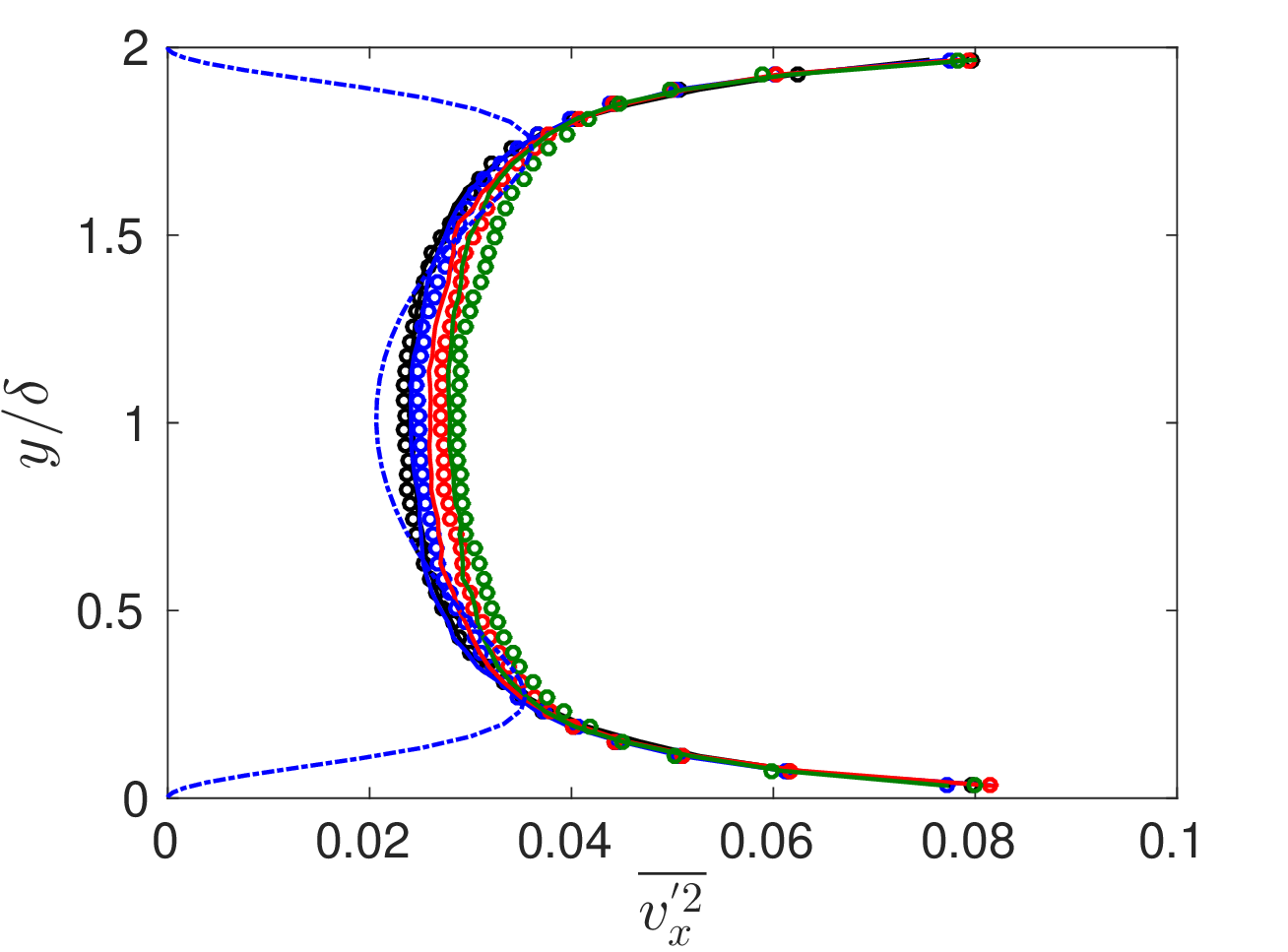}
    \caption*{(a)}
\end{subfigure}
\begin{subfigure}{0.49\textwidth}
 	\includegraphics[width=1.0\textwidth]{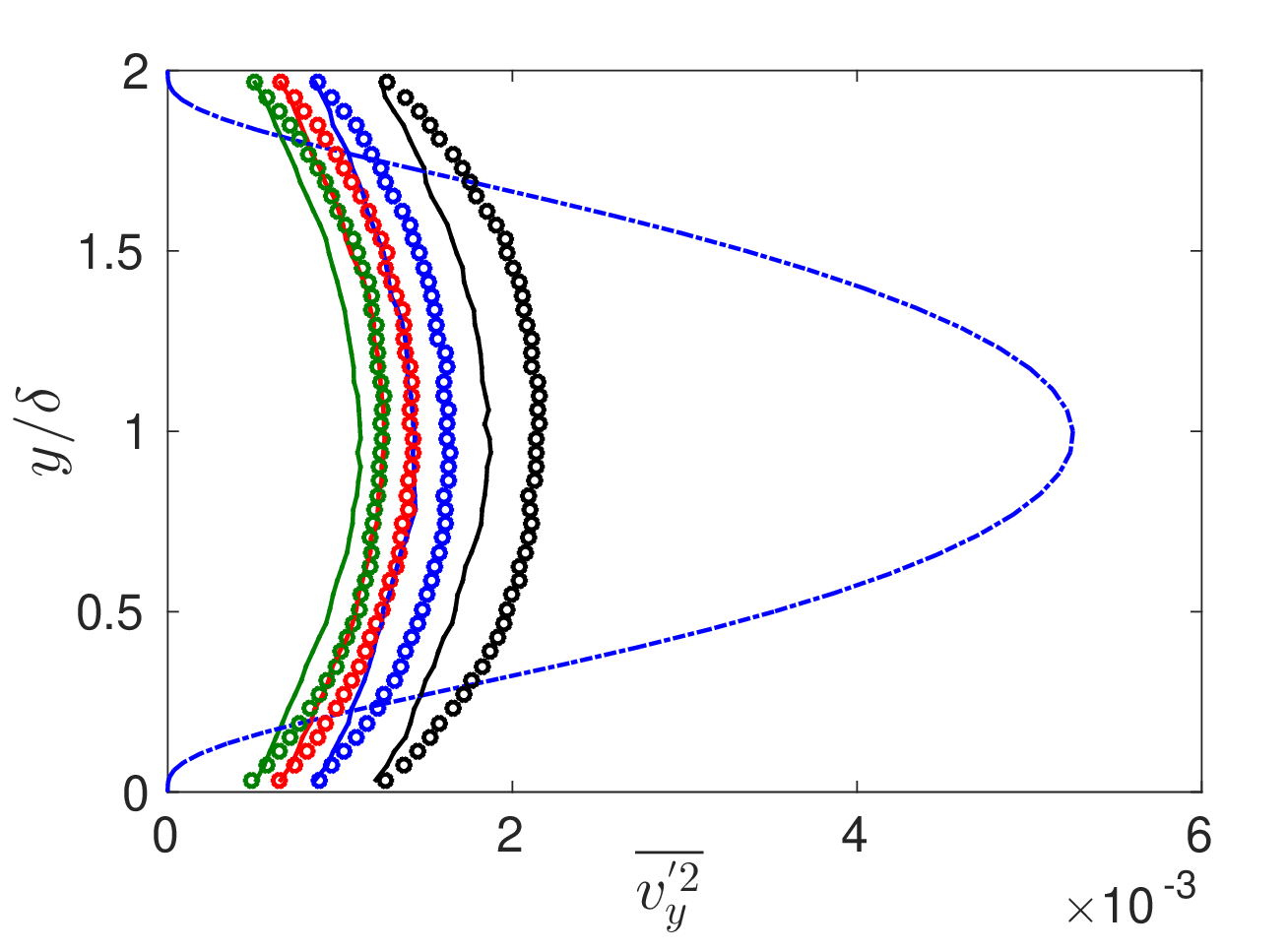}
 	\caption*{(b)}
 	\end{subfigure}
 	\begin{subfigure}{0.49\textwidth}
 	\includegraphics[width=1\textwidth]{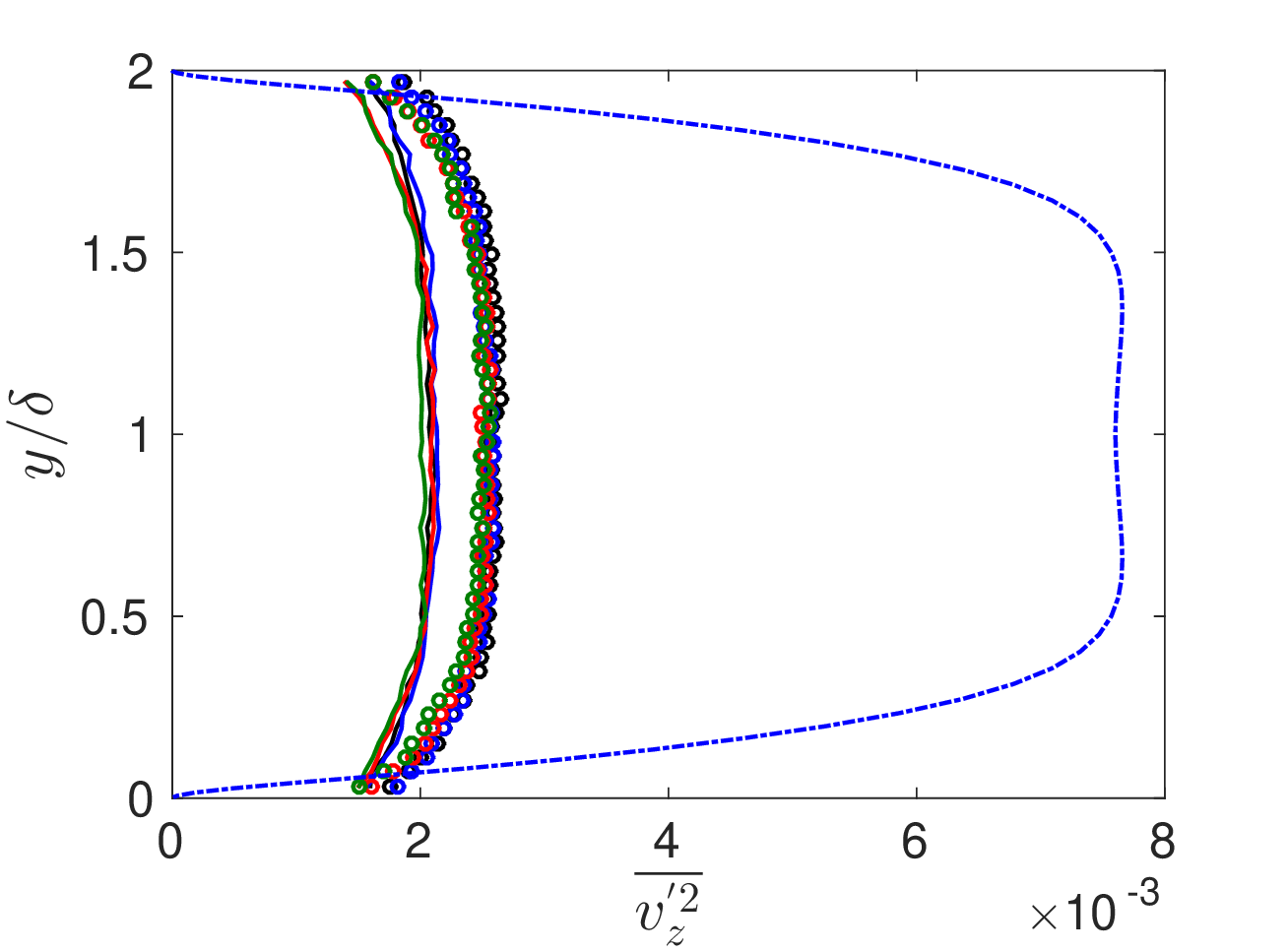}
  	\caption*{(c)}
  	\end{subfigure}
  	\begin{subfigure}{0.49\textwidth}
  		\includegraphics[width=1.0\textwidth]{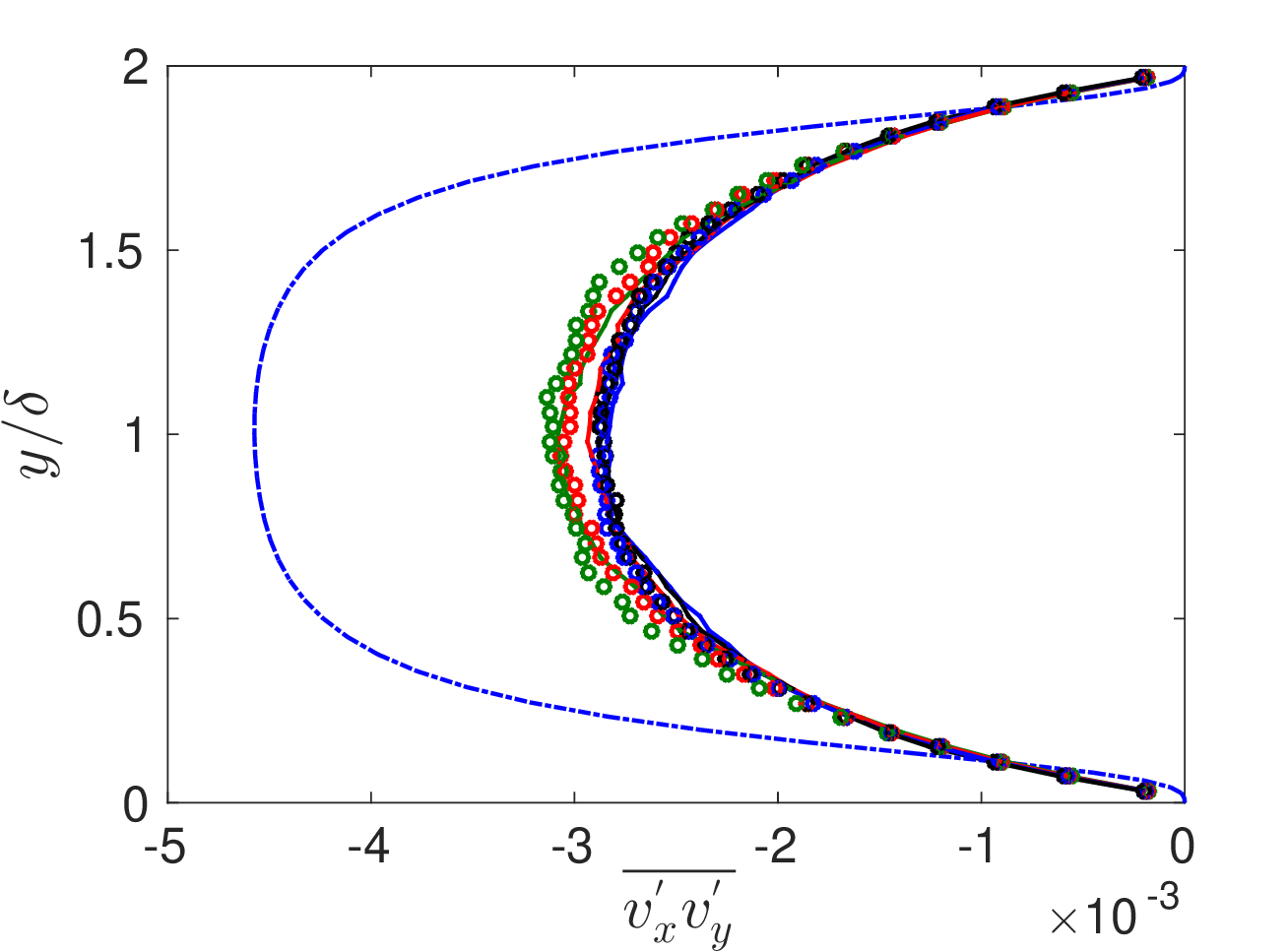}
  	\caption*{(d)}
  	\end{subfigure}
	\caption{Variation of particle mean square velocity $\overline{v_i' v_j'}$ along channel-width $y/\delta$ (a) mean square velocity along x $\overline{v_x'^2}$, (b) mean square velocity along y $\overline{v_y'^2}$, (c) mean square velocity along z $\overline{v_z'^2}$ and (d) $-\overline{v_x' v_y'}$ for smooth particles with $\beta_{pp} = \beta_{pw} = -1$
	and different values of the normal coefficient of restitution $e$. 
	The lines are the results
  from DNS simulations that resolve all the turbulence length scales without any sub-grid modeling, and the symbols 
  are the results of the F3T model. The different colours are {\bf ---} $e=1.0$, \textcolor{blue}{\bf ---} $e=0.9$,
  \textcolor{red}{\bf ---} $e=0.8$ and \textcolor{green}{\bf ---} $e=0.7$, \textcolor{blue}{- $\cdot$ - $\cdot$ -} fluid profile.
	}
	\label{fig:e_stress}
\end{figure}

The second moments of the particle velocity fluctuations are shown in figures~\ref{fig:e_stress} (a) to (c). 
The mean square velocity in the stream wise direction is a maximum at the wall and it decreases towards the center. This is in
in contrast to the stream-wise fluid mean square velocity (figure~\ref{fig:fluidms}), which is zero at the wall and shows 
a maximum value at $y^+ \sim 15$. 
The particle velocity fluctuations in the stream-wise direction is larger than the fluid velocity fluctuations,
indicating that particle collisions and cross-stream migration play a significant role in generating stream-wise fluctuations.
The particle velocity fluctuations in the cross-stream and span-wise directions, shown in figure \ref{fig:e_stress} (b) and (c),
are about one order of magnitude smaller than those in the stream-wise direction; these have a maximum at the center and minima
at the walls. The particle velocity fluctuations in the cross-stream and span-wise directions are also smaller than the 
fluid velocity fluctuations in these directions.

Figure \ref{fig:e_stress} shows that the F3TS predictions for the stream-wise mean square velocities are in quantitative agreement
with the DNS results; the F3TS simulations correctly capture the maxima at the walls and the minimum in the center. The mean square 
velocities in the cross-stream and span-wise directions are in qualitative agreement, and the F3TS simulations correctly capture
the minima at the walls and the maximum at the center. However, there is a variation of about 20\% in the maximum values of the 
mean square velocities perpendicular to the flow direction at the center of the channel. 
This could be because the ratio $(\tau_v/\tau_f) \approx 2.5$ is a little low, where \( \tau_{f} \) is the fluid velocity decorrelation time---it was shown in \citep{goswami2010particle}
that the predictions of fluctuating force simulations are accurate for $(\tau_v/\tau_f)\geq3$. 
It should be noted that the mean square velocities in the cross-stream
and span-wise directions are an order of magnitude smaller than those in the stream-wise direction, and therefore the errors in these
quantities are expected to be larger relative to the mean.
It is interesting to note that a decrease in the coefficient of restitution results in a slight decrease in the mean
square velocities in the stream-wise and span-wise directions (figures \ref{fig:e_stress} (a) and (c)), but there is a
significant decrease in the mean square velocity in the cross-stream direction (figure \ref{fig:e_stress} (b)). This
is due to the damping of the cross-stream velocity fluctuations by inelastic particle-wall collisions. 
All of these variations due to inelastic collisions are quantitatively captured by the F3TS simulations.

\begin{figure}
	\begin{subfigure}{0.49\textwidth}
\centering
    \includegraphics[width=1\textwidth]{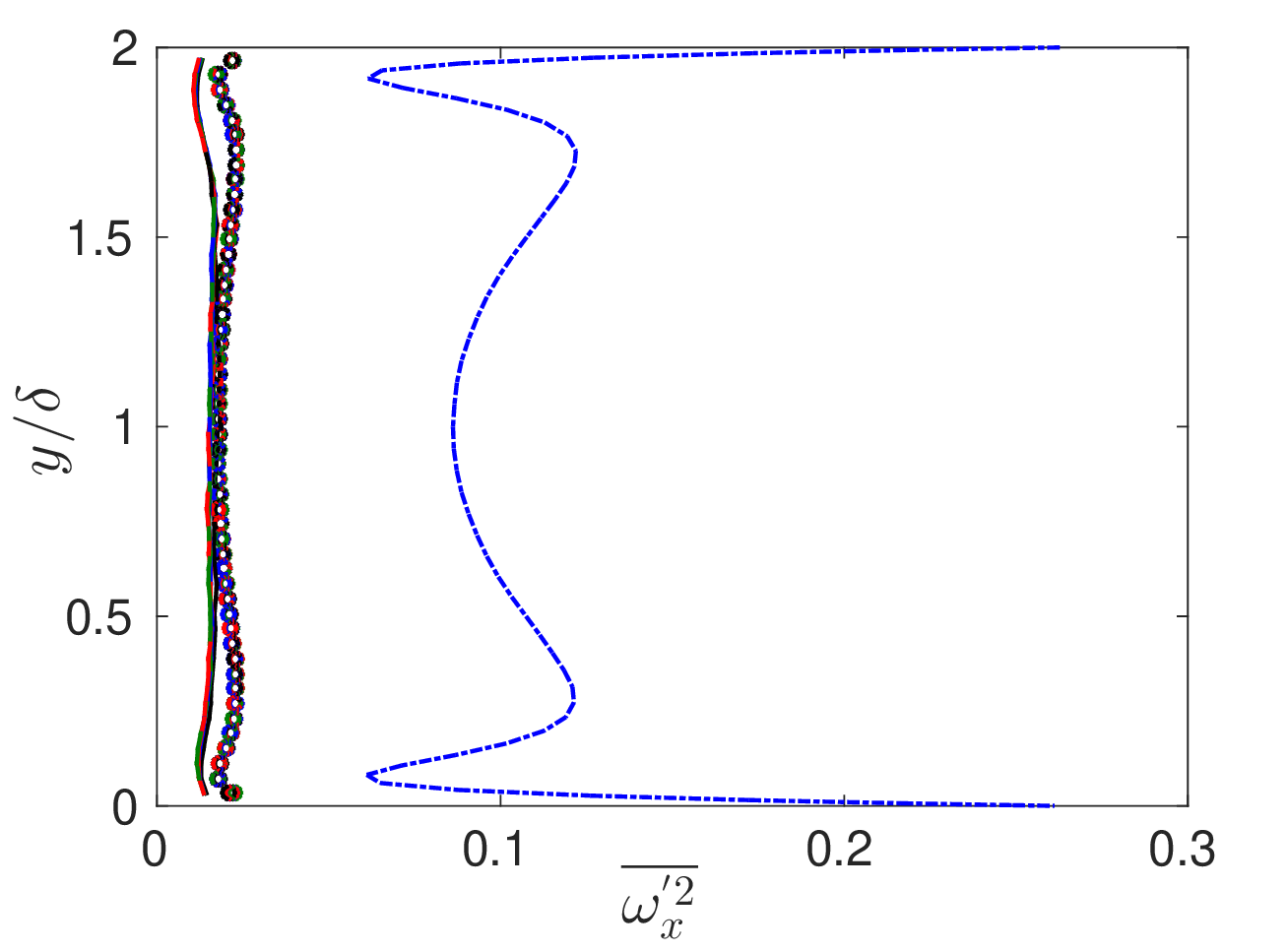}
    \caption*{(a)}
\end{subfigure}
\begin{subfigure}{0.49\textwidth}
 	\includegraphics[width=1\textwidth]{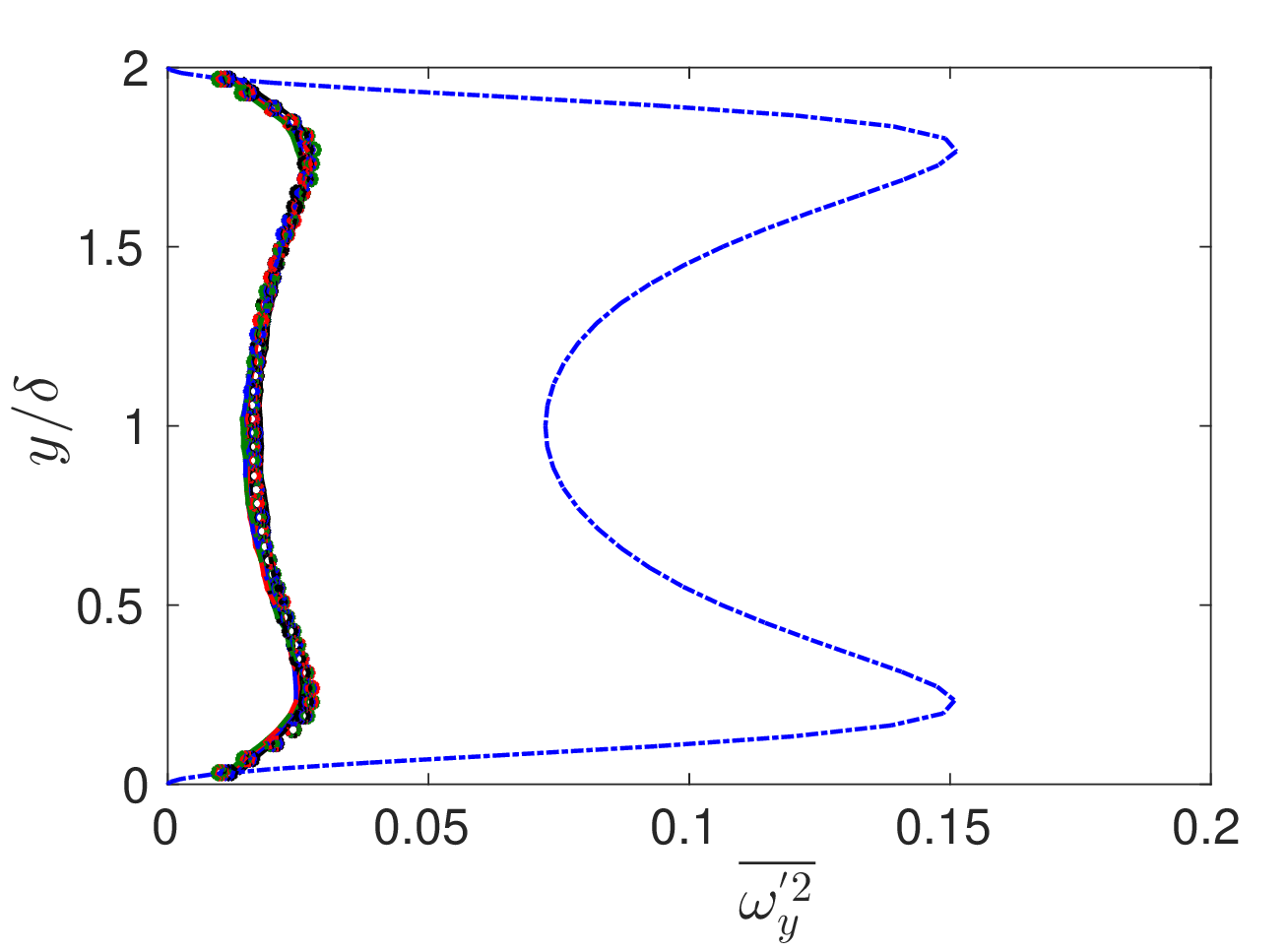}
 	\caption*{(b)}
 	\end{subfigure}
 	\begin{subfigure}{0.49\textwidth}
 	\centering
 	\includegraphics[width=1\textwidth]{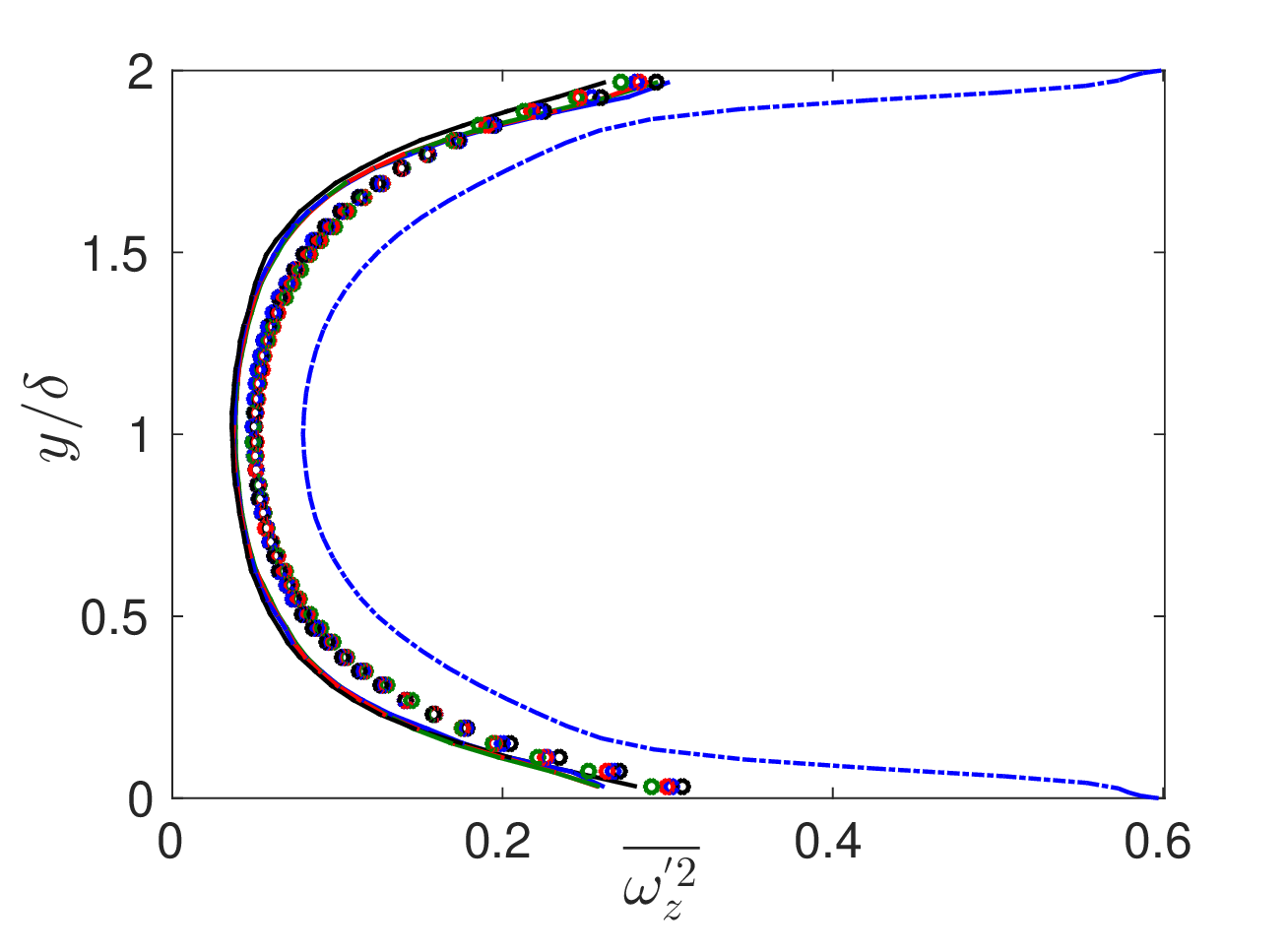}
  	\caption*{(c)}
  	\end{subfigure}
  	\begin{subfigure}{0.49\textwidth}
 	\centering
 	\includegraphics[width=1\textwidth]{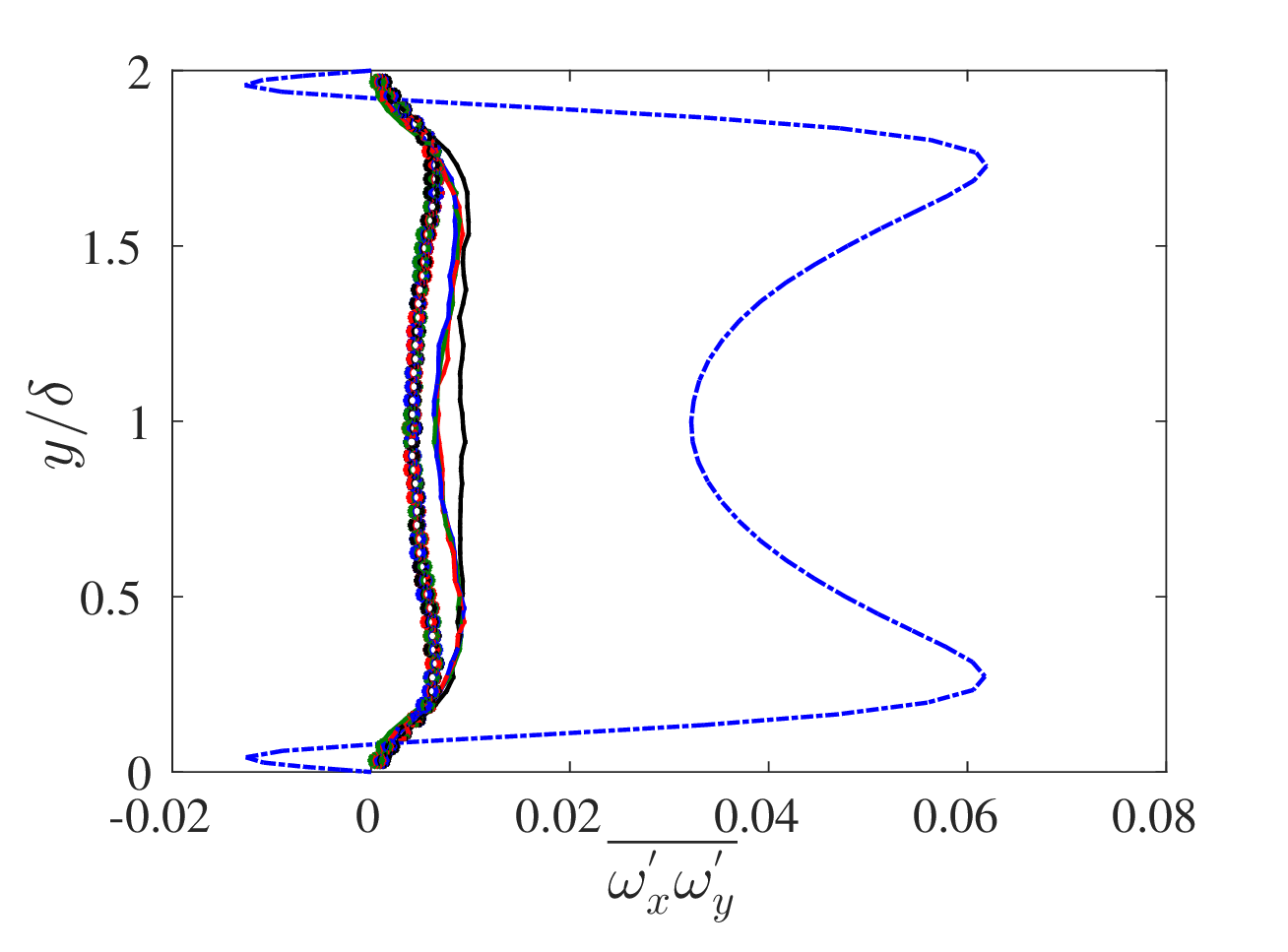}
  	\caption*{(d)}
  	\end{subfigure}
	\caption{Variation of particle mean square angular velocity $\overline{\omega_i'^2}$ along channel-width $y/\delta$ (a) mean square angular velocity along z $\overline{\omega_x'^2}$, (b) mean square angular velocity along y $\overline{\omega_y'^2}$, (c) mean square angular velocity along x $\overline{\omega_z'^2}$ and (d) $\overline{\omega_x'\omega_y'}$  for smooth particles with $\beta_{pp} = \beta_{pw} = -1$
	and different values of the normal coefficient of restitution $e$. The lines are the results
  from DNS simulations that resolve all the turbulence length scales without any sub-grid modeling, and the symbols 
  are the results of the F3T model. The different colours are {\bf ---} $e=1.0$, \textcolor{blue}{\bf ---} $e=0.9$,
  \textcolor{red}{\bf ---} $e=0.8$ and \textcolor{green}{\bf ---} $e=0.7$, \textcolor{blue}{- $\cdot$ - $\cdot$ -} fluid profile.
	\label{fig:e_ang_stress}}
\end{figure}

The mean square of the angular velocity fluctuations are shown in figure \ref{fig:e_ang_stress}. 
Figure~\ref{fig:e_ang_stress} (c) shows that mean square angular velocity along span wise ($z$) direction, 
$\overline{\omega_z^{\prime 2}}$ is
the largest contribution to the mean square angular velocity, and this is an order of magnitude higher than the other two components 
$\overline{\omega_y^{\prime 2}}$ and $\overline{\omega_x^{\prime 2}}$.
The span-wise mean square angular velocity 
$\overline{\omega_z^{\prime 2}}$
is highest at walls and decreases towards the center. This component exhibits very little variation with the particle coefficient of restitution,
and the F3TS results are in good agreement with the DNS results for this case. The mean square angular velocity in the wall-normal direction,
$\overline{\omega_y^{\prime 2}}$, exhibits unusual behaviour with local minima and the center of channel, and a maximum at $y^+ \sim 15$. 
This unusual behaviour is also correctly captured by the F3TS simulations, and the results of the F3TS simulations are in good agreement
with the DNS simulations. The mean square angular velocity in the stream-wise direction, 
$\overline{\omega_x^{\prime 2}}$, is small in 
magnitude compared to the other components, and it exhibits very little variation across the channel in comparison to the other two 
components. The order of magnitude of this component is correctly captured by the F3TS simulations, though there is a difference in 
the numerical predictions. Thus, the two largest components of the mean square angular velocity are quantitatively predicted with 
good accuracy by the F3TS, but the F3TS simulations under-predicts the smallest component or the mean square angular velocity. 


	\begin{figure}
	\begin{subfigure}{1.0\textwidth}
	\centering
	\includegraphics[width=0.65\textwidth]{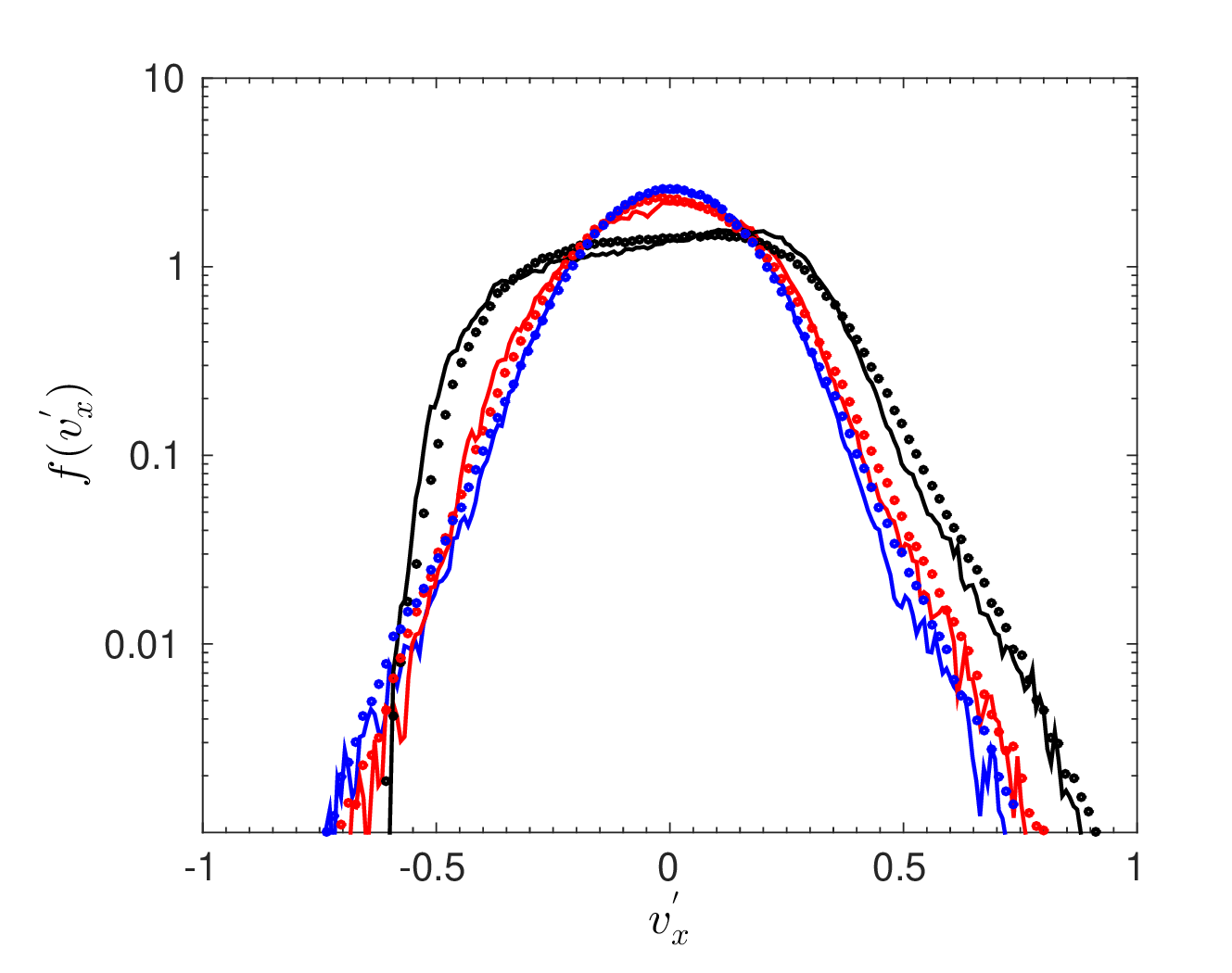}
	\caption*{(a)}
	\end{subfigure}
	\begin{subfigure}{0.49\textwidth}
		\includegraphics[width=1\textwidth]{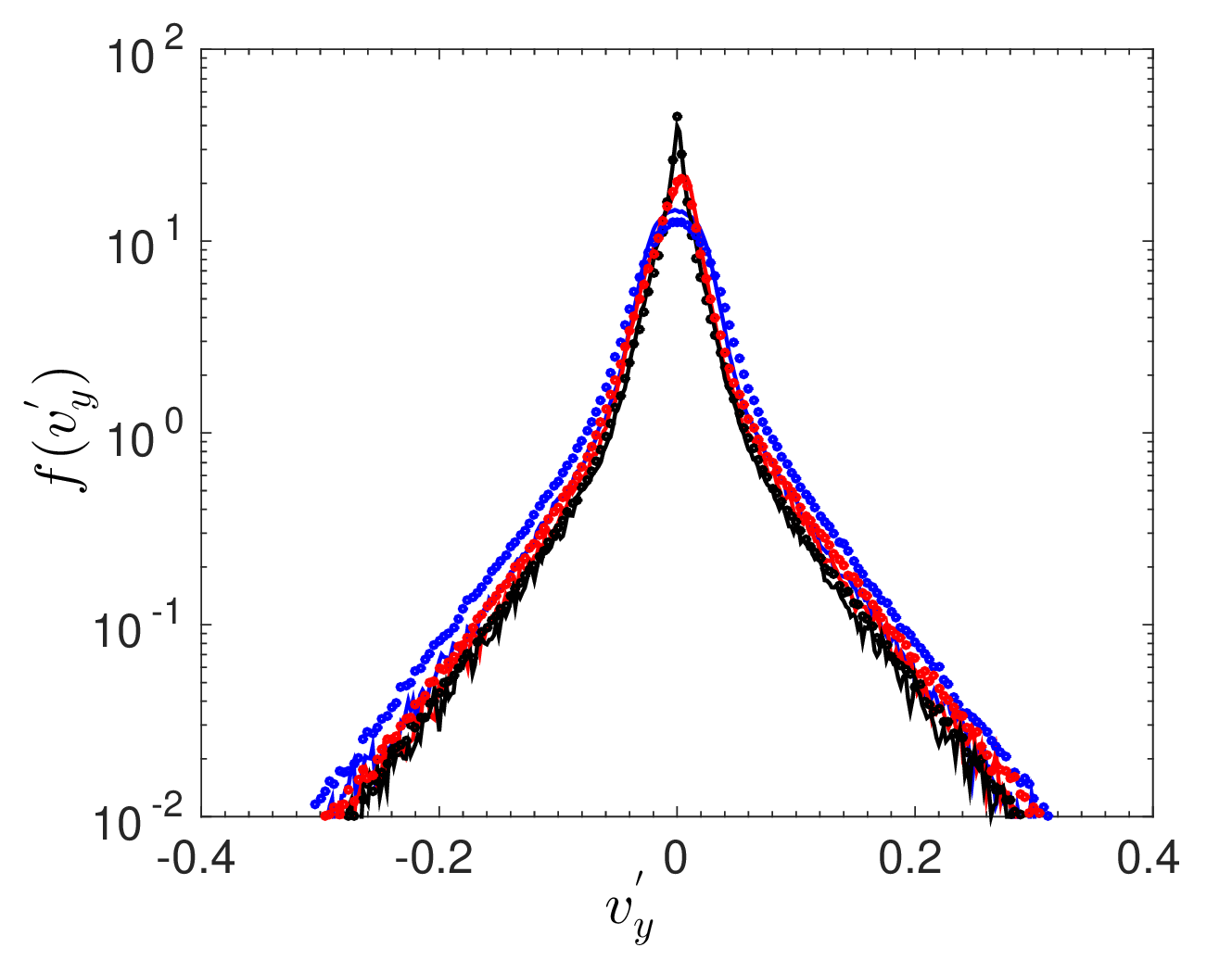}
		\caption*{(b)}
	\end{subfigure}
	\begin{subfigure}{0.49\textwidth}
		\centering
		\includegraphics[width=1\textwidth]{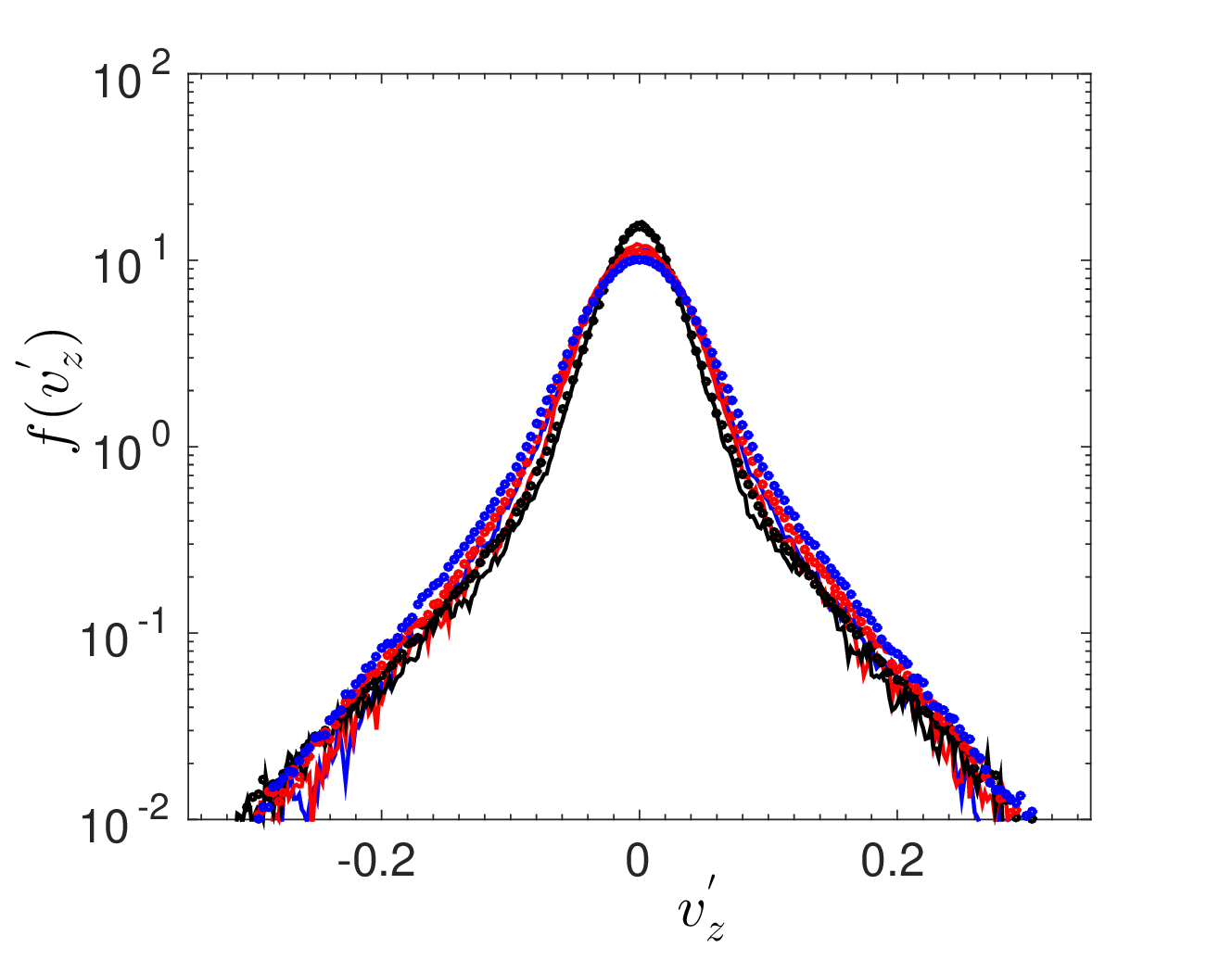}
		\caption*{(c)}
	\end{subfigure}
	\caption{Particle velocity distribution function along three different directions (a) $f(v_x')$,(b) $f(v_y')$ and (c) $f(v_z')$.  Each pdf is computed at three different '$y$' positions in Couette: at the center $y^+=52.7$ or $y/\delta=1.0$ \textcolor{blue}{\bf ---}, at the middle $y^+=19.5$ or $y/\delta=0.37$ \textcolor{red}{\bf ---} 
	and near the wall  $y^+=4.74$ or $y/\delta=0.09$ {\bf ---} 
	 for smooth particles with $\beta_{pp} = \beta_{pw} = -1$
	and different values of the normal coefficient of restitution $e$. The lines are the results
  from DNS simulations that resolve all the turbulence length scales without any sub-grid modeling, and the points
  are the results of the F3T model. }	
	\label{fig:vel_dist}
	\end{figure} 

The distribution functions from F3TS simulations and now compared those with the DNS results. It should be noted that the 
distributions of the acceleration and angular acceleration are assumed to be Gaussian in the F3TS simulations, but the 
fluid velocity and vorticity 
distributions are clearly not Gaussian in the DNS simulations, as shown in figure \ref{fig:fluid_vel_dist} and
\ref{fig:fluid_vort_dist}. The particle velocity and angular velocity distributions are shown at three different 
zones in the channel, in order to
examine the variation in the form of the distribution with location, and to determine the accuracy of the F3TS simulations
at different cross-stream locations. These distributions are averaged over a distance $0.16 \delta$ at the center 
of the channel and $0.08 \delta$ at other locations, where \( \delta \) is the half-width of the channel.
There is very little variation in the distribution functions with the coefficient of
restitution for smooth particles, and so the present discussion is restricted to $e=1$.

Figure \ref{fig:vel_dist} (a), (b) and (c) show the velocity distribution functions 
along the $x$, $y$ and $z$ direction respectively at various locations in the wall normal direction. There are qualitative differences
in the distribution functions in the three directions. The distribution function $f(v_x')$ in the stream-wise direction exhibits
a significant dependence on position; the distribution function is flatter with a positive skewness at the 
wall, but is closer to a Gaussian at the center. The distribution functions in the other two directions are symmetric, 
Gaussian close to the center, but have long non-Gaussian tails for higher velocities. The variance in the cross-stream and span-wise
distributions are smaller near the wall and larger at the center. The F3TS predictions for each of these distributions
is found to be in quantitative agreement with the DNS simulations, including the change in the form of the distribution
function with location for the stream-wise velocity fluctuations, the variance at the center and the non-Gaussian tails for
the cross-stream and span-wise directions. 
%
	\begin{figure}		
	
	\begin{subfigure}{1.0\textwidth}
		\centering
		\includegraphics[width=0.65\textwidth]{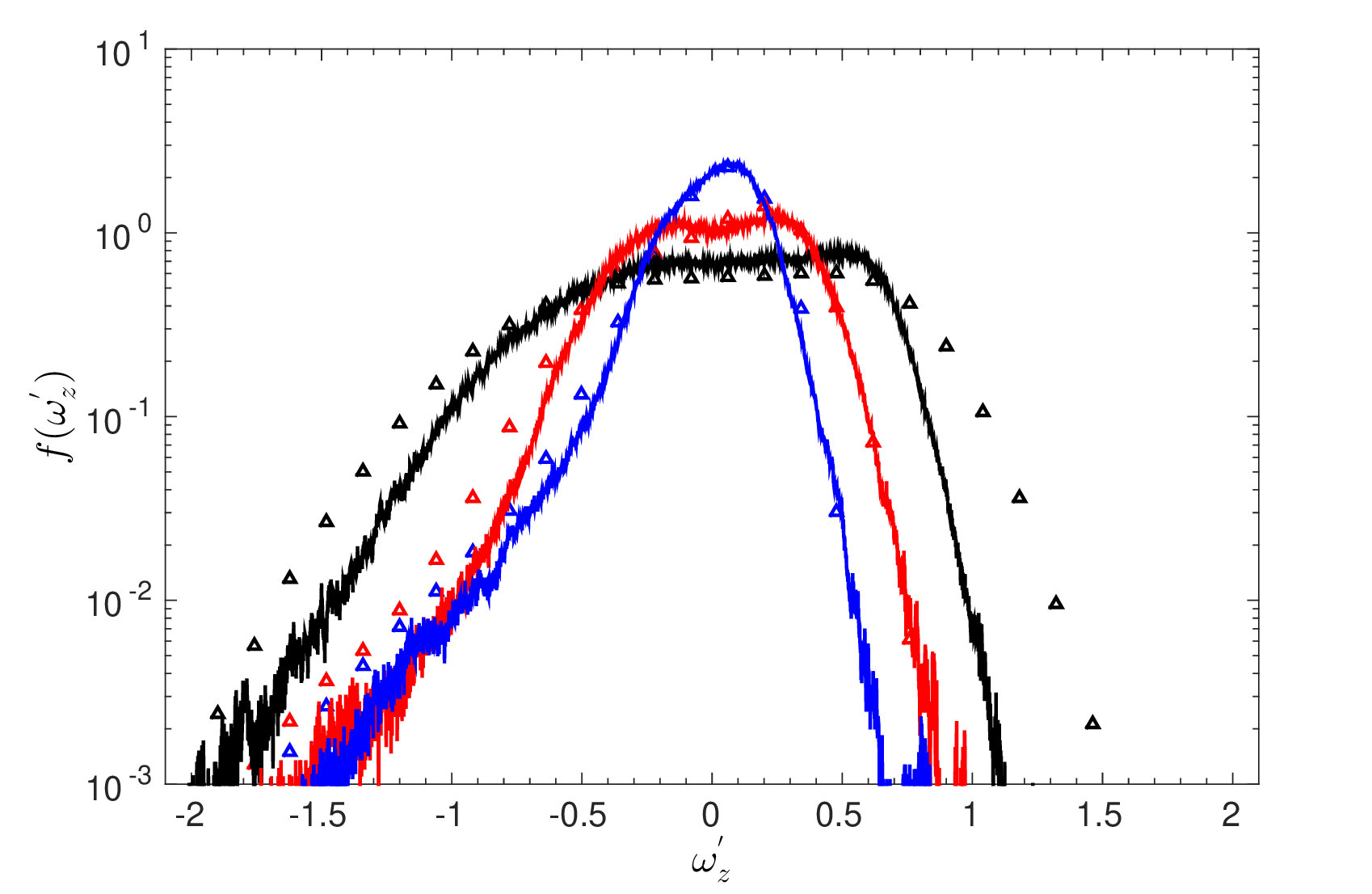}
		\caption*{(a)}
	\end{subfigure}
	\begin{subfigure}{0.49\textwidth}
		\includegraphics[width=1\textwidth]{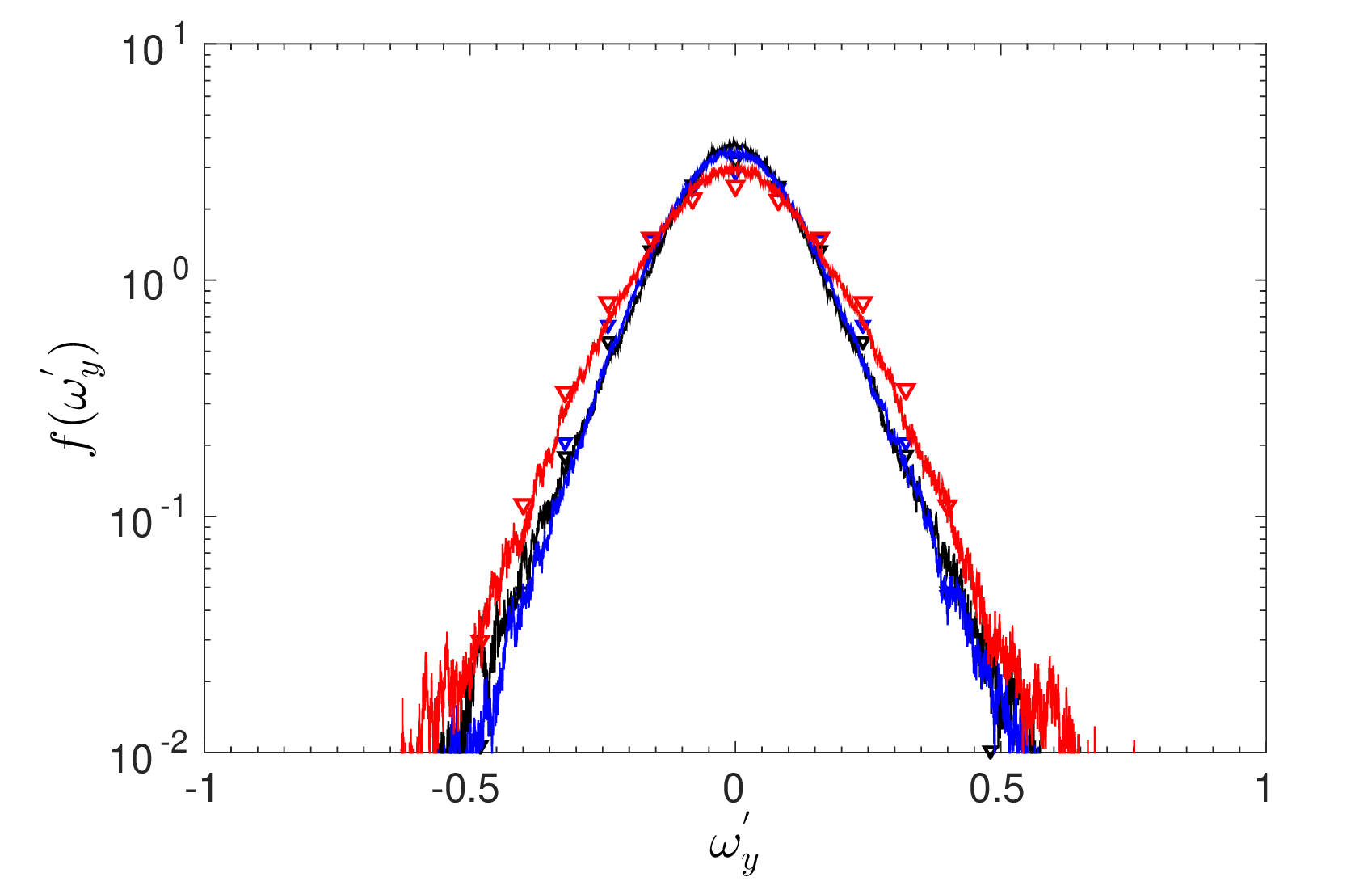}
		\caption*{(b)}
	\end{subfigure}
	\begin{subfigure}{0.49\textwidth}
		\centering
		\includegraphics[width=1\textwidth]{smooth_elastic_rot_vel_dist_z.eps}
		\caption*{(c)}
	\end{subfigure}
	\caption{Particle angular velocity distribution function along three different directions (a) $f(\omega_z')$,(b) $f(\omega_x')$ and (c) $f(\omega_y')$;  Each pdf is computed at three different 'y' positions in Couette: at the center $y^+=52.7$ or $y/\delta=1.0$ \textcolor{blue}{\bf ---}, at the middle $y^+=19.5$ or $y/\delta=0.37$ \textcolor{red}{\bf ---}, and near the wall  $y^+=4.74$ or $y/\delta=0.09$ {\bf ---}; lines represent DNS results and symbols represent model predictions
	 for smooth particles with $\beta_{pp} = \beta_{pw} = -1$
	and different values of the normal coefficient of restitution $e$.
	The lines are the results
  from DNS simulations that resolve all the turbulence length scales without any sub-grid modeling, and the points
  are the results of the F3T model.}	
	\label{fig:vel_rot_dist}
\end{figure} 

Figures \ref{fig:vel_rot_dist} (a), (b) and (c) show angular velocity distribution functions along $z$, $x$ and $y$ directions respectively at different cross-stream positions. The forms of the angular velocity distributions also exhibit qualitative
variations with direction and location across the channel. Here, the distribution function for the span-wise angular velocity fluctuations,
$f(\omega_z')$, exhibits significant variations with cross-stream location; the distribution is broader near the wall with negative skewness
and becomes more symmetric at the center. The distribution functions for the other two components of the angular velocity show 
non-Gaussian tails. In all cases, the F3TS simulations quantitatively predict the forms of the distribution function, and their
dependence on the cross-stream location and the direction of the angular velocity fluctuations.

The effect of inelastic collisions on the linear and angular velocity distributions have also been studied, 
and it is found that the shape of the distribution function is not affected by inelasticity. The agreement 
between the results of the F3TS and DNS simulations is similar to that in figures \ref{fig:vel_dist} and 
\ref{fig:vel_rot_dist}; these are not shown here.


\subsection{Effect of Roughness on Particle Statistics}
\label{sec:rough}
The effect of particle/wall surface roughness on the particle phase velocity statistics is examined here. Three different cases
are considered.
\begin{enumerate} \item
Smooth particle-particle and particle-wall collisions ($e=1$, $\beta_{pp} = \beta_{pw} = -1 $ in
the collision law \ref{eq:collrule2}). 
\item Rough particle-particle collisions and smooth particle-wall collisions ($e=1$, $\beta_{pp} = 1, \beta_{pw} = -1 $ in the 
collision law \ref{eq:collrule2}). Here, there is a change in the angular velocity of the particles in particle-particle
collisions, but there is no change in particle-wall collisions. In this case, there is no tangential force exerted on the 
particle phase in a particle-wall collision.
\item Rough particle-particle and particle-wall collisions ($e=1$, $\beta_{pp} = \beta_{pw} = 1 $ in the 
collision law \ref{eq:collrule2}). Here, there is a tangential force exerted by the walls on the particles.
\end{enumerate}
We have also carried out smooth particle-particle and rough particle-wall collisions
($\beta_{pp} = \beta_{pw} = 1 $ in the collision law \ref{eq:collrule2}); these are not quantitatively
different from (iii).

\begin{figure}
 \begin{subfigure}{0.49\textwidth}
    \includegraphics[width=1\textwidth]{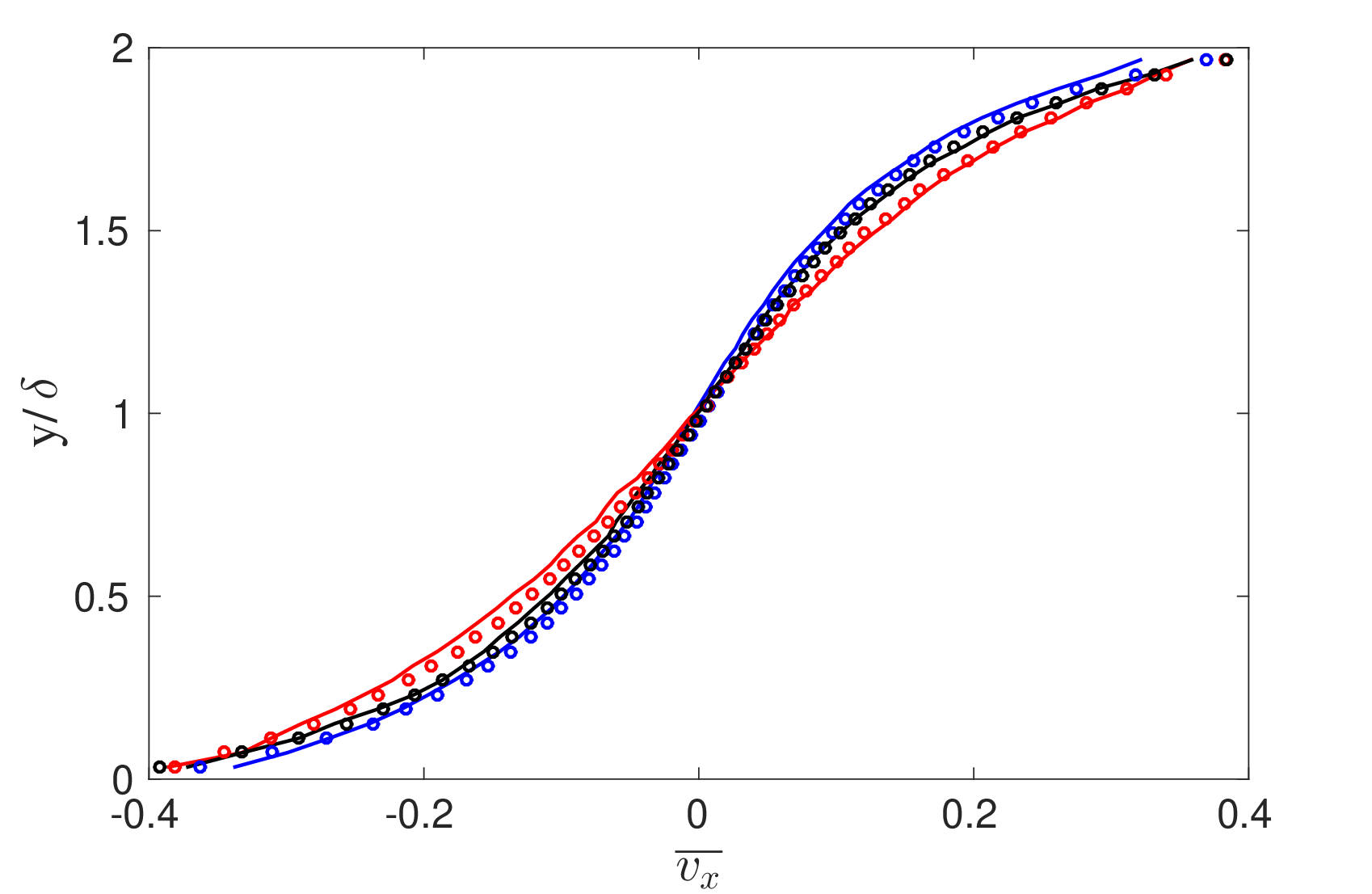}
    \caption*{(a)}
\end{subfigure}
\begin{subfigure}{0.49\textwidth}
 	\includegraphics[width=1\textwidth]{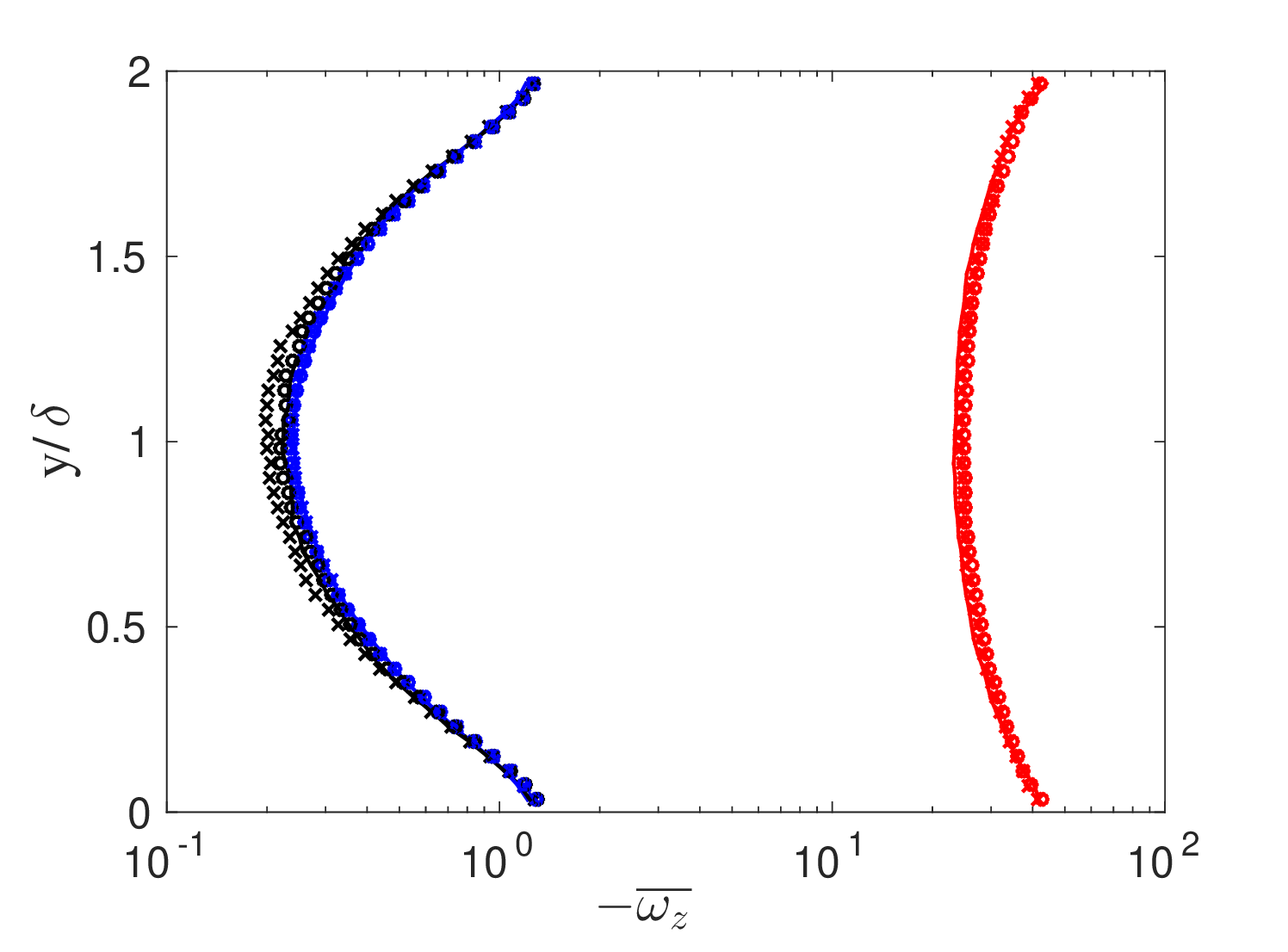}
 	\caption*{(b)}
 	\end{subfigure}
 	\begin{subfigure}{1.0\textwidth}
 	\centering
 	\includegraphics[width=0.65\textwidth]{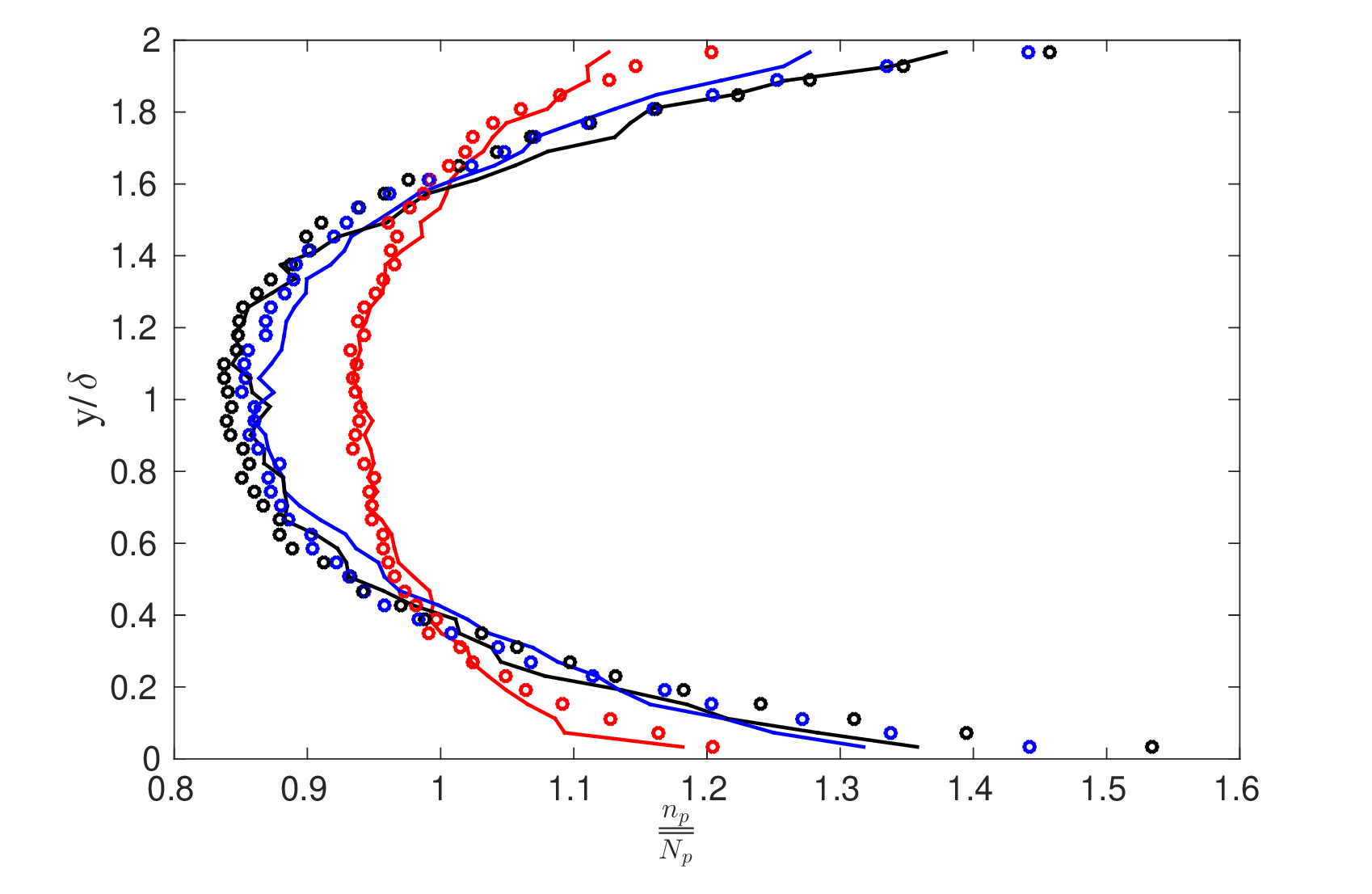}
  	\caption*{(c)}
  	\end{subfigure}
 	\caption{Effect of roughness on (a) mean particle velocity $\overline{v_x}$, (b) mean particle angular velocity $\overline{\omega_z}$ and (c) particle concentration $\frac{n_p}{\overline{n_p}}$ at different wall-normal positions ($y/\delta$). The lines are the results
  from DNS simulations that resolve all the turbulence length scales without any sub-grid modeling, and the symbols
  are the results of the F3T model. The colours are \textcolor{blue}{\bf ---} $\beta_{pp}=
 \beta_{pw}=-1$, \textcolor{red}{\bf ---} $\beta_{pp}=
 \beta_{pw}=1$, and {\bf ---} $\beta_{pp}=1$,  $\beta_{pw}=-1$.}
 	\label{wall_beta_mean_part}     
 \end{figure}

The mean velocity, mean vorticity and particle number density profiles are shown in figure \ref{wall_beta_mean_part}.
Figure \ref{wall_beta_mean_part} (a) shows that there is no discernible change in the strain rate between smooth and
rough particle-particle collisions, indicating that particle roughness has little effect on the mean velocity profile.
However, there is a small but perceptible increase in the strain rate when the particle-wall collisions are rough. The
F3TS simulations do quantitatively capture these small variations in the mean velocity profile. Figure 
\ref{wall_beta_mean_part} (b) shows that there is a large increase in the mean angular velocity due to 
the roughness of particle-wall collisions. 
The mean angular velocity for rough particle-wall collisions is more than an
order of magnitude higher than that for smooth particle-wall collisions. It should be noted that for smooth particle-wall
collisions, the particle angular velocity is only affected by the torque due to fluid vorticity and that due to 
particle-particle collisions. The results in \ref{wall_beta_mean_part} (b) indicate that the change in angular
momentum in rough particle-particle collisions does not alter the average angular velocity, but there is a significant
increase in the mean angular momentum due to rough wall-particle collisions. The large increase in the magnitude
 of the angular velocity is caused by the slip for the particle mean velocity at the wall. 
 When a particle collides with the wall, the increase in the angular velocity is
 $O(U/d_p)$ for rough inelastic particles (see equation \ref{eq:collrule5}). This large increase in the angular velocity in particle-wall collisions results in a large increase in the mean angular velocity across the channel. In comparison, there is no 
 change in the particle angular velocity in a smooth particle-wall collision.

The particle number density, shown in \ref{wall_beta_mean_part} (c), is higher lower at the center and higher close to
the walls. There is not much change in the number density profile when the particle-particle collisions are 
rough or smooth, but there is a distinct change in the number density profile when the particle-wall are rough. The F3TS
simulations are able to quantitatively reproduce the variation in the number density due to rough particle-wall
collisions. 

\begin{figure}
 \begin{subfigure}{0.49\textwidth}
  \includegraphics[width=1.0\textwidth]{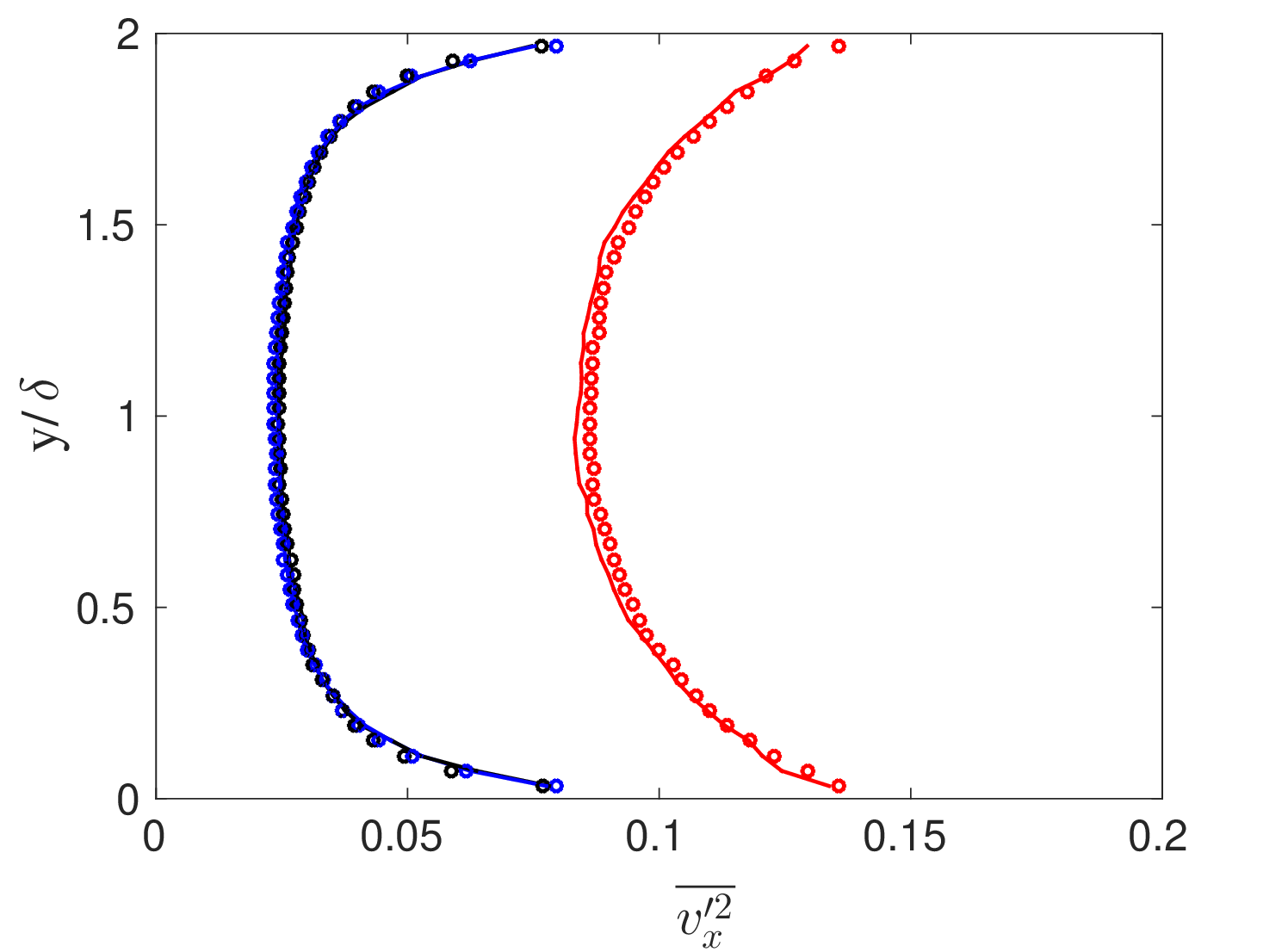}
    \caption*{(a)}
\end{subfigure}
\begin{subfigure}{0.49\textwidth}
 	\includegraphics[width=1.0\textwidth]{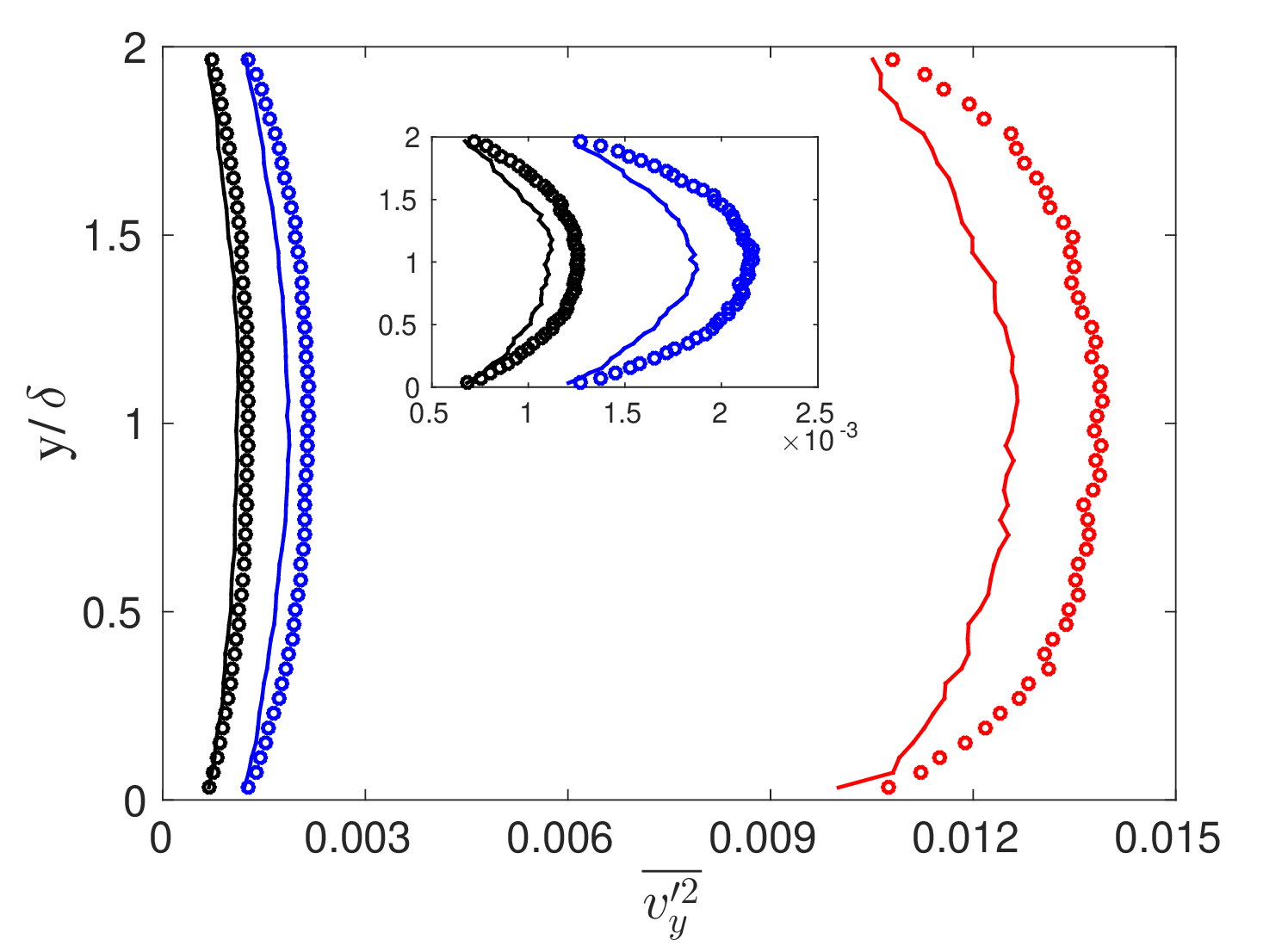}
 	\caption*{(b)}
 	\end{subfigure}
 	\begin{subfigure}{0.49\textwidth}
 	\includegraphics[width=1.0\textwidth]{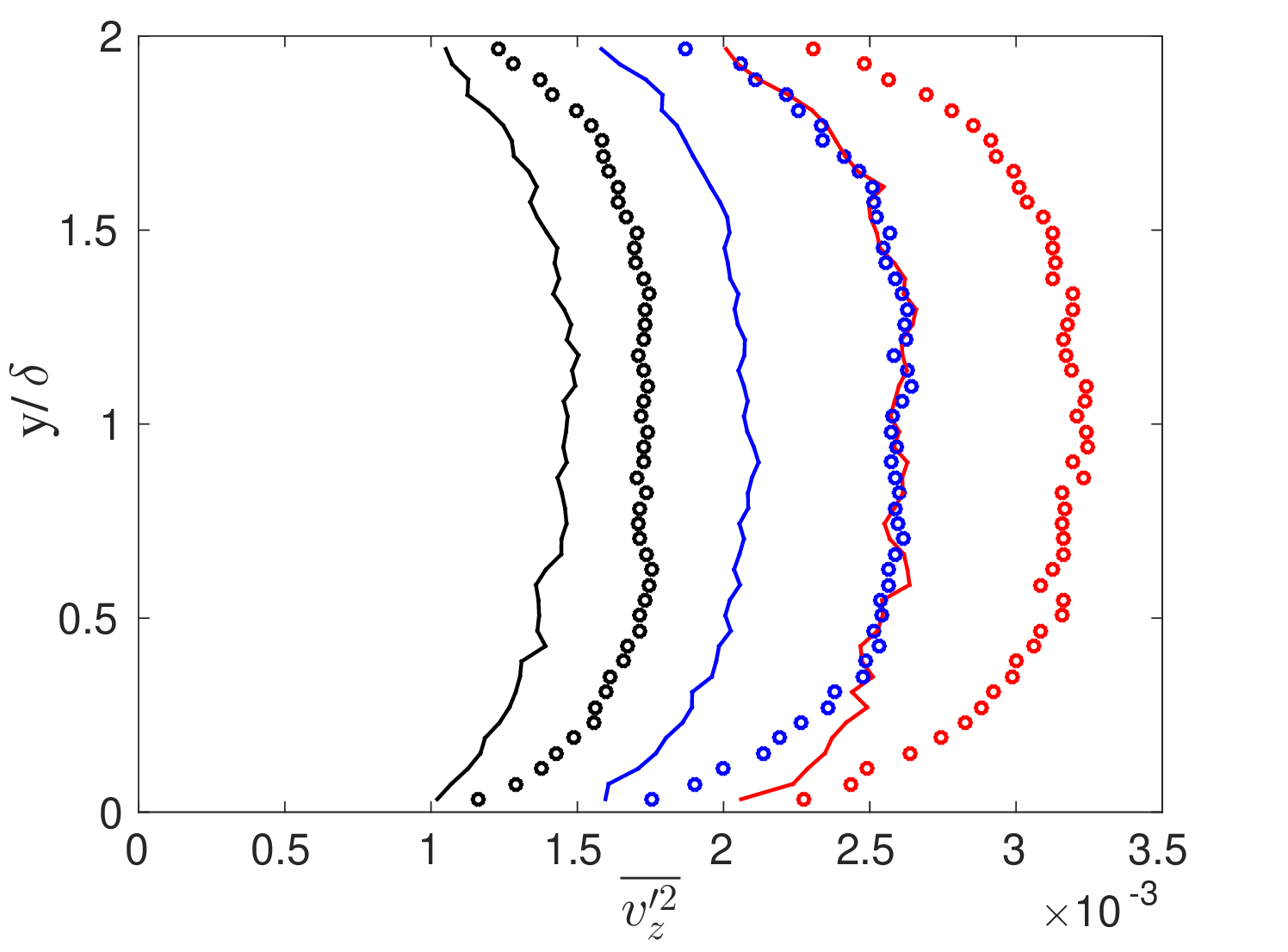}
  	\caption*{(c)}
  	\end{subfigure}
\begin{subfigure}{0.49\textwidth}
 	\includegraphics[width=1.0\textwidth]{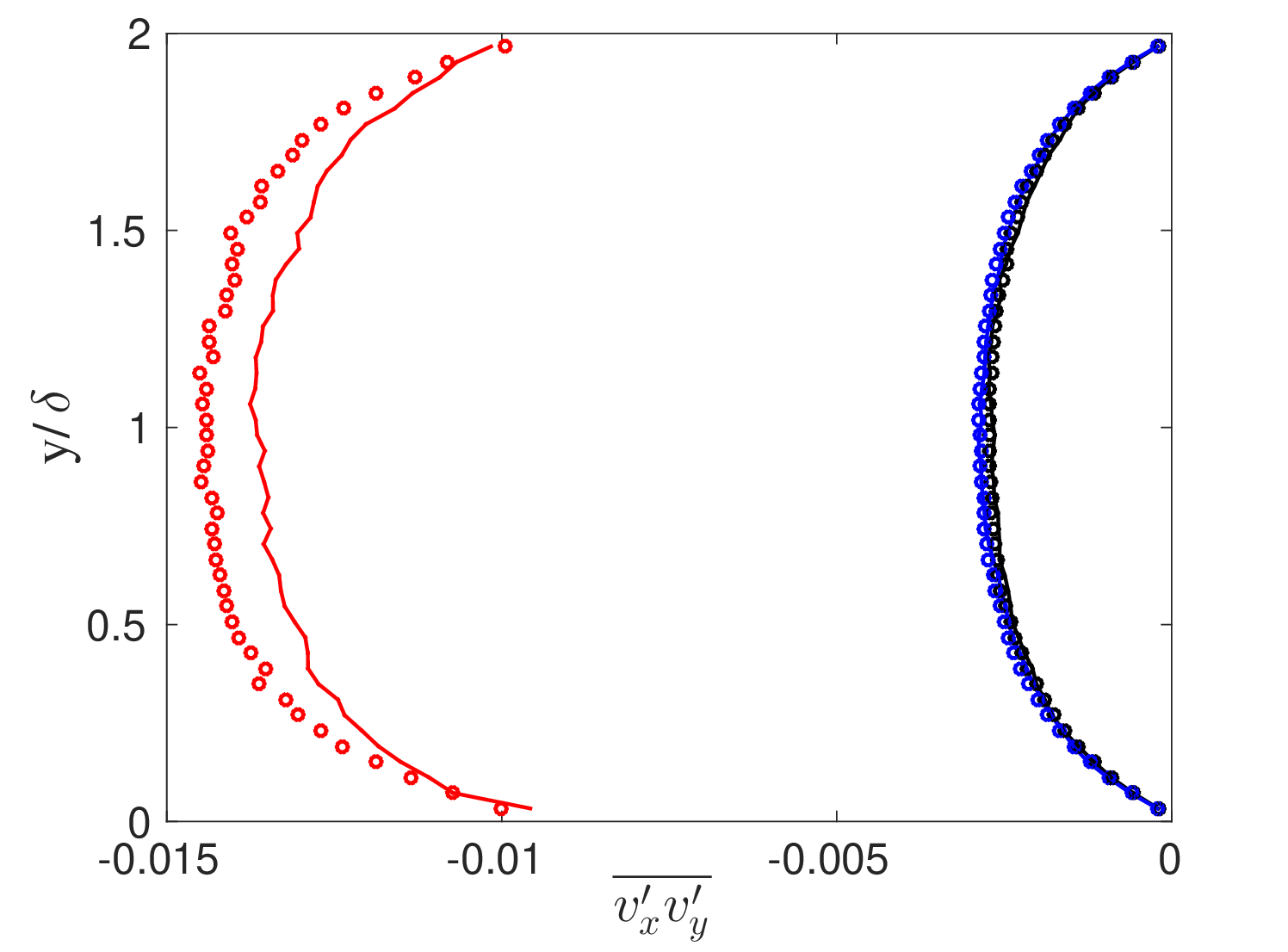}
 	\caption*{(d)}
 	\end{subfigure}
  	\caption{Effect of roughness on (a) $\overline{v_x'^2}$, (b) $\overline{v_z'^2}$ and (c) $\overline{v_y'^2}$ at different wall-normal positions ($y/\delta$) and (d) $\mbox{} - \overline{v_x' v_y'}$. 
  	The lines are the results
  from DNS simulations that resolve all the turbulence length scales without any sub-grid modeling, and the points
  are the results of the F3T model. The colours are \textcolor{blue}{\bf ---} $\beta_{pp}=
 \beta_{pw}=-1$, \textcolor{red}{\bf ---} $\beta_{pp}=
 \beta_{pw}=1$, and {\bf ---} $\beta_{pp}=1$,  $\beta_{pw}=-1$.
  	}
 	\label{wall_beta_part_ms}     
 \end{figure}

The second moments of the particle velocity fluctuations are shown as a function of cross-stream position in figure
\ref{wall_beta_part_ms}. 
It is observed that there is a significant effect of particle-wall roughness on the mean square velocity.
The stream-wise mean square velocity $\overline{v_x^{\prime 2}}$ (figure \ref{wall_beta_part_ms} (a))
increases by a factor of $2$ when the particle-wall collisions are rough, but there is little difference 
between rough and smooth particle-particle collisions. The  cross-stream mean square velocity $\overline{v_y^{\prime 2}}$
does increase by a factor of $2$ for rough particle-particle collisions in comparison to smooth particle-particle
collisions, but there is an increase by an order of magnitude when the particle-wall collisions are made rough.
There is also a significant, but smaller, increase in the span-wise mean square velocity $\overline{v_z^{\prime 2}}$
due to particle-particle and particle-wall roughness. The F3TS results are in quantitative agreement with the 
DNS results for $\overline{v_x^{\prime 2}}$ and $\overline{v_y^{\prime 2}}$, but there is some
numerical difference for $\overline{v_z^{\prime 2}}$ even though the trends are correctly reproduced.
It should be noted that the magnitude of $\overline{v_z^{\prime 2}}$ is much smaller than the other
two components, and so the error in the F3TS prediction for the mean square velocity is small. There is
an increase by over an order of magnitude in the particle Reynolds stress, $- \overline{v_x' v_y'}$, as 
shown in figure \ref{wall_beta_part_ms}.

\begin{figure}
 \begin{subfigure}{0.49\textwidth}
  \includegraphics[width=1.0\textwidth]{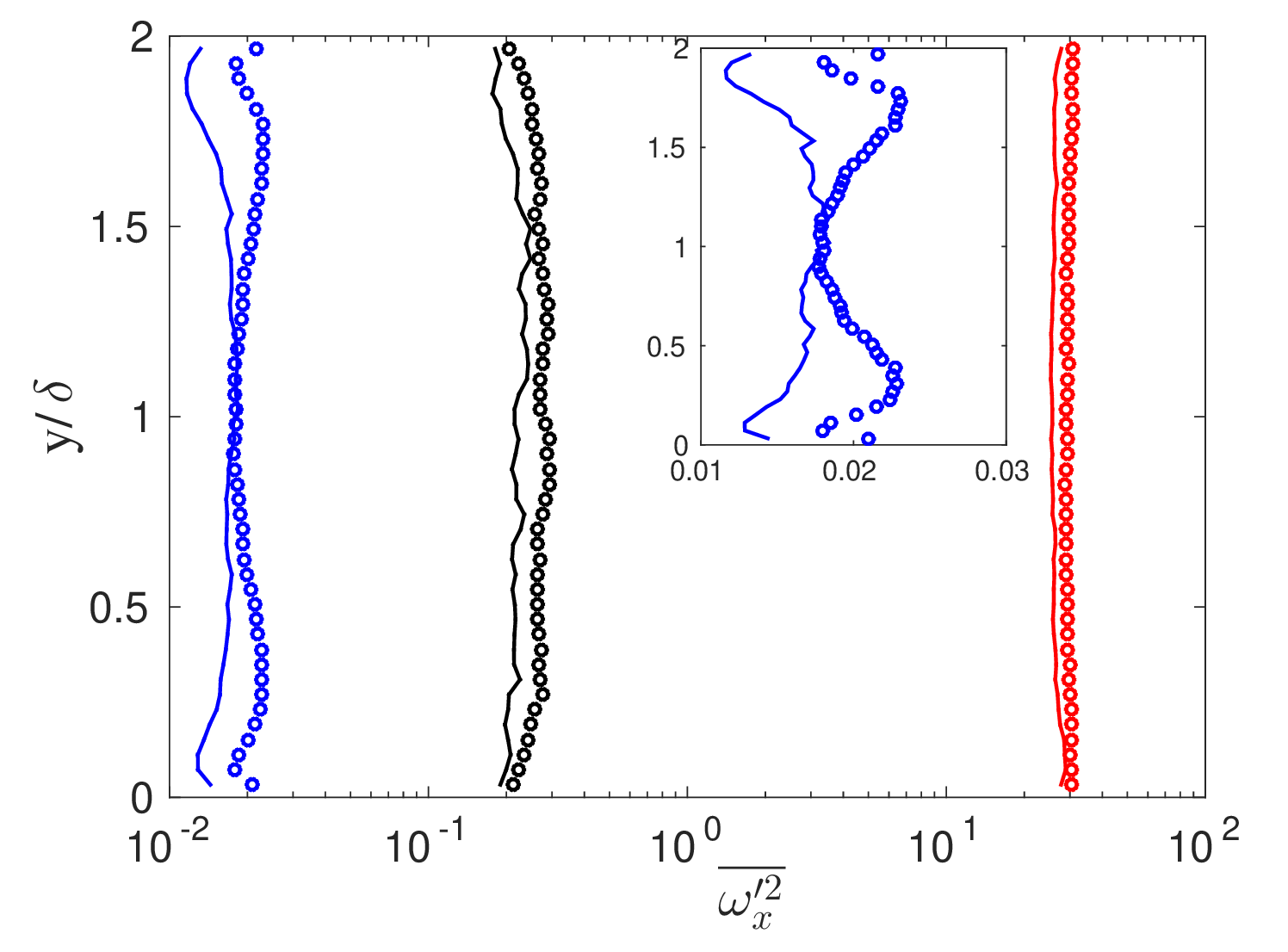}
    \caption*{(a)}
\end{subfigure}
\begin{subfigure}{0.49\textwidth}
 	\includegraphics[width=1.0\textwidth]{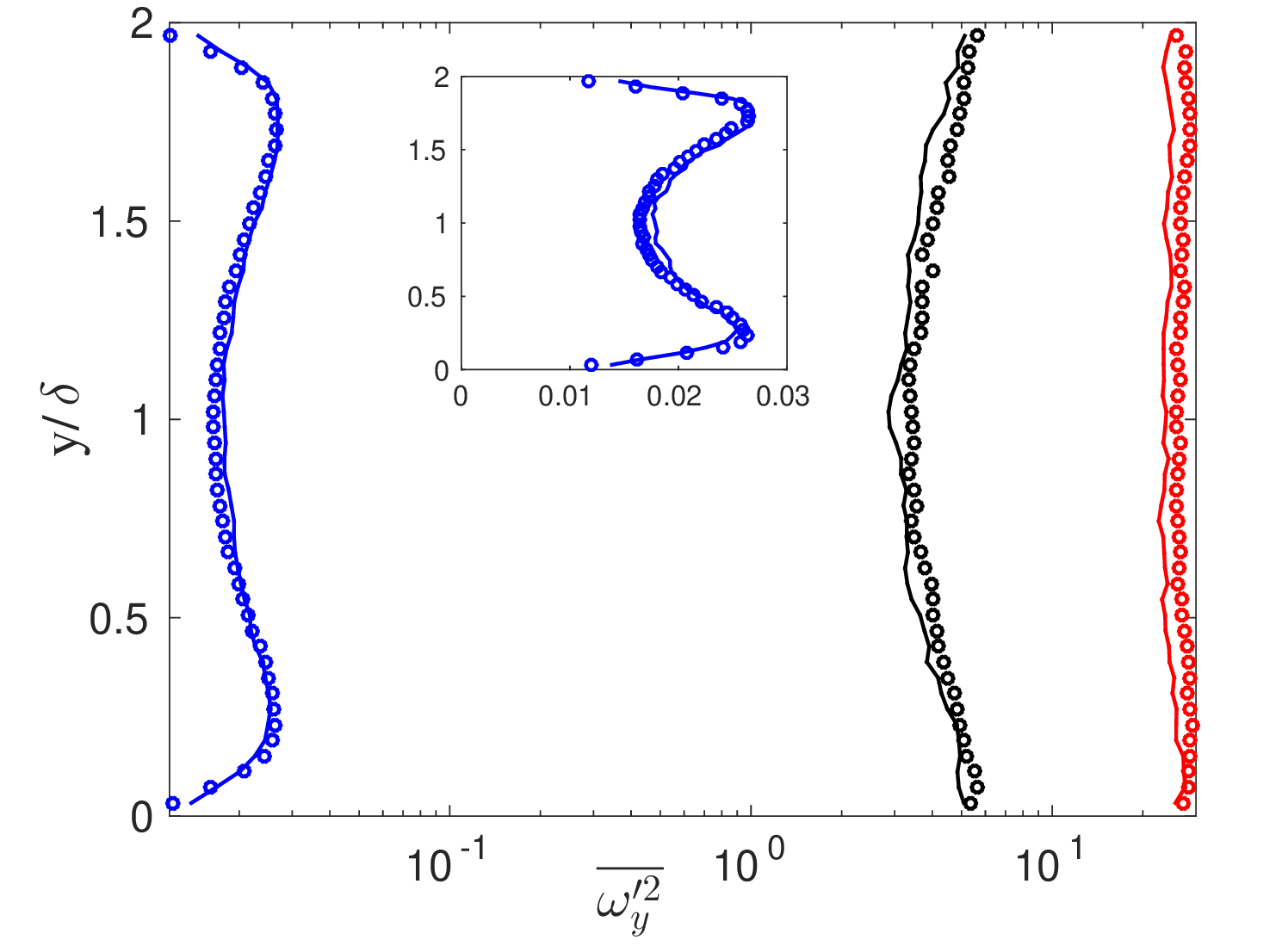}
 	\caption*{(b)}
 	\end{subfigure}
 	\begin{subfigure}{0.49\textwidth}
 	\includegraphics[width=1.0\textwidth]{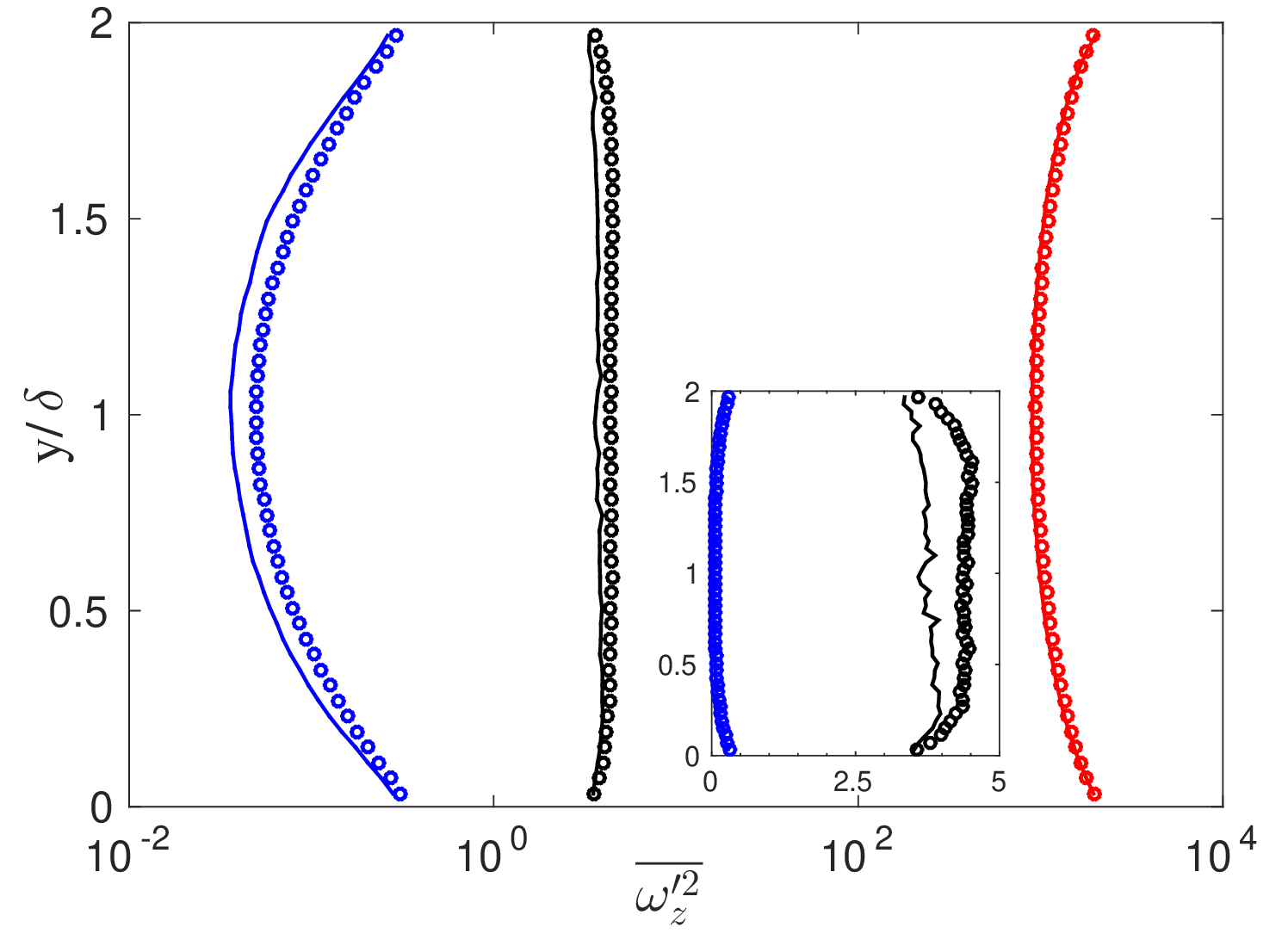}
  	\caption*{(c)}
  	\end{subfigure}
  	\begin{subfigure}{0.49\textwidth}
 	\includegraphics[width=1.0\textwidth]{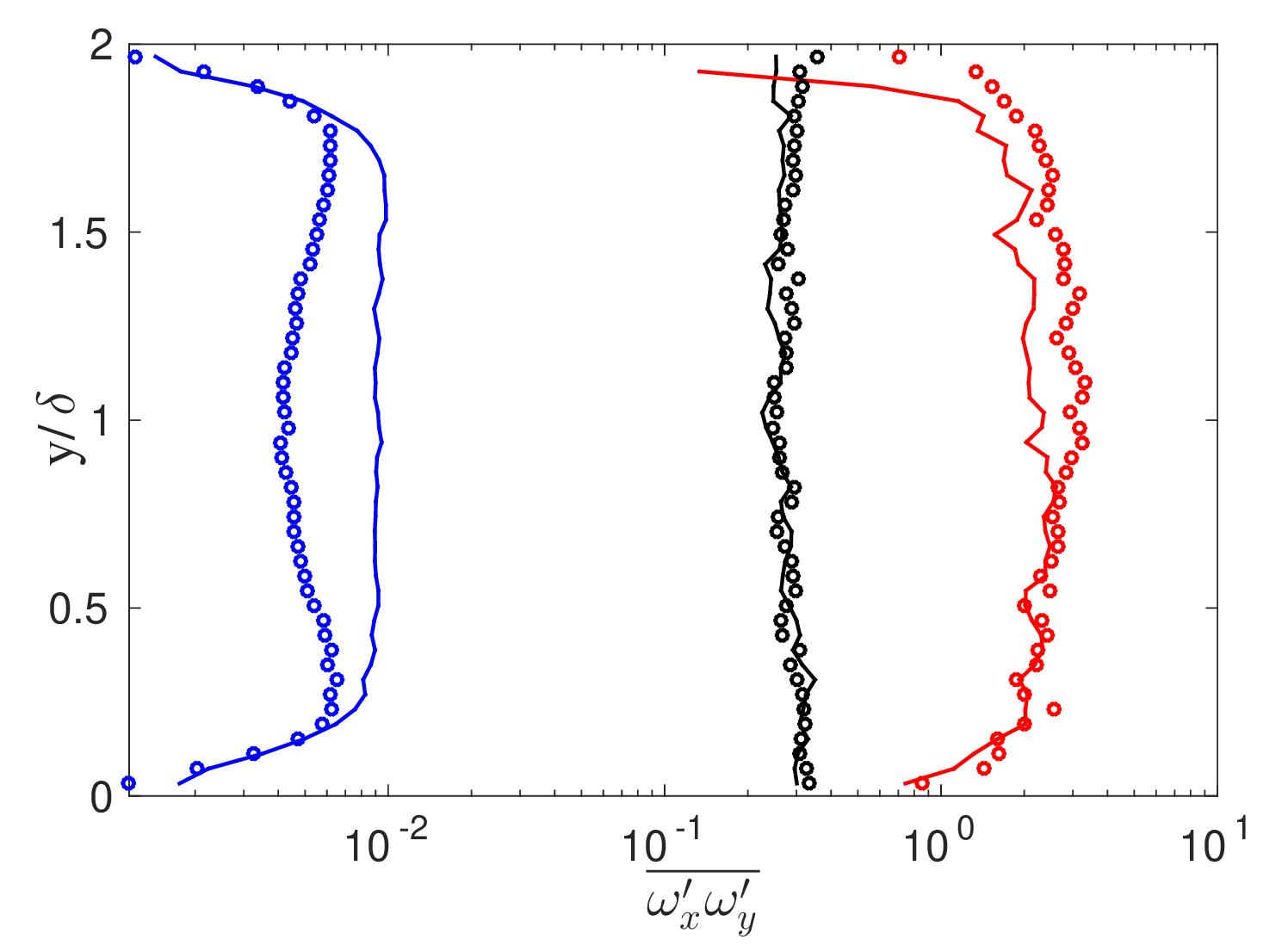}
  	\caption*{(d)}
  	\end{subfigure}
 	\caption{Effect of roughness on (a) $\overline{\omega_x'^2}$, (b) $\overline{\omega_y'^2}$, (c) $\overline{\omega_z'^2}$ and (d) $\overline{\omega_x' \omega_y'}$ at different wall-normal positions ($y/\delta$).
 	The lines are the results
  from DNS simulations that resolve all the turbulence length scales without any sub-grid modeling, and the points
  are the results of the F3T model. The colours are \textcolor{blue}{\bf ---} $\beta_{pp}=
 \beta_{pw}=-1$, \textcolor{red}{\bf ---} $\beta_{pp}=
 \beta_{pw}=1$, and {\bf ---} $\beta_{pp}=1$,  $\beta_{pw}=-1$.
 	\label{wall_beta_part_rot_ms}}
 \end{figure}

The most dramatic effect of roughness is observed in the mean square of the angular velocity in
figure \ref{wall_beta_part_rot_ms}. In comparison to the fluctuations in the linear velocities, there is 
significantly higher anisotropy in the fluctuations in the angular velocities, and the span-wise angular
velocities are up to two orders of magnitude higher than the other two components.
In comparison to smooth particle-particle and particle-wall collisions,
figure \ref{wall_beta_part_rot_ms} (c) shows that $\overline{\omega_z^{\prime 2}}$ increases by a factor of $10$ 
for rough particle-particle and smooth particle-wall collisions, and a factor of $10^4$ for rough particle-particle
and particle-wall collisions. Figure \ref{wall_beta_part_rot_ms} (a) and (b) show that $\overline{\omega_x^{\prime 2}}$
and $\overline{\omega_y^{\prime 2}}$ increase by a factor of about $50$ for rough particle-particle and smooth 
particle-wall collisions, and about $250$ for rough particle-particle and particle-wall collisions.
The qualitative variations in the mean square of the angular velocity are quantitatively captured by the 
F3TS simulations, even for the smallest component $\overline{\omega_y^{\prime 2}}$ which is two orders
of magnitude smaller than $\overline{\omega_z^{\prime 2}}$.

\begin{figure}
\begin{subfigure}{1.0\textwidth}
\centerline{\includegraphics[width=0.6\textwidth]{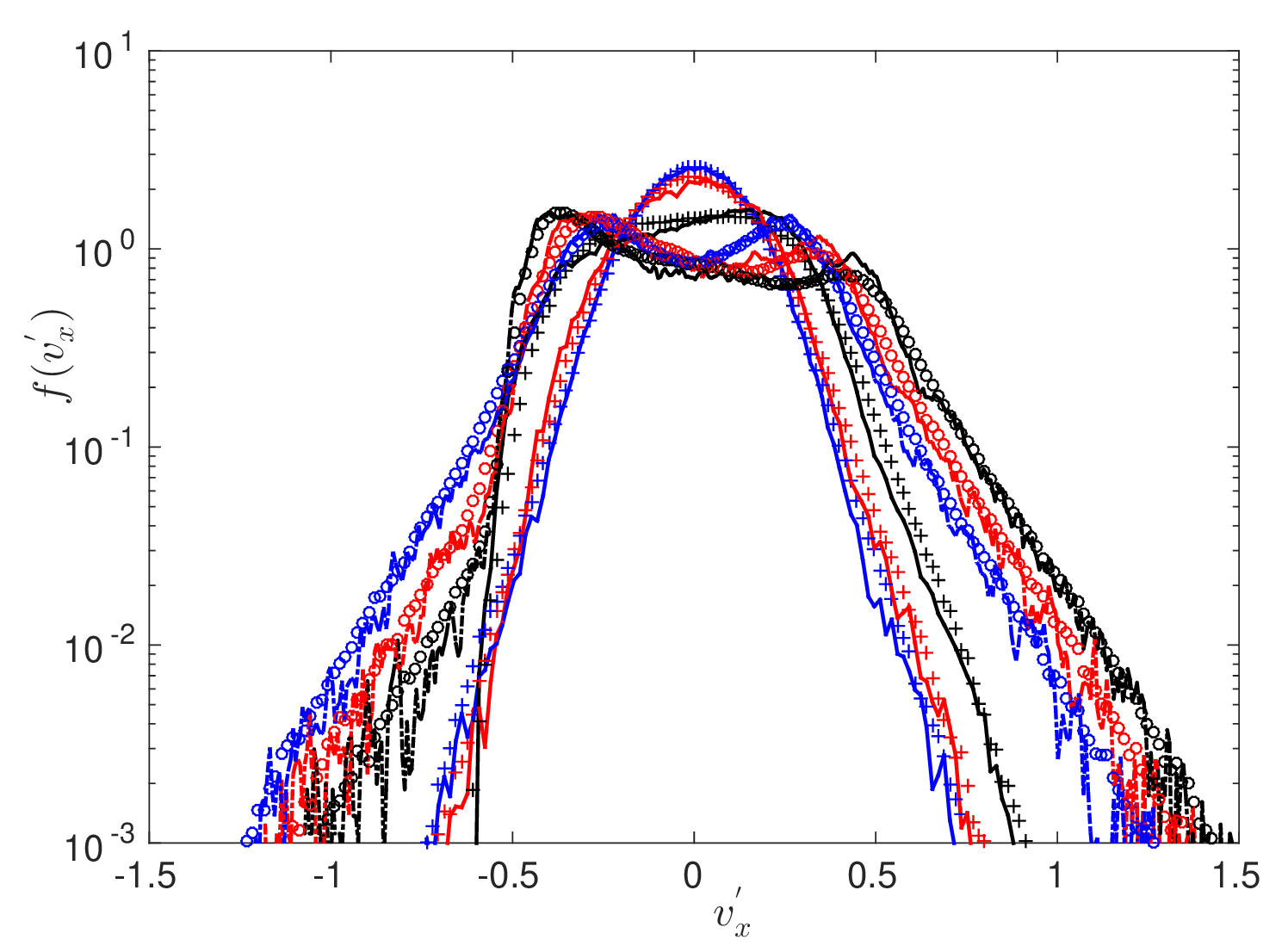}}
\caption*{(a)}
\end{subfigure}
\begin{subfigure}{0.49\textwidth}
\includegraphics[width=1.0\textwidth]{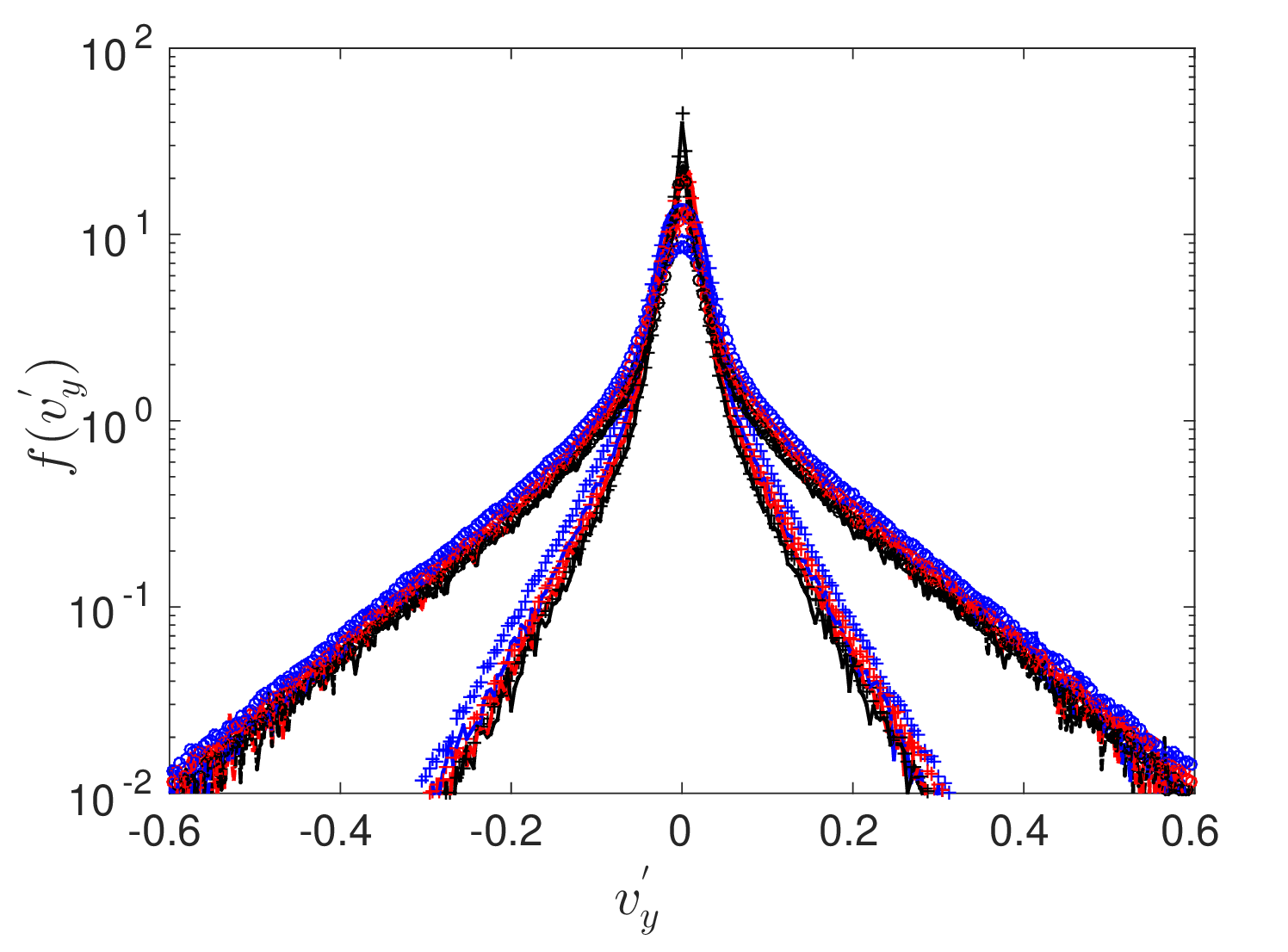}
\caption*{(b)}
\end{subfigure}
\begin{subfigure}{0.49\textwidth}
\includegraphics[width=1.0\textwidth]{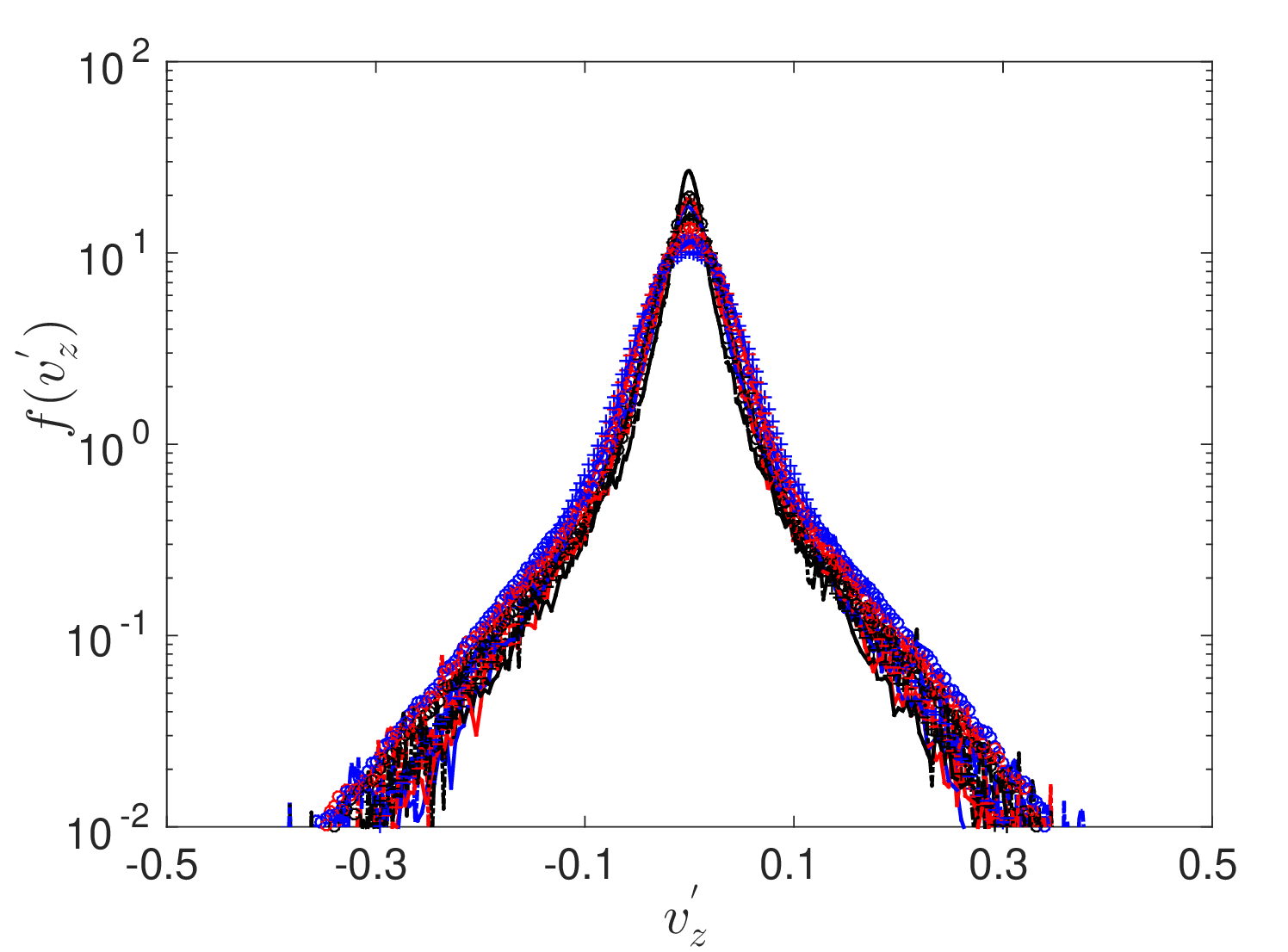}
\caption*{(c)}
\end{subfigure}
\caption{Particle velocity distribution function (a) $f(v_x)$, (b) $f(v_y)$ and (c) $f(v_z)$ computed (i) at the center $y/\delta=1.0$ \textcolor{blue}{\bf ---}, (ii) at the middle $y/\delta=0.37$ \textcolor{red}{\bf ---}, and (iii) near the wall $y/\delta=0.09$ {\bf ---}. 
Four cases are compared at each position: (1) $\beta=-1$ DNS '$-$' lines, (2) $\beta=-1$ F3TS '$+$' symbol, (3) $\beta=+1$ DNS '$-\cdot$' lines and (4) $\beta=+1$ F3TS '$o$' symbol. 
}
\label{fig:beta_vel_dist}
\end{figure}

The velocity distribution functions for the translational velocity fluctuations are shown in 
figure (\ref{fig:beta_vel_dist} (a)) to (\ref{fig:beta_vel_dist} (c)). 
In comparison to smooth particles, the distribution function for 
the stream-wise velocity for rough particles is more flat at the center with two maxima at non-zero 
fluctuating velocities and a very slowly decaying tail. The
conversion between the translational and rotational energies in particle-particle and 
particle-wall collisions due to particle roughness results in a bimodal distribution with
maxima at non-zero fluctuating velocities. The asymmetry observed in $f(v_x')$, is similar to
that for a granular Poiseuille Flow in
presence of rough walls \citet{vijayakumar2007velocity,alam2010velocity}. 
The bimodal distribution for $f(v_x')$ is symmetric about zero velocity at the center. 
The reason for weak bimodality is as
follows. When the particles move from an initial location at the center towards the
lower wall with negative velocity, the mean velocity decreases and
consequently the velocity fluctuation at the final location increases.
This induces a positive fluctuation. When the particle collides with the lower wall,
a negative velocity fluctuation is induced due to wall roughness. Thus,
the bimodal distribution is due to the cross-stream migration of the 
particles and the wall roughness.

The distributions for the cross-stream and span-wise
velocities, $f(v_y')$ and $f(v_z')$, shown in figure
\ref{fig:beta_vel_dist}(b) and (c), are Gaussian at the center with long exponential tails for high
velocity. In contrast to the distributions for smooth particle-particle and particle-wall
collisions in figure \ref{fig:vel_dist}, the tails of the distribution function decay much slower
for rough collisions. The reason is as follows. When there are large angular velocity
fluctuations, a collision between particles with large angular velocities could result
in one or both of the particles acquiring a large linear velocity. Similarly, the rough 
collision of a rapidly spinning particle with the wall could also result in a large
angular velocity. Due to this, the distributions for the cross-stream and span-wise
velocities decay far slower than the equivalent distributions for smooth particles.

\begin{figure}
\begin{subfigure}{0.49\textwidth}
\includegraphics[width=1.0\textwidth]{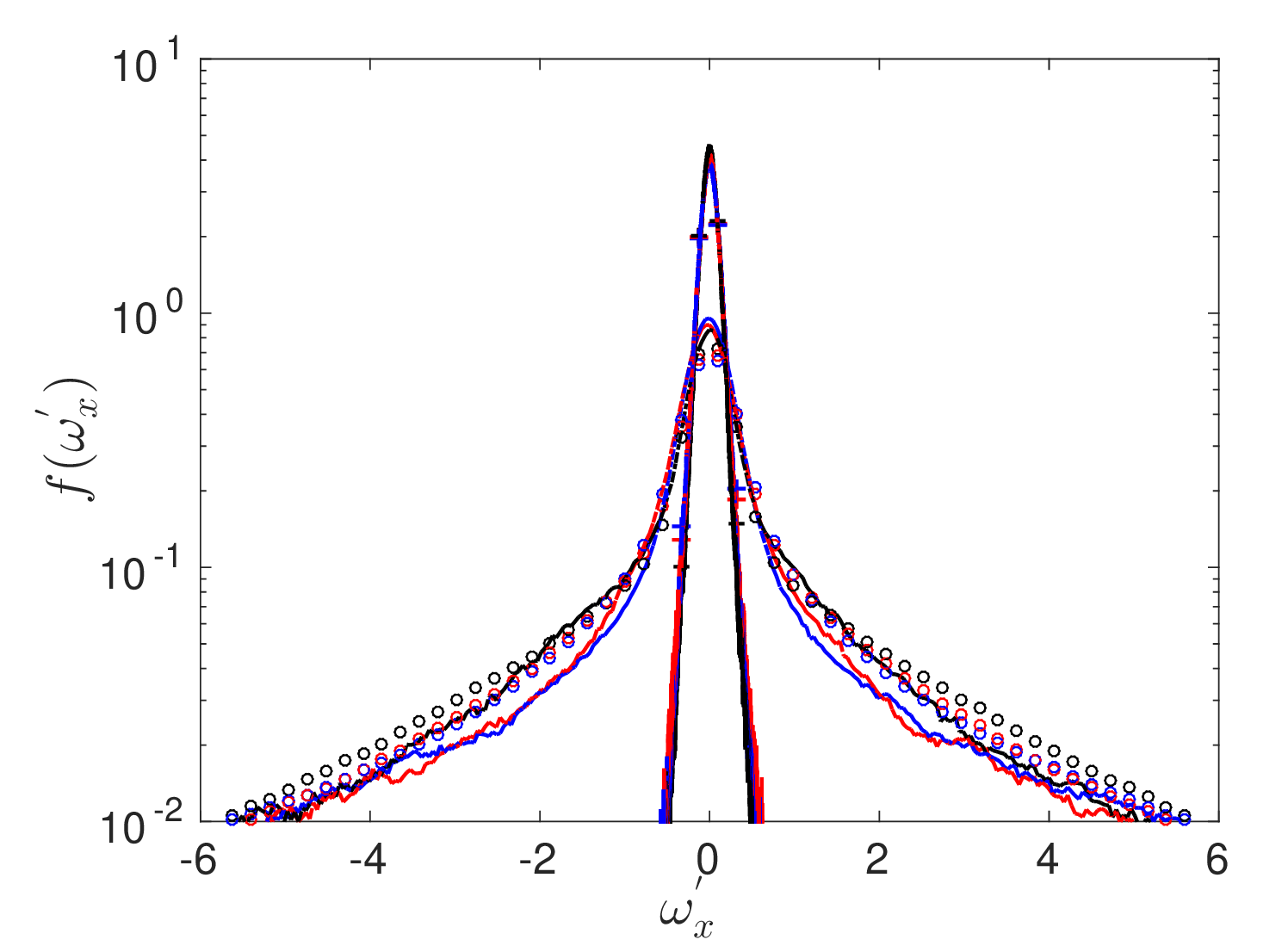}
\caption*{(a)}
\end{subfigure}
\begin{subfigure}{0.49\textwidth}
\includegraphics[width=1.0\textwidth]{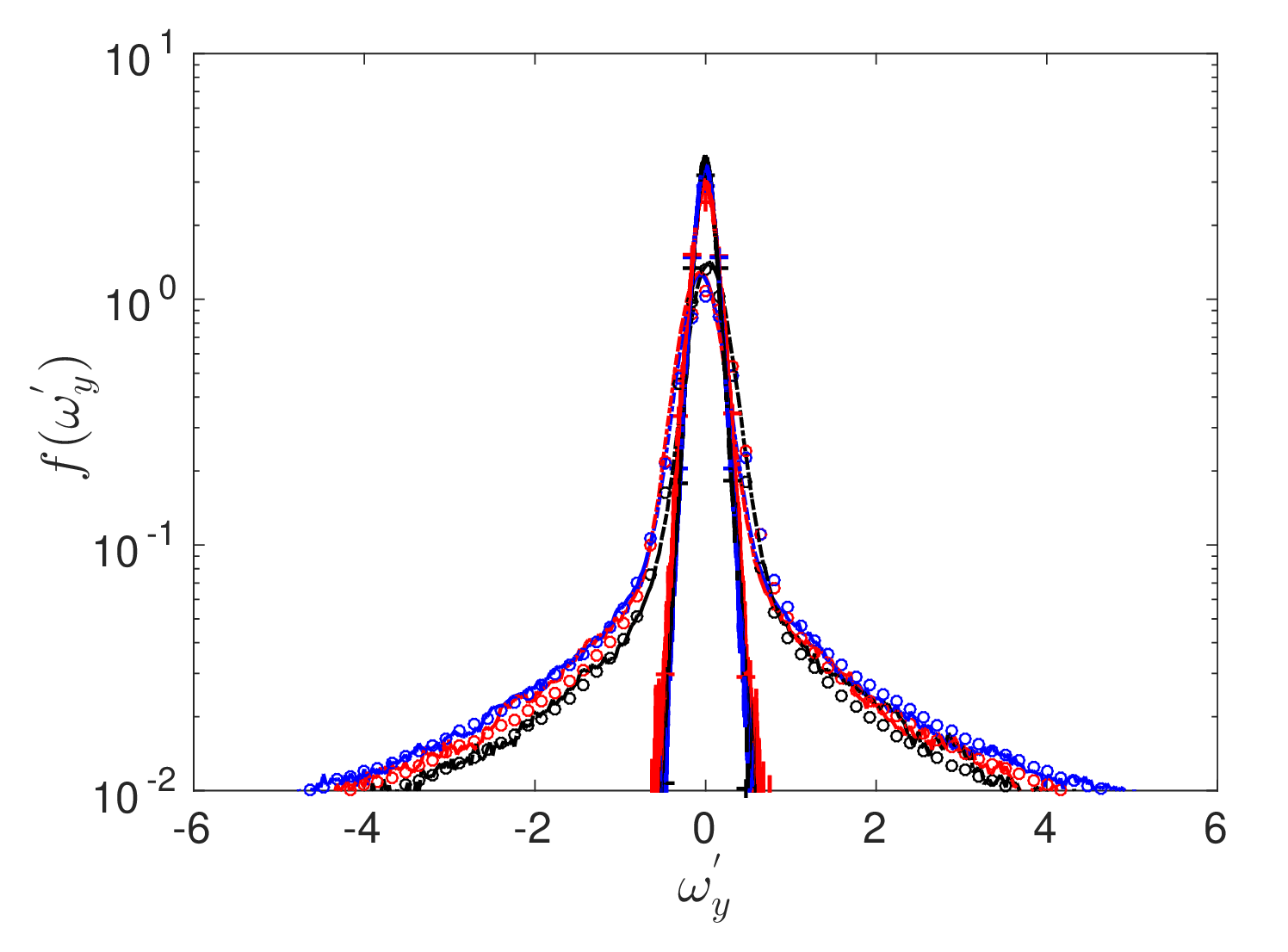}
\caption*{(b)}
\end{subfigure}
\begin{subfigure}{1.0\textwidth}
\centering
\includegraphics[width=0.6\textwidth]{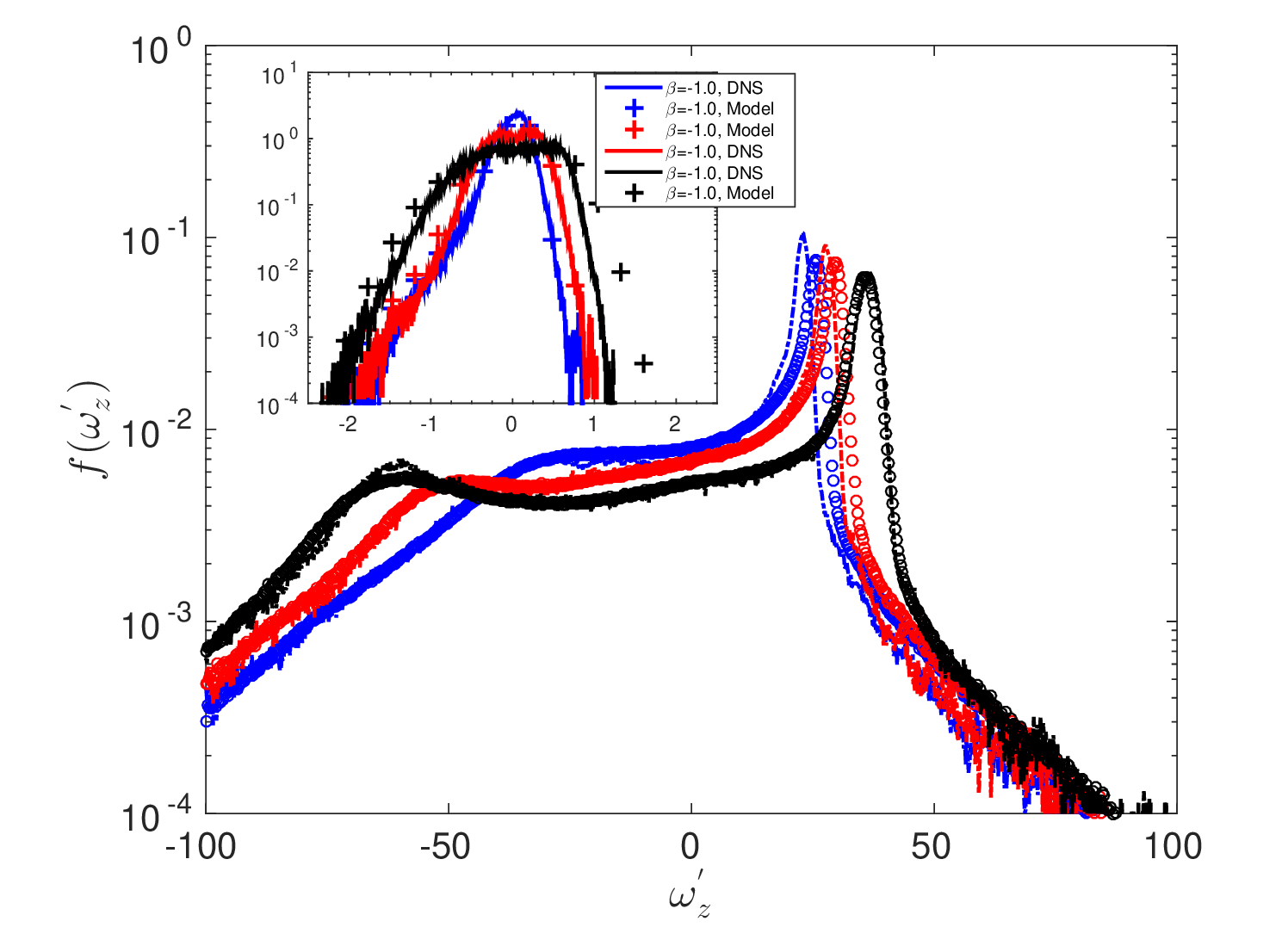}
\caption*{(c)}
\end{subfigure}
\caption{Particle rotational velocity distribution function (a) $f(\omega_x^{'})$, (b) $f(\omega_y^{'})$ and (c) $f({\omega_z^{'}})$ computed (i) at the center $y/\delta=1.0$ \textcolor{blue}{\bf ---}, (ii) at the middle $y/\delta=0.37$ \textcolor{red}{\bf ---} and (iii) near the wall $y/\delta=0.09$ {\bf ---}. 
Four cases are compared in each position: (1) $\beta=-1$ DNS '$-$' lines, (2) $\beta=-1$ F3TS '$+$' symbol, (3) $\beta=+1$ DNS '$-\cdot$' lines and (4) $\beta=+1$ F3TS '$o$' symbol.
}
\label{fig:beta_rot_vel_dist}
\end{figure}




There is a dramatic change in the form of the angular velocity distribution function when particle roughness is included.
The distribution function for the span-wise angular velocity is shown in figure \ref{fig:beta_rot_vel_dist} (a). The distribution
function is asymmetric with a positive maximum and a clear negative skewness 
in the angular velocity distribution, and a rapid 
decay of the distribution function for positive angular velocity. Though the maximum in the distribution is
at positive \( \omega_z^\prime \), it should be noted that the average angular velocity \( \overline{\omega}_z \)
is negative (see figure \ref{fig:e_mean_part_stats}(b)). When expressed in terms of the absolute angular velocity, the maximum
occurs at a negative value of \( \omega_z \). This maximum is because  both walls rotate the particles in the 
clockwise direction, thereby inducing a negative angular velocity. The location of the 
maximum of the distribution function shifts towards 
higher angular velocity as we move from the center to the wall. 
The other two components of particle angular velocity fluctuation, 
$f(\omega_x')$ and $f(\omega_y')$, are symmetric with slowly decaying tails as shown in 
figures \ref{fig:beta_rot_vel_dist} (a) and \ref{fig:beta_rot_vel_dist} (b). The decay at high velocity is slower for rough
particles in comparison to smooth particles. In all the cases, the distribution functions are well predicted by F3TS simulations.

\begin{figure}
     \centering
     \includegraphics[width=1.0\textwidth]{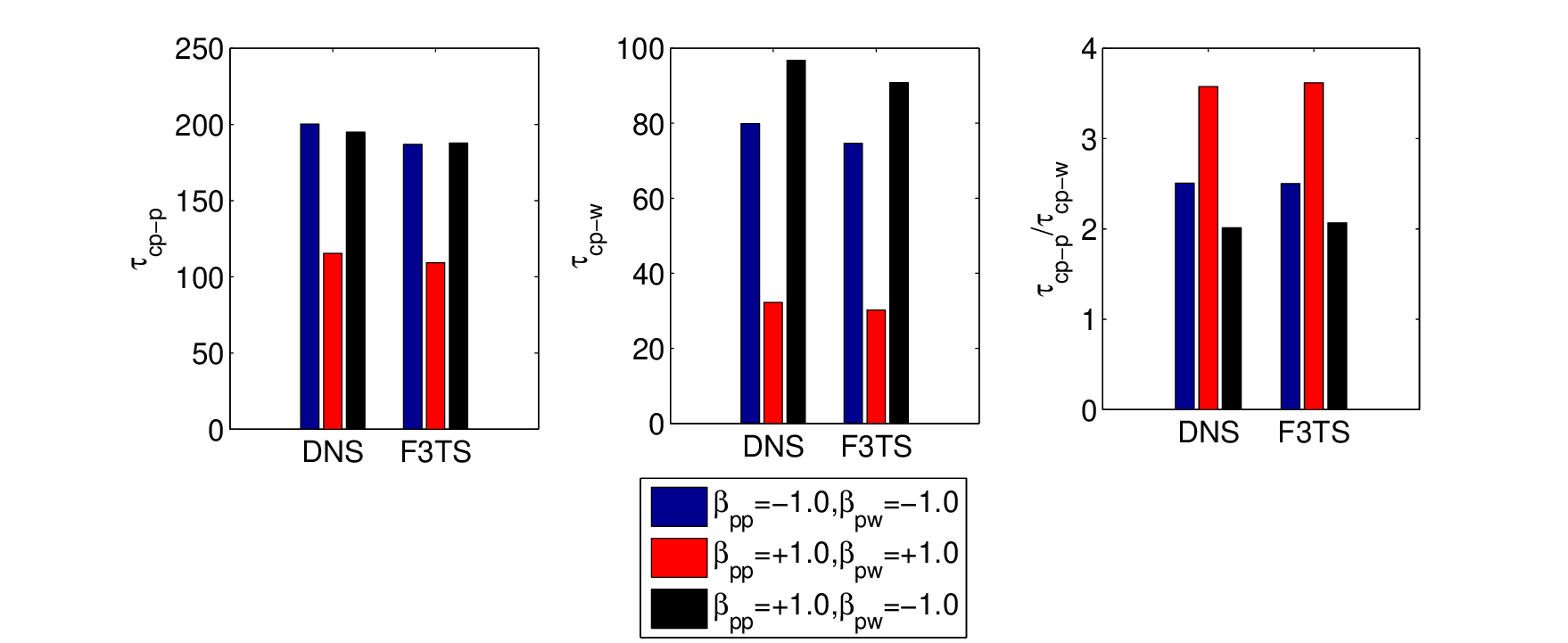}
     \caption{Comparison of collision statistics (a) average inter-particle collision time $\tau_{cp-p}$, (b) average wall-particle collision time $\tau_{cp-w}$ and (c) ratio of average inter-particle collision time to average wall-particle collision time $\tau_{cp-p}/\tau_{cp-w}$, obtained from DNS simulations that resolve all the turbulence
     length scales without sub-grid modeling (labeled DNS), and the F3T model (labeled F3TS). The collision times
     are all scaled by the flow time scale $(\delta/U)$. The ratio of particle relaxation time to flow time scale is 50.}
     \label{fig:coll_freq_bar}
 \end{figure}

From figure \ref{wall_beta_part_ms}, it is observed that
the magnitudes of velocity fluctuations for $\overline{v_y'^2}$ and $\overline{v_z'^2}$ are highest in the case of $\beta_{pw}=+1.0$, $\beta_{pp}=+1.0$ and lowest when $\beta_{pw}=-1.0$, $\beta_{pp}=+1.0$. This ordering is
correlated to the average particle-particle and particle-wall collision times shown in fig. \ref{fig:coll_freq_bar}(a) 
and (b). Here, the highest wall-particle collision frequency (i.e. lowest average $\tau_{cp_w}$) is observed for $\beta_{pw}=+1.0$, $\beta_{pp}=+1.0$, and the lowest wall-particle collision frequency (i.e. highest average $\tau_{cp_w}$) for $\beta_{pw}=-1.0$, $\beta_{pp}=+1.0$. Particle roughness increases the particle rotational velocity fluctuations,
and there is an exchange between rotational and translational energy in particle-particle and particle-wall
collisions. Due to this, both the collision frequency and the cross-stream velocity fluctuations are correlated.
Figure \ref{fig:coll_freq_bar} also shows that the collision times for both particle-particle and particle-
wall collisions are accurately predicted by the fluctuating force simulations.

 From the above discussions it is clear that particle roughness affects both the translational and 
rotational velocity statistics even when the volume fraction of the particles in very low. In presence of roughness, 
the rotational velocity statistics are mainly controlled by particle-particle and particle-wall interactions, and are not significantly affected by the fluid torque 
on the particles. All the second moments of translational and rotational velocity fluctuations, the distributions
of the fluctuating velocities and the particle-particle and particle-wall collision times are well predicted by the F3TS simulation method.
A common feature of the PDFs for the fluctuating velocity and vorticity is the asymmetry in the PDFs for
the stream-wise velocity and the span-wise vorticity, which have non-zero average, and the long tails in 
the distributions for the other PDFs which are symmetric. The long tails are indicative of generation of
large fluctuating velocities due to particle-particle or particle-wall collisions. The tails decay more
slowly for rough particle-wall collisions in comparison to smooth particle-wall collisions, because large
fluctuations in angular velocity are generated by particle-wall collisions and the collisions between
rapidly spinning particles results in large linear velocity fluctuations.

\section{Conclusion}
There are two main objectives of the present study on the Couette flow of a turbulent particle-gas suspension. 
The first was to include the effect of inelasticity and particle rotation on the dynamics
of the particle phase in a turbulent particle-gas suspension. Here, we have examined the effect of rough particle-particle and
particle-wall collisions, and the torque exerted on the particles due to the fluid vorticity. The second objective was to 
examine the applicability of the F3T model, where the force and torque are modeled as Gaussian white noise with mean and 
variance determined from direct numerical simulations (DNS) of the unladen flow. Since the focus is on 
examining the effect of turbulence on the dynamics of the particle phase, the turbulence modification due to the forces
exerted by the particles is not included.

The particle-particle and particle-wall interactions are modeled as instantaneous collisions between hard inelastic
spheres in section \ref{sec:coll}. In a collision, the relative velocity is resolved into components 
tangential and normal to the surface of contact. The post-collisional relative velocity normal to the surface is $-e$ 
times the pre-collisional velocity,
and the post-collisional relative velocity tangential to the surface is $- \beta$ times the pre-collisional velocity.
For smooth elastic particles, $\beta = -1$ and $e = 1$---the relative velocity normal to the surface is reversed in
a collision, while the relative velocity tangential to the surface is unchanged. In this case, the angular velocity 
of the particle does not change in a collision. For rough elastic particles, $\beta = 1$
and $e = 1$---the relative velocity normal and tangential to the surface are reversed.

The force and torque on the particles due to the fluid is modeled as Gaussian white noise in the limit where the correlation
time of the fluid velocity fluctuations is much smaller than the collision time or the viscous relaxation time of the 
particles. This leads to Langevin equations for the particle linear and angular velocity, equations \ref{eq:vel} and 
\ref{eq:angvel}, where the acceleration and angular acceleration due to fluid drag is modeled using as a linear 
relaxation with time constants $\tau_v$ and $\tau_r$ respectively, and
the variance of the noise is related to the velocity and vorticity autocorrelation functions. The variance of the 
noise correlations is related to the diffusion tensors in the Boltzmann-Fokker-Planck equation for the velocity and
vorticity distributions, equation \ref{eq:boltzmann}. The diffusion tensors are anisotropic, with different variances in 
different directions, and some of the off-diagonal elements are non-zero due to correlations in the velocity and vorticity
fluctuations in the stream-wise and cross-stream directions. The diffusion tensors also depend on cross-stream position
due to the variation of the fluid velocity fluctuations with cross-stream position.
The elements of the diffusion tensor are determined from
the time correlation function for the velocity and vorticity fluctuations in equations \ref{eq3.1} and \ref{eq3.2}.

This correlation time for the velocity fluctuations was examined in section \ref{sec:Fluid}. It was found that
with two exceptions, all the correlation functions decay exponentially in time in an Eulerian reference frame, and the 
correlation time is smaller than the viscous relaxation time of the particles. The two exceptions are the 
stream-wise velocity autocorrelation function and the span-wise vorticity autocorrelation function, which exhibit a
stretched-exponential decay, and the correlation time is comparable to the viscous relaxation time. Despite the slow
decay of the stream-wise velocity and span-wise vorticity in an Eulerian reference frame, the decay in a Lagrangian
reference frame is found to be exponential, due to the motion of the particle. Therefore, the terms in the diffusion
tensor were calculated using an upper time cut-off in the integrals for the fluid velocity correlation functions, which was
$7$ times the time for the autocorrelation function to decrease to $\mbox{e}^{-1}$ times its value at equal time.
The effect of the upper time cut-off was found to be small---the rotational diffusion coefficients increased by about 
15\% when the upper time cut-off was doubled from $1.33 \tau_r$ to $2.67 \tau_r$.

The other important approximation made in the F3TS is that the distribution of the random force and torque is Gaussian.
This is equivalent to assuming that the distribution of the fluid velocity and vorticity is Gaussian.
The actual forms of the distributions, shown in figures \ref{fig:fluid_vel_dist} and \ref{fig:fluid_vort_dist}, 
are clearly not Gaussian. We examine the extent to which the assumption of a Gaussian distribution for the fluid
fluctuations affects the results for the particle velocity and angular velocity distributions.

The F3TS simulations, where the fluctuating force and torque are modeled as random Gaussian noise, are compared with
the simulations where turbulent fluctuating velocity and vorticity from DNS simulations are used to determine
the force and torque on a particle. The quantities compared include the profiles of the mean velocity and angular
velocity, the second moments of the velocity and angular velocity fluctuations and the distribution functions for
the particle velocity and angular velocity in the different directions.

The important conclusions for the studies on smooth particles, discussed in section \ref{sec:inelastic}, are,
\begin{enumerate}
 \item The mean velocity and angular velocity profiles do not vary significantly between elastic and 
 inelastic particles. There is a small but measurable change in the number density profiles---the 
 number density is higher at the walls and lower at the center when the coefficient of restitution decreases.
 These trends are accurately captured by the F3TS simulations.
 \item The particle mean square velocity in the stream-wise direction is higher than that for the fluid,
 while that in the other two directions is lower. These do not depend on the coefficient of restitution, 
 with the exception of the cross-stream mean square velocity which decreases with coefficient of restitution
 due to inelastic particle-wall collisions. The F3TS simulations are in quantitative agreement with DNS for
 the stream-wise fluctuations and the Reynolds stress. There is some quantitative difference for the cross-stream
 and span-wise fluctuations which are much smaller in magnitude, but the qualitative variation of these
 moments is accurately predicted by F3TS.
 \item The particle angular velocity fluctuations are smaller than the fluctuations of one half of the 
 fluid vorticity, because the particle angular velocity is driven by the fluid vorticity fluctuations.
 These do not vary as the coefficient of restitution is decreased. The F3TS simulations quantitatively
 predict the magnitude and the profiles of the angular velocity fluctuations.
 \item The distribution of the stream-wise velocity and span-wise vorticity fluctuations are highly
 non-Gaussian, they exhibit skewness and they depend on cross-stream location. The F3TS simulations
 quantitatively predict the nature of the distribution functions and their variation with cross-stream
 location.
 \item The distribution functions for the other components of the velocity and angular velocity are
 Gaussian near the center, but they have exponential tails. The predictions of the F3TS simulations
 are in quantitative agreement with DNS for the distribution functions and their variation with
 cross-stream location.
\end{enumerate}

The important conclusions for the studies on rough particles, presented in section \ref{sec:rough}, are
\begin{enumerate}
 \item Though the mean velocity profiles for rough particle-wall collisions are not very different from
 those for smooth particle-wall collisions, the magnitude of the mean angular velocity decreases by a 
 factor of $10$. This is due to the slip in the mean velocity between the particles and the wall, which 
 results in an impulse that acts in the direction opposite to the rotation caused by the flow. There is
 also a change in the number density profile---the number density for rough particles is smaller at the 
 center and larger at the walls in comparison to smooth particles and walls.
 \item The mean square of the span-wise angular velocity fluctuations for rough particle-wall and 
 particle-particle collisions are two orders of magnitude larger than those for smooth collisions.
 The mean square of the stream-wise and cross-stream angular velocity fluctuations for rough
 particle-wall and particle-particle collisions are three orders of magnitude larger than those
 for smooth collisions. When the particle-particle collisions are rough and particle-wall collisions
 are smooth, there is an increase by 1-2 orders of magnitude in comparison to smooth particle-particle
 collisions.
 \item The mean square of the velocity fluctuations are also larger for rough particles in comparison
 to smooth particles. When particle-particle and particle-wall collisions are rough, the mean square
 of the fluctuating velocities in the stream-wise and span-wise directions are higher by a factor of $2$,
 and those in the cross-stream direction are higher by a factor of $10$, in comparison to smooth
 particles and walls.
 \item The F3TS results are in quantitative agreement with the DNS results for the number density,
 mean velocity and mean angular velocity profiles, and for the mean square of the stream-wise
 velocity and span-wise angular velocity fluctuations. The magnitudes and the trends of the other moments of
 the velocity and angular velocity fluctuations are correctly captured, though there are quantitative
 differences of about 10-20\%.
 \item The velocity distribution function in the stream-wise direction and the angular velocity distribution in
 the span-wise direction are highly non-Gaussian, and are very different from those for smooth particles. 
 The stream-wise velocity distribution is bimodal, and the span-wise angular velocity distribution is highly
 skewed with a sharp maximum. The velocity and angular velocity distributions in the other directions are
 Gaussian at the center and have long exponential tails. The decay of the exponential tails is much slower
 for rough particles in comparison to smooth particles.
 \item The F3TS simulations quantitatively predict all the velocity distribution functions, including the 
 bimodal stream-wise velocity distribution, the highly skewed span-wise angular velocity distribution and
 the exponential tails in the distributions for the other components of the velocity and angular velocity.
\end{enumerate}

In summary, the present study reveals the importance of accurate descriptions of particle-particle and, especially, particle-wall
collisions for accurate predictions of the dynamics of the particle phase in particle-gas suspensions. It also shows that the 
properties of the particle phase are accurately predicted by the F3T model, despite the approximations made in the model.

\appendix
\label{App}
\section{Validation of DNS}
\label{sec:validation}
\begin{figure}
	\includegraphics[width=1.0\textwidth]{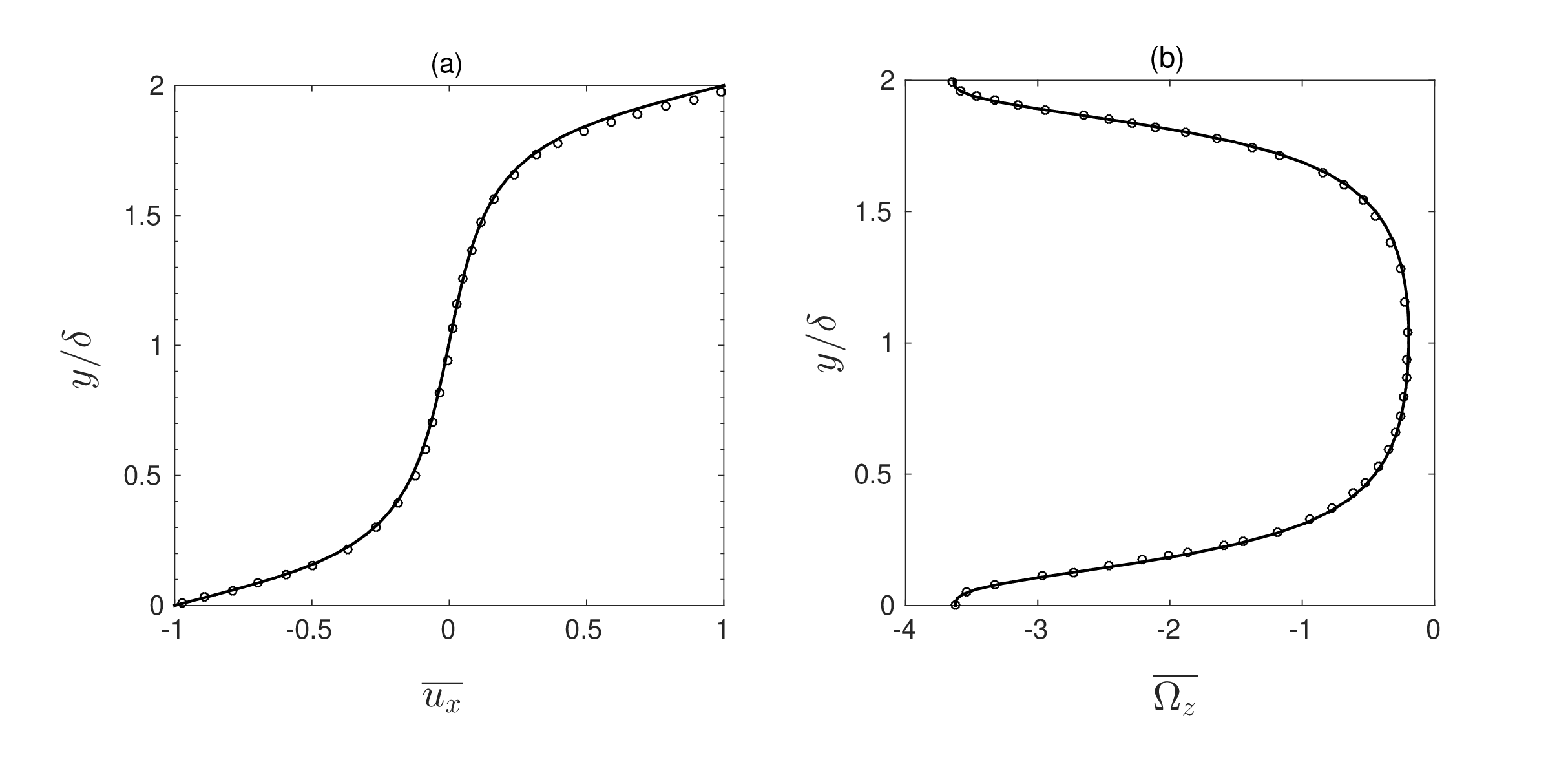}
	\caption{Variation of (a) fluid phase mean velocity $\overline{u}_x$ and  (b) mean vorticity $\overline{\Omega}_z$. 
	The thick lines '-' represent the present DNS simulations which are validated by DNS results of \citep{komminaho1996very}, represented by symbols 'o'. 
	}
	\label{fig:mean_fluid_vel}
\end{figure}

\begin{figure}
\begin{subfigure}{0.4\linewidth}
{\includegraphics[width=1.0\textwidth]{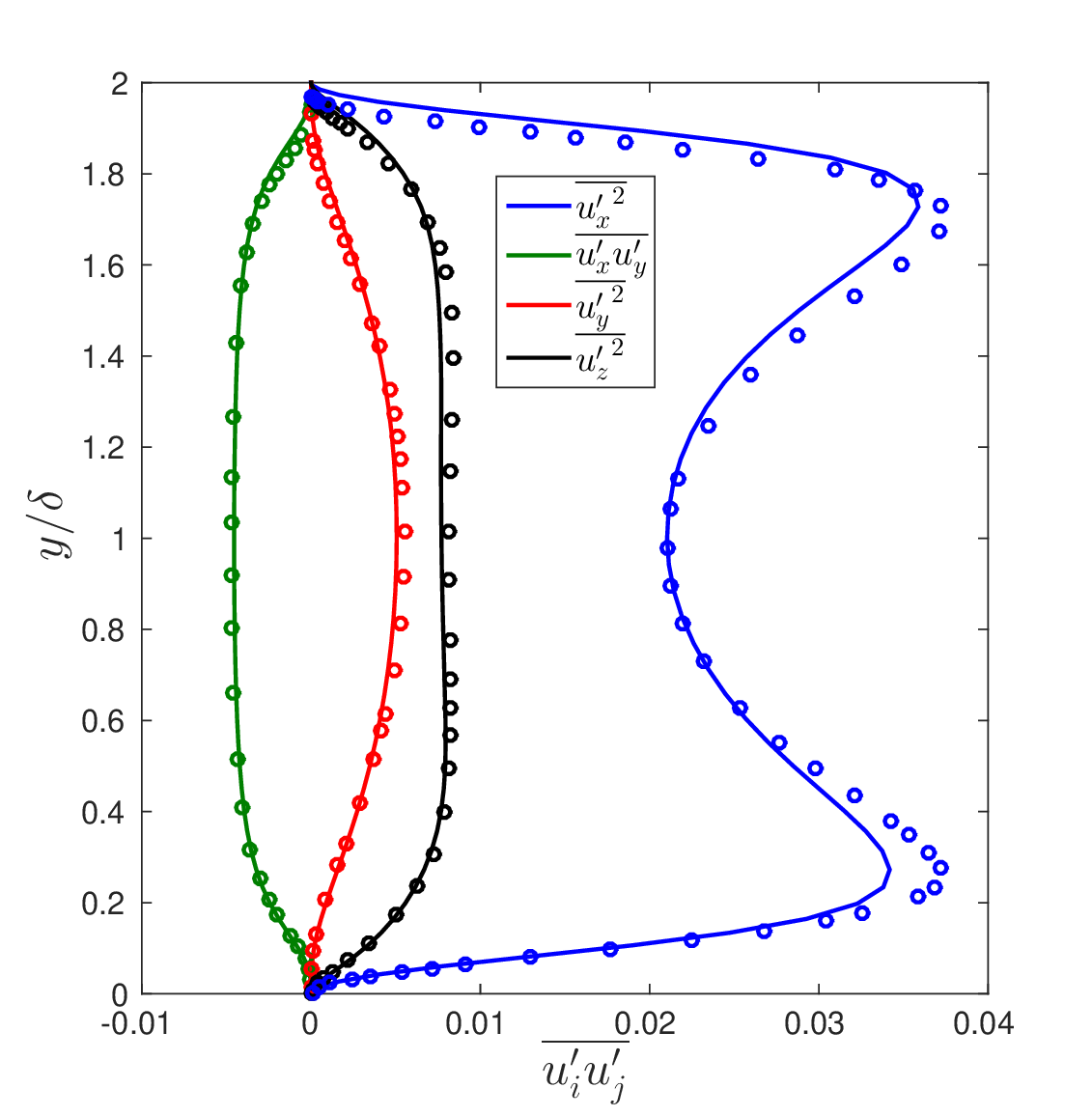}}
	\caption*{(a)}
\end{subfigure}
\begin{subfigure}{0.5\linewidth}
{\includegraphics[width=1.0\textwidth]{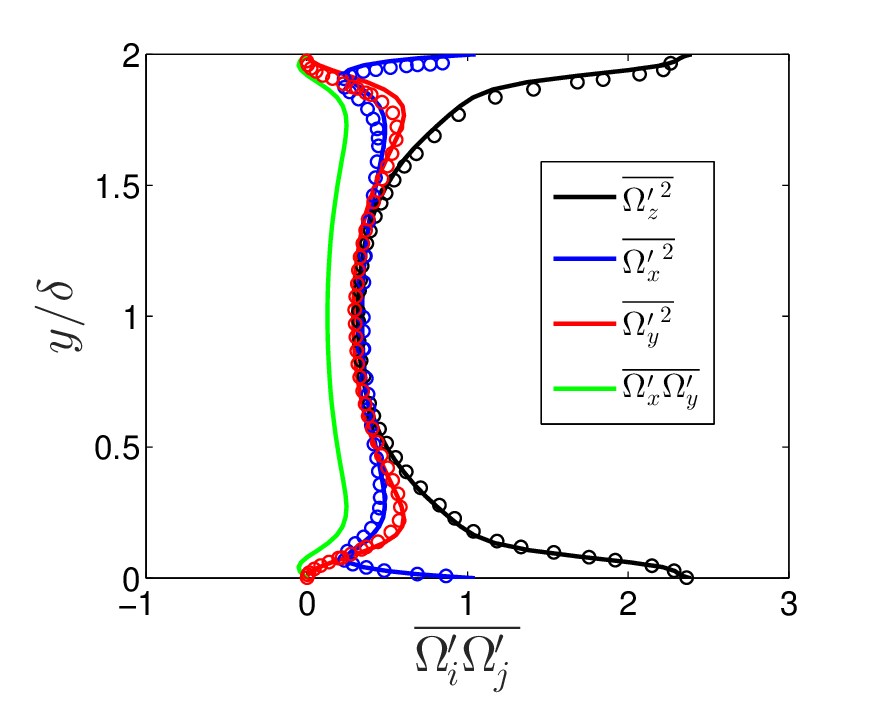}}
\caption*{(b)}
\end{subfigure}
	\caption{Variation of fluid mean square statistics along channel-width $y^+$ (a) second moment of velocity $\overline{u_i' u_{j}'}$, (b) mean square angular velocity along $\overline{\Omega_i' \Omega_j'}$. The thick lines '-' represent the present DNS simulations which are validated by DNS results of \citep{komminaho1996very}, represented by symbols 'o'. 
	}
	\label{fig:fluidms}
\end{figure}

The velocity statistics of the present simulations have been validated against the simulations of 
\citep{komminaho1996very}. The mean and root mean square of the fluid velocity are shown in
in figures~\ref{fig:mean_fluid_vel} and \ref{fig:fluidms}. The mean velocity in the flow direction, figure \ref{fig:mean_fluid_vel} (a), is antisymmetric about the channel center-line, while the mean vorticity, \ref{fig:mean_fluid_vel} (b) in 
the span-wise direction is symmetric with a maximum at the center. The stream-wise
mean square velocity, shown in figure \ref{fig:fluidms} (a), exhibits the characteristic near-wall maximum, while the cross-stream and span-wise velocities are much smaller in magnitude. The Reynolds stress is almost a constant at the center of the channel, as shown in figure \ref{fig:fluidms} (a).

 The mean square of
the fluid velocity and angular velocity fluctuations are shown in figure \ref{fig:fluidms}(a) and
(b). As expected, the mean square of the vorticity fluctuations in the span-wise direction is the
largest in magnitude, while those in the other two directions are smaller. The vorticity component 
$\Omega_y = (\partial u_x/\partial z) - (\partial u_z/\partial x)$ is necessarily zero at the wall because
of the no-slip condition, but the mean square vorticity $\overline{\Omega_x^2}$ shows a 
large increase at the wall. 

\section{Fluctuating torque formulation}
Fluctuation torque in the x, y, and z direction are computed using the following relations.
\begin{equation}
	\label{eq5.39}
	T_x=\frac{\sqrt{2D^{\omega}_{{xx}}}}{\sqrt{\Delta t}}\zeta_1
\end{equation} 





\begin{equation}
	\label{eq5.45}
	T_y=\frac{\sqrt{2D^{\omega}_{{yy}}}}{\sqrt{\Delta t}}\Bigg[\frac{D^{\omega}_{{xy}}}{\sqrt{D^{\omega}_{{xx}}D^{\omega}_{{yy}}}}\zeta_1+ \sqrt{1-\frac{(D^{\omega}_{xy})^2}{D^{\omega}_{{xx}}D^{\omega}_{{yy}}}}\zeta_2\Bigg]
\end{equation}\\

\begin{eqnarray} 
	\label{eq5.54}
	T_z & = & \frac{\sqrt{2D^{\omega}_{{zz}}}}{\sqrt{\Delta t}}\left[\frac{D^{\omega}_{{xz}}}{\sqrt{D^{\omega}_{{xx}}D^{\omega}_{{zz}}}}\zeta_1+{\frac{\frac{D^{\omega}_{{yz}}}{\sqrt{D^{\omega}_{{yy}}D^{\omega}_{{zz}}}}-\frac{D^{\omega}_{{xz}}}{\sqrt{D^{\omega}_{{xx}}D^{\omega}_{{zz}}}}\frac{D^{\omega}_{{xy}}}{\sqrt{D^{\omega}_{{xx}}D^{\omega}_{{yy}}}}}{\sqrt{1-\frac{(D^{\omega}_{xy})^2}{D^{\omega}_{{xx}}D^{\omega}_{{yy}}}}}}\zeta_2+ \right. \nonumber \\
	&& \mbox{}\left.{\sqrt{1-\frac{(D^{\omega}_{xz})^2}{D^{\omega}_{{xx}}D^{\omega}_{{zz}}}-\frac{\left[\frac{D^{\omega}_{{yz}}}{\sqrt{D^{\omega}_{{yy}}D^{\omega}_{{zz}}}}-\frac{D^{\omega}_{{xz}}}{\sqrt{D^{\omega}_{{xx}}D^{\omega}_{{zz}}}}\frac{D^{\omega}_{{xy}}}{\sqrt{D^{\omega}_{{xx}}D^{\omega}_{{yy}}}}\right]^2}{\left[1-\frac{(D^{\omega}_{xy})^2}{D^{\omega}_{{xx}}D^{\omega}_{{yy}}} \right]}}}\zeta_3 \right] \end{eqnarray} 
	In the above equations $\zeta_1$, $\zeta_2$ , and  $\zeta_3$ are the random deviates with zero mean and unit variance.

\section{Integral times for the diffusivities}
\label{sec:integraltimes}
 The Eulerian and Lagrangian integral times, which are the ratios of the integrals of
 the autocorrelation functions (equations \ref{eq3.1}-\ref{eq3.2}), are shown in table 
 \ref{table:1} for the 
 stream-wise velocity and the span-wise vorticity, and \ref{table:2} for other components
 of the autocorrelation function. The Lagrangian correlations were extracted from one-way coupled DNS 
 with perfectly smooth ($\beta=-1.0$) and elastic particles ($e=1.0$) of density 2000 kg/m$^3$ ($\mbox{St}\sim50$ based on fluid-integral time-scale $\delta/\bar{u}$) in the dilute limit ($\phi=0.0001$).
 In order to avoid the simulation artifact due to the migration of particles from adjacent zones, the Particle-Lagrangian correlations could only be analysed over a finite time-period (for angular velocity correlations $\tau=30$ or $\tau/\tau_r=2$ and $\tau=50$ or or $\tau/\tau_v=1$ for velocity correlations)
beyond which the tails become noisy. 

The integral times in the Lagrangian frame are well fitted using an exponential function. The integral times in the Eulerian reference frame are also
 exponential, with the exception of $R_{xx}$ and $R^{\Omega}_{zz}$, which are stretched 
 exponentials, as discussed in section \ref{sec:Fluid}. 
 The Lagrangian correlations were calculated over 3000 time frames,
with a minimum of 50 particles per frame in each zone. All the correlations were fitted and tabulated in table 
\ref{table:1} and \ref{table:2}. The upper limit for the velocity autocorrelation function
is set to $\tau=50$ ($\delta/U$ unit) or $\tau/\tau_v=1$, which is about 7 times the value at which correlation function decays to $\mbox{e}^{-1}$. In case of vorticity correlation function the upper limit is set to to $\tau=20$ or $\tau/\tau_v=1.33$, which is also about 7 times the value of the time at which the autocorrelation function decays
to $\mbox{e}^{-1}$.
\begin{table}
\scriptsize
	\centering
	\begin{tabular}{||c | c | c | c | c | c | c | c||} 
	\hline
	Correlation & $y/\delta$ & \multicolumn{1}{|p{2cm}|}{\centering Time-scale from Particle\\-Lagrangian Correlation} & \multicolumn{1}{|p{2cm}|}{\centering Fitting Function of \\ Particle-Lagrangian \\ Correlation} & $R^2$ &  \multicolumn{1}{|p{2cm}|}{\centering Fitting Function of \\ Eulerian Correlation} & $R^2$ & \multicolumn{1}{|p{2cm}|}{\centering Time-scale from \\Eulerian Correlation}  \\ 
				\hline 
  $R_{uxx}$ & 0.1-0.2 & 11.275
 & $\exp{(-0.1045\tau)}$ & 0.9956 & $\exp{(-0.09556\tau^{0.3237})}
$ & 0.96 & 14.22
 \\  
  $R_{uxx}$ & 0.2-0.4 & 10.3221
 & $\exp{(-0.08868\tau)}$ & 0.9956 & $\exp{(-0.4098\tau^{0.3417})}
$ & 0.965 & 16.44 \\  
  $R_{uxx}$ & 0.4-0.6 & 10.7666
 & $\exp{(-0.09007\tau)}$ & 0.9958 & $\exp{(-0.3483\tau^{0.3643})}
$ & 0.966 & 18.12 \\ 
  $R_{uxx}$ & 0.6-1.0 & 10.1488
 & $\exp{(-0.354\tau)}$ & 0.9978 & $\exp{(-0.2944\tau^{0.3911})}
$ & 0.967 & 19.66 \\ 
  $R^\Omega_{ zz}$ & 0.1-0.2 & 6.0584 & $\exp{(-0.1045\tau)}$ & 0.976 & $\exp{(-0.951\tau^{0.261})}
$ & 0.978 & 4.12 \\ 
	\hline
	\end{tabular}
	\caption{Stretched Exponential Fitting of Eulerian Correlations }
	\label{table:1}
\end{table}

\begin{table}
	\centering
	\begin{tabular}{||c | c | c ||} 
		\hline
		Correlation & $y/\delta$ zone  & \multicolumn{1}{|p{2cm}|}{\centering Time-scale from \\Eulerian Correlation} \\ [0.5ex] 
		\hline\hline
		$R_{uxx}$ & 0.1-0.2  & 14.22 \\ 
		$R_{uxx}$ & 0.2-0.4  & 16.44 \\
		$R_{uxx}$ & 0.4-0.6 & 18.12 \\
		$R_{uxx}$ & 0.6-1.0  & 19.66 \\ \hline
		$R_{uyy}$ & 0.1-0.2   & 2.053 \\ 
		$R_{uyy}$ & 0.2-0.4  & 3.172  \\
		$R_{uyy}$ & 0.4-0.6 & 4.96  \\
		$R_{uyy}$ & 0.6-1.0  & 7.92 \\ \hline
		$R_{uzz}$ & 0.1-0.2  & 5.181  \\ 
		$R_{uzz}$ & 0.2-0.4  & 7.289  \\
		$R_{uzz}$ & 0.4-0.6  & 8.143  \\
		$R_{uzz}$ & 0.6-1.0  & 7.391 \\ \hline
		$R^\Omega_{ xx}$ & 0.1-0.2  & 1.997  \\
		$R^\Omega_{ xx}$ & 0.2-0.4  & 2.164  \\ 
		$R^\Omega_{ xx}$ & 0.4-0.6 & 2.625  \\
		$R^\Omega_{ xx}$ & 0.6-1.0  & 3.271  \\ \hline
		$R^\Omega_{ yy}$ & 0.1-0.2  & 2.883  \\
		$R^\Omega_{ yy}$ & 0.2-0.4  & 3.163  \\ 
		$R^\Omega_{ yy}$ & 0.4-0.6   & 3.41  \\
		$R^\Omega_{ yy}$ & 0.6-1.0  & 3.88  \\ \hline
		$R^\Omega_{ zz}$ & 0.1-0.2  & 4.12 \\
		$R^\Omega_{ zz}$ & 0.2-0.4   & 2.21   \\ 
		$R^\Omega_{ zz}$ & 0.4-0.6  & 2.21   \\
		$R^\Omega_{ zz}$ & 0.6-1.0  & 2.69   \\[1ex] \hline 
			\end{tabular}
	\caption{Time-scales from correlations in Eulerian reference frames}
	\label{table:2}
\end{table}

\noindent {\bf Acknowledgements:}
The authors thank Science and Engineering Research Board, Government of India, for funding
through research grant CRG/2018/002968. VK thanks Science and Engineering Research Board, Government 
of India, for funding through research grant JBR/2021/000021/SSC, and Synopsys for financial support.

\noindent {\bf Declaration of interests:} The authors report no conflict of interest.
\bibliography{prf_ref}

%
%
%
%

\end{document}